\documentclass[journal,comsoc,onecolumn,12pt]{IEEEtran}

\renewcommand{\baselinestretch}{1.6}

\usepackage[T1]{fontenc}
\usepackage{graphicx} 
\usepackage{subfig}
\usepackage{amssymb,amsfonts,amsthm}
\usepackage{amsmath}
\usepackage{array,booktabs,multirow}
\usepackage{bm}
\usepackage{cite}
\usepackage[cmintegrals]{newtxmath}

\newtheorem{theorem}{Theorem}

\newtheorem{remark}{Remark}

\newcommand{\rel}{ {\sf R} }
\newcommand{\eqdef}{ := }
\newcommand{\egd}{\emph{e.g.}, }
\newcommand{\ied}{\emph{i.e.}, }
\newcommand{\oa}{the OA }
\newcommand\floor[1]{\lfloor#1\rfloor}

\newcommand\indep{\protect\mathpalette{\protect\independenT}{\perp}}
\def\independenT#1#2{\mathrel{\rlap{$#1#2$}\mkern2mu{#1#2}}}
\newcommand{\overbar}[1]{\mkern 1.5mu\overline{\mkern-1.5mu#1\mkern-1.5mu}\mkern 1.5mu}

\allowdisplaybreaks[4]

\makeatletter
\def\hlinewd#1{%
	\noalign{\ifnum0=`}\fi\hrule \@height #1 \futurelet
	\reserved@a\@xhline}
\makeatother

\begin{document}
%
\title{Dual-Band Fading Multiple Access Relay Channels}
%
%
%

\author{Subhajit~Majhi~and~Patrick~Mitran,\\
	 \IEEEauthorblockA{\small{The Department of Electrical and Computer Engineering, University of Waterloo, Waterloo, ON N2L 3G1, Canada \\ (e-mail: smajhi@uwaterloo.ca; pmitran@uwaterloo.ca).}}
        \thanks{The work was supported by the Natural Sciences and Engineering Research Council of Canada.}%
}

\maketitle

		

\vspace*{-15mm}
\begin{abstract}
		Relay cooperation and integrated microwave and millimeter-wave (mm-wave) dual-band communication are likely to play key roles in 5G. In this paper, we study a two-user uplink scenario in such dual-bands, modeled as a multiple-access relay channel (MARC), where two sources communicate to a destination  assisted by a relay. However, unlike the microwave band, transmitters in the mm-wave band must employ highly directional antenna arrays to combat the ill effects of severe path-loss and small wavelength. The resulting mm-wave links are point-to-point and highly directional, and are thus used to complement the microwave band by transmitting to a specific receiver. For such MARCs, the capacity is partially characterized for sources that are \emph{near} the relay in a \emph{joint} sense over both bands. We then study the impact of the mm-wave spectrum on the performance of such MARCs by characterizing the transmit power allocation scheme for  phase faded mm-wave links that maximizes the \emph{sum-rate} under a total power budget.  The resulting scheme adapts the link transmission powers to channel conditions by transmitting in different \emph{modes}, and all such modes and corresponding conditions are characterized. Finally, we study the properties of the optimal link powers  and derive practical insights.
\end{abstract}


%
\IEEEpeerreviewmaketitle

\section{Introduction}
%
%
%
%
 Fueled by the ever increasing demand for bandwidth-hungry applications, global wireless traffic is expected to continue its rapid growth \cite{itu_m2083}. However, due to scarce microwave bandwidth (\ied sub-$6$ GHz spectrum) current 4G technologies are unlikely to be able to support the anticipated massive growth in traffic \cite{4G_5}. To tackle this challenge, several new technologies are being studied to be potentially incorporated into 5G standards. Among these, a key technology is to integrate the vast bandwidth in the $28-300$ GHz frequency range, referred to as the millimeter wave (mm-wave) band, with  sub-$6$ GHz spectrum \cite{docomo,queue_mm,survey2015}, and provide cellular access jointly over these two bands.

 	Transmission in the mm-wave band differs from that in the conventional microwave band in that omnidirectional mm-wave transmission suffers from much higher power loss and absorption. Thus, a transmitter must use beamforming via highly directional antenna arrays to reach a receiver \cite{brady_beamspace}. Due to  the small wavelength at mm-wave frequencies and large path loss, beamforming typically creates links that have a strong line-of-sight (LoS) component and only a few, if any, weak multi-path components. Such mm-wave links are inherently point-to-point, and are well modeled as AWGN links \cite{directed_1,caire_121_relay,multibeam_transmitter}. 
 Although mm-wave links support high data rates due to their large bandwidths, they provide limited coverage,  whereas microwave links typically provide reliable coverage and support only moderate data rates. Thus, in a dual-band setting, these two bands mutually complement each other: conventional traffic and control information can be reliably communicated in the microwave band, and high data-rate traffic can be communicated via the mm-wave links \cite{directed_8,mm&microwave,docomo,hetnet_3,majhi1,s16060892,queue_mm,survey2015,mm&microwave2}.
 
 In future 5G networks, access via dual microwave and mm-wave bands will likely be a key technology, and hence they have been subject to much investigation recently. For example, studies  as in \cite{directed_8,hetnet_3,s16060892,mm&microwave} focus on improving network layer metrics such as the number of served users, throughput, and link reliability, etc., while studies as in \cite{joint_opn_2,resource_3,resource_4} focus on improving physical layer metrics such as the achievable rates and outage probability. Moreover, the emergence of dual-band modems from Intel \cite{intel} and Qualcomm \cite{qualcomm}, and practical demonstrations such as that in the $3$ GHz-$30$ GHz dual-bands in \cite{queue_mm} clearly illustrate the immense potential of such networks.	However, few studies have been reported on the information-theoretic limits of multi-user dual-band networks \cite{majhi_IC_journal}, which are crucial in identifying the limits of achievable rates, simplified encoding schemes, etc., in practical dual-band networks.  For example, the study on the two-user interference channel over such integrated dual-bands \cite{majhi1} has shown that forwarding interference to the non-designated receivers through the mm-wave links can improve achievable rates considerably.  Moreover, relay cooperation, which already plays a key role in microwave networks, will likely play a vital role in such dual-band networks as well, especially to offset impairments such as blockage in the mm-wave band  \cite{mmwave_relay,hetnet_3,caire_121_relay, gaussian_121_2}.
 
 Thus motivated, we study the two-user Gaussian multiple-access relay channel (MARC) over dual microwave and mm-wave bands, which models uplink scenarios, \egd fixed wireless access \cite{kramer_relay} which is expected to eliminate last mile wired connections  to end users. In this case, the base station will communicate with a  fixed access-point that is equipped with the hardware necessary for dual-band communication including mm-wave beamforming, which will likely be located outside a building and will provide high data rate access to users inside the building (end users).  As such, the dual-band MARC can model relay-assisted uplink from two such fixed access-points located in nearby buildings. In the future, when mobile handsets are equipped with dual-band communication capable hardware, the dual-band MARC can also model relay-aided cellular uplink from mobile users.
 
 In this MARC, two sources communicate to a destination with the help of a relay over dual microwave and mm-wave bands. In the microwave band, transmissions from both sources are \emph{superimposed} at the relay and at the destination as in a conventional MARC (c-MARC)  \cite{kramer_relay}. In contrast, since mm-wave links are highly directional \cite{directed_1}, when a transmitter in the mm-wave band transmits specifically to the relay or the destination, the resulting mm-wave link causes minimal to no interference to the unintended receiver \cite{Multibeam_1,multibeam_transmitter}. In fact, a mm-wave transmitter can create two  \emph{parallel} non-interfering links via beamforming, and then communicate with both relay and destination \emph{simultaneously} \cite{Multibeam_1,brady_beamspace,multibeam_transmitter}. Therefore, in this work a mm-wave transmitter is modeled as being able to create two such parallel non-interfering AWGN links to simultaneously transmit to the relay and the destination, while a mm-wave receiver is modeled as being able to simultaneously receive transmissions from multiple mm-wave transmitters via \emph{separate} mm-wave links \cite{Multibeam_receive} with negligible inter-link interference. 
 
 It is natural to ask whether a user (or source) in the mm-wave band should transmit to the relay, the destination, or both. Depending on whether each of the two sources transmits to only the relay, only the destination, both, or none,  $16$ different models are possible. The general model that includes all microwave and mm-wave links is referred to as the  destination-and-relay-linked MARC (DR-MARC), where the two sources ($\mathsf{S}_1$ and $\mathsf{S}_2$) simultaneously communicate to the destination  ($\mathsf{D}$) via the mm-wave $\mathsf{S}_1$-$\mathsf{D}$ and $\mathsf{S}_2$-$\mathsf{D}$  \emph{direct} links as well as to the relay ($\mathsf{R}$)  via the mm-wave $\mathsf{S}_1$-$\mathsf{R}$ and $\mathsf{S}_2$-$\mathsf{R}$  \emph{relay} links. Since all other models with varying mm-wave link connectivity can be obtained from the DR-MARC by setting the relevant transmit powers to zero, they are not defined explicitly. However, the model where transmit powers in the mm-wave direct links are set to zero is an important one and referred to as the relay-linked MARC (R-MARC).

 In addition to mm-wave links, the dual-band MARC also consists of an underlying conventional microwave band c-MARC. The capacity of such an individual c-MARC was partially characterized under phase and Rayleigh fading \cite{kramer_relay,dabora_MARC}, and therefore, we assume that the dual-band MARC is subject to a general ergodic fading where the phase of the fading coefficients are i.i.d. uniform  in $[0, 2\pi)$, similar to  phase and Rayleigh fading. The general fading contains phase and Rayleigh fading as special cases, and  can model a  range of channel impairments. For example, phase fading models the effect of oscillator phase noise in high-speed time-invariant communications \cite{phase_noise}, the effect of phase-change due to slight transmitter-receiver misalignment in LoS dominant links \cite{patick_relay}, etc.,  while Rayleigh fading models the effect of rich scattering \cite{rayleigh}. 
 
 In \cite{kramer_relay}, the conventional c-MARC was classified into the \emph{near c-MARC} and the \emph{far c-MARC} cases. In the near c-MARC, the sources are \emph{near} the relay in that the source-relay channels are \emph{stronger} than the source-destination channels in the sense of \cite[Theorem~9]{kramer_relay}, and thus the capacity of the near c-MARC was characterized. Naturally, the far c-MARC case is complementary to the near  case. Here, we similarly classify the dual-band MARCs (DR-MARC and R-MARC) based on whether the underlying c-MARC in the microwave band is a near or a far c-MARC in the sense of \cite{kramer_relay}.
 
 First, we consider the DR-MARC where the sources simultaneously transmit in both the mm-wave relay and mm-wave direct links. We show that irrespective of whether the underlying c-MARC is a near or a far c-MARC, its capacity can be \emph{decomposed} into the capacity of the \emph{underlying} R-MARC (that consists of the c-MARC and the two mm-wave relay links) and the two mm-wave direct links. Hence, it is sufficient to focus on the R-MARC.  The capacity of the R-MARC with near underlying c-MARC are characterized under the same conditions as in \cite{kramer_relay} and thus does not need additional conditions on the mm-wave links. Therefore, we focus primarily on R-MARCs with \emph{far} underlying c-MARC where the mm-wave links play a key role, and for such  R-MARCs, we find sufficient channel conditions under which its capacity is characterized by an achievable scheme.

 The DR-MARC is a building block for future dual-band multiuser networks. Since, its performance will be significantly affected by the mm-wave links due to their large bandwidths \cite{majhi_IC_journal, resource_4, mm&microwave}, it is useful to understand how allocating the mm-wave band resources optimizes the performance, similar to other multiuser networks \cite{majhi_IC_journal,caire_121_relay,lgr_mai_vu}.  Hence, to quantify the impact of the mm-wave spectrum on the performance of the DR-MARC, we study the power allocation strategy for the mm-wave direct and relay links (subject to a power budget) that maximizes the achievable sum-rate. 
 
 The contributions of this paper is summarized as follows. \begin{itemize}
 	\item We decompose the capacity of the DR-MARC into the capacity of the underlying R-MARC and two direct links. This shows that irrespective of whether the underlying c-MARC is a near or a far c-MARC, operating the R-MARC independently of the direct links  is optimal. 
 	\item We derive an achievable region for the R-MARC. Then, for R-MARCs with \emph{far} underlying c-MARC, we obtain sufficient conditions under which this achievable scheme is capacity achieving. 
 	
 	\item We characterize the optimal power allocation scheme (OA) for the mm-wave direct and relay links that maximizes the sum-rate achievable on the DR-MARC with the aforementioned achievable scheme. For intuition, we partition the range of the total power budget ($P$) into several \emph{link gain regimes} (LGR) based on whether $P$ satisfies certain channel conditions, and show that \oa allocates link powers in different \emph{modes} in each LGR. We obtain all such LGRs and modes of power allocation which reveal useful insights.
 	
 	We observe that for DR-MARCs with \emph{near} underlying c-MARC, \oa allocates $P$ entirely to the direct links for all $P \geq 0$. However, for DR-MARCs with \emph{far} underlying c-MARC, we observe the following:

 	$\quad\rm{(i)}$ when $P$ is \emph{smaller} than a certain \emph{saturation threshold} ($\mathsf{P}_{\sf sat}$), for the direct and relay links of each source, \oa allocates powers following a Waterfilling (WF) approach. Specifically, for sufficiently small $P$, \oa allocates $P$ entirely to the strongest of the direct and relay links of a source, and as $P$ increases, power is eventually allocated  to the remaining links.  Thus, for $P < \mathsf{P}_{\sf sat}$, each link-power either increases piecewise linearly with $P$, or remains zero.
 	
 	$\quad\rm{(ii)}$ when $P \geq \mathsf{P}_{\sf sat}$, \emph{saturation} occurs where the relay link powers are \emph{constrained} to satisfy a certain saturation condition. As $P$ increases beyond $\mathsf{P}_{\sf sat}$, the direct link powers grow unbounded with $P$, while the relay link powers vary with $P$ as follows. There exists a threshold  $\mathsf{P}_{\sf fin} \geq \mathsf{P}_{\sf sat}$, such that  $\sf (a)$ if one relay link is \emph{significantly stronger} than the other (in a sense to be defined later), then for all $P \geq \mathsf{P}_{\sf fin}$, power in the stronger relay link remains fixed at a constant level and that in the weaker relay link at zero, and $\sf (b)$ if the relay link is only \emph{stronger} but not significantly stronger,  for all $P \geq \mathsf{P}_{\sf fin}$, power in the stronger and the weaker relay links monotonically increase and decrease respectively, and approach constant levels.
 	
 	$\quad\rm{(iii)}$ if the mm-wave bandwidth is large and the power received at the destination from the relay via the mm-wave link is also large, allocating power as in the WF-like solution is optimal for all practical $P$, and saturation only occurs for large values of $P$.
 \end{itemize}

 This paper is organized as follows. The system model is defined in Section II. The results on the DR-MARC and the R-MARC are presented in Section III and Section IV respectively. The optimum sum-rate problem is presented in Section V, while in Section VI insights are derived from the link gain regimes. Finally, conclusions are drawn in Section VII.
 
 \textit{Notation}: The sets of real, non-negative real and complex numbers are denoted by $\mathbb{R}, \mathbb{R}_+$ and $ \mathbb{C}$. Vectors are generally denoted in bold (\egd $\mathbf{p}$) with $\mathbf{p} \succeq \mathbf{0}$ denoting that each $p_i \in \mathbb{R}_+$. Random variables (RVs) and their realizations are denoted by upper and lower cases (\egd $X$ and $x$). Specifically, $Z \sim \mathcal{CN}(0, \sigma^2)$ denotes a  circularly symmetric complex Gaussian (CSCG) RV with mean $0$ and variance $\sigma^2$, and $\Theta \sim \mathcal{U}[0, 2\pi)$ denotes a uniformly distributed RV in $[0, 2\pi)$. Also, $\mathbb{E}[.]$ denotes expectation, while $\floor{x}$ denotes the greatest integer no larger than $x$, and $\mathsf{C}(p) \eqdef \log(1+p)$.
 
\section{System Model} 

\begin{figure*}[!t]  
	\captionsetup[subfloat]{captionskip=0mm}
	\centering    
	\subfloat[]{\includegraphics[width=4.6cm, height=3.8 cm]{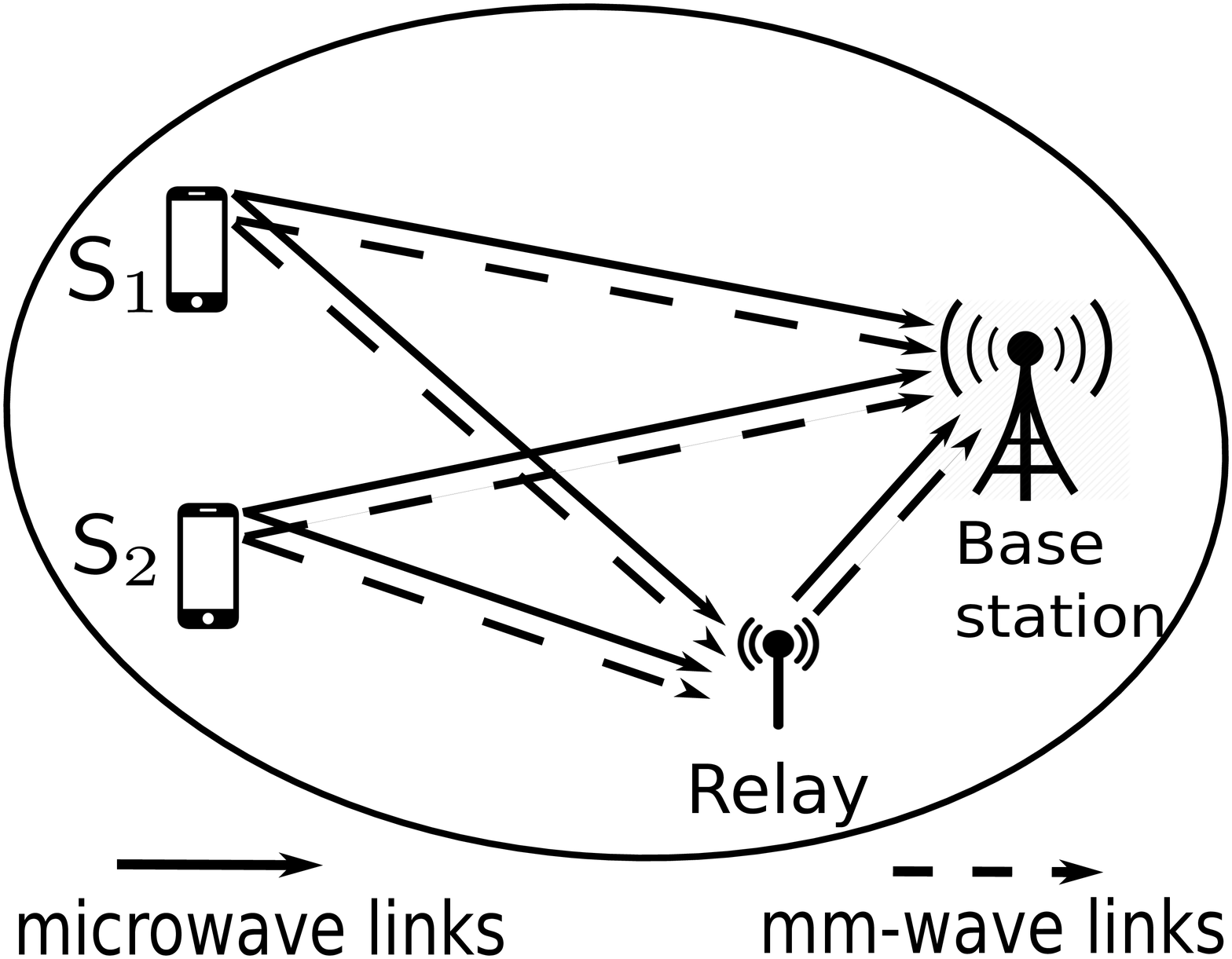}%
		\label{system_model:a}} \hfil \hspace*{3mm}
	\subfloat[]{\includegraphics[width=3.5cm, height=4 cm]{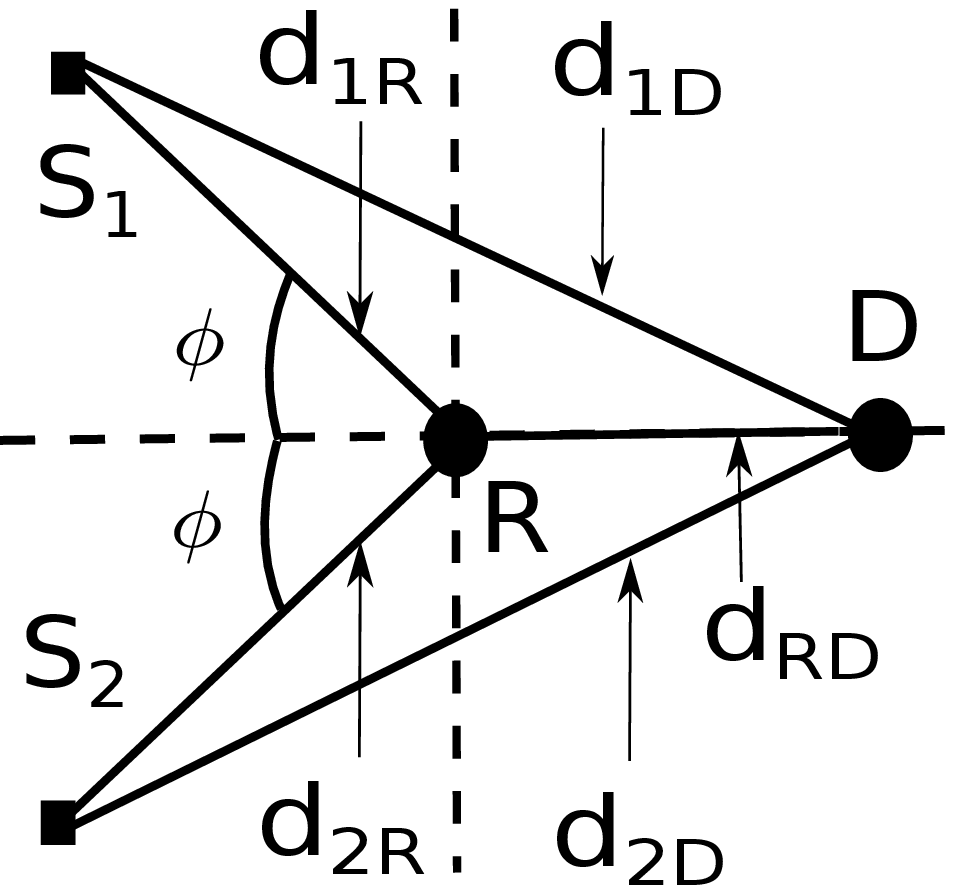}%
		\label{fig2:a}}  \hfil \hspace*{3mm}
	\subfloat[]{\includegraphics[width=8.5cm,height=4.5 cm]{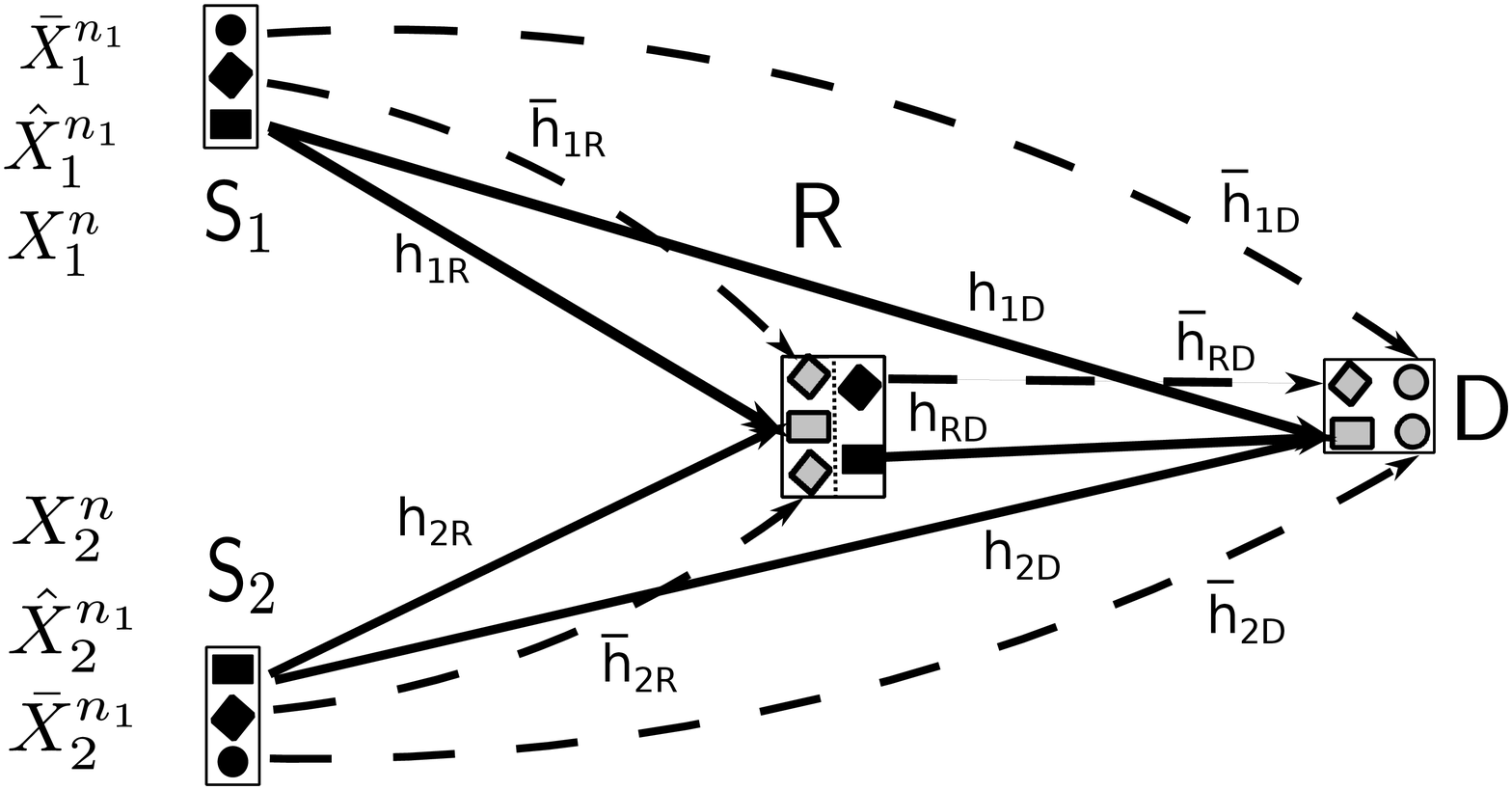}%
		\label{system_model:b}}
	\caption{\small (a) Example of the DR-MARC in a cellular uplink. (b) A 2-D geometry of the DR-MARC where the relay and the destination are located on the x-axis, and the sources are located symmetrically on either side of the  x-axis.  The distance between nodes $m$ and $t$ is denoted by ${\sf d}_{mt}$ where $m  \in  \{1,2,\mathsf{R}\}, t \in \{\mathsf{D},\mathsf{R}\}, m  \neq   t$. (c) System model of the Gaussian DR-MARC: solid line and dashed line denote microwave band and mm-wave band transmissions respectively.} 
	\label{system_model:fig}
\end{figure*}

We consider a relay-assisted two-user uplink scenario as in Fig.~\ref{system_model:a} which is modeled as the DR-MARC as in Fig.~\ref{system_model:b}. Note that a bandwidth mismatch factor (BMF) $\alpha$ may exist between the two bands such that for $n$ accesses of the microwave  band, the mm-wave  band is accessed $n_1(n) \eqdef \floor{\alpha n}$ times. To communicate a message $M_k$ from source $\mathsf{S}_k$,  it is encoded into three codewords, ${X}_k^{n}(M_k),\hat{X}_k^{n_1}(M_k)$ and $ \bar{X}_k^{n_1}(M_k)$, of lengths $n, n_1$ and $n_1$ respectively. Then,  ${X}_k^{n}(M_k)$ is transmitted towards $\mathsf{D}$ by using the microwave (first) channel $n$ times, and due to the nature of this band, ${X}_1^{n}(M_1)$ and ${X}_2^{n}(M_2)$ superimpose at $\mathsf{D}$ and at $\mathsf{R}$ as in the c-MARC \cite{kramer_relay}. Meanwhile, in the mm-wave (second) band, $\hat{X}_k^{n_1}(M_k)$ is transmitted to $\mathsf{R}$ through the $\mathsf{S}_k$-$\mathsf{R}$ relay link and $\bar{X}_k^{n_1}(M_k)$ to $\mathsf{D}$ through the $\mathsf{S}_k$-$\mathsf{D}$ direct link simultaneously by using the links $n_1$ times. The relay aids by creating codewords $X_{\mathsf{R}}^n$ and $\bar{X}_{\mathsf{R}}^{n_1}$ from its received signals and transmitting them to $\mathsf{D}$ in both bands.

We now define the channel model of the Gaussian DR-MARC. As in \cite{kramer_relay}, in the first band, the channel outputs at $\mathsf{D}$ and $\mathsf{R}$ at the $i$-th use of the band are  given by \begin{align}
{Y}_{\mathsf{D},i} &= H_{1\mathsf{D},i} {X}_{1,i} + H_{2\mathsf{D},i}{X}_{2,i} + H_{\mathsf{R}\mathsf{D},i} X_{\mathsf{R},i} + {Z}_{\mathsf{D},i} \\ 
{Y}_{\mathsf{R},i}& = H_{1\mathsf{R},i} {X}_{1,i} + H_{2\mathsf{R},i}{X}_{2,i} + {Z}_{\mathsf{R},i},   \quad i=1,\ldots,n, \end{align}  where $H_{mt,i} \in \mathbb{C}$ are channel fading coefficients from the transmitter at node $m$ to the receiver at $t$, $m \neq t, m \in \{1,2,\mathsf{R}\}, t \in \{\mathsf{R},\mathsf{D}\}$, and input $X_{m,i} \in \mathbb{C}$ are block power constrained, $\frac{1}{n} \sum_{i=1}^{n} \mathbb{E}[|X_{m,i}|^2] \leq P_m, m \in \{1,2,\mathsf{R}\}$. Also, the noise RVs are ${Z}_{\mathsf{R},i} \sim \mathcal{CN}(0,1)$, i.i.d., and ${Z}_{\mathsf{D},i} \sim \mathcal{CN}(0,1)$, i.i.d.

In the second band, the outputs of the $\mathsf{S}_k$-$\mathsf{R}$ relay links at the relay $\mathsf{R}$ are modeled as  
\begin{align}  
 {\bar{Y}}_{k\mathsf{R},\ell}  =  \bar{H}_{k\mathsf{R},\ell} \hat{X}_{k,\ell} + \bar{Z}_{k\mathsf{R},\ell},   \quad k \in \{1,2\}, \quad   \ell=1,\ldots,n_1, 
\end{align} and the outputs of the $\mathsf{S}_k$-$\mathsf{D}$ direct links and the $\mathsf{R}$-$\mathsf{D}$ link at  $\mathsf{D}$ are modeled respectively as 
\begin{align}  
& {\bar{Y}}_{m\mathsf{D},\ell} = \bar{H}_{m\mathsf{D},\ell} {\bar{X}}_{m,\ell} + {\bar{Z}}_{m\mathsf{D},\ell},    \;\; m \!\in \!\{1,2,\mathsf{R}\}, \;\;  \ell\!=\!1,\ldots,n_1, 
\end{align} where $\bar{H}_{k\mathsf{R},\ell}$ are the fading coefficients of the $\mathsf{S}_k$-$\mathsf{R}$ relay links, while $\bar{H}_{k\mathsf{D},\ell}$ and $\bar{H}_{\mathsf{R}\mathsf{D},\ell}$ are the same for the $\mathsf{S}_k$-$\mathsf{D}$ direct links and the $\mathsf{R}$-$\mathsf{D}$ mm-wave link respectively. The input symbols, $\hat{X}_{k,\ell} \in \mathbb{C}$ and $\bar{X}_{m,\ell}  \in \mathbb{C}$, are block power constrained as follows: $\frac{1}{n_1} \sum_{\ell=1}^{n_1} \mathbb{E}[|\hat{X}_{k,\ell}|^2] \leq \hat{P}_k$, $\frac{1}{n_1} \sum_{\ell=1}^{n_1} \mathbb{E}[|\bar{X}_{m,\ell}|^2] \leq \bar{P}_m$. Also, the noise RVs are ${\bar{Z}}_{k\mathsf{R},\ell} \sim \mathcal{CN}(0,1), k \in \{1,2\}$, i.i.d., and ${\bar{Z}}_{m\mathsf{D},\ell} \sim \mathcal{CN}(0,1), m \in \{1,2,\mathsf{R}\}$, i.i.d.

We assume that the DR-MARC is subject to an ergodic fading process where, across channel uses, the phase of the fading coefficients are $\sim \mathcal{U}[0, 2\pi)$  i.i.d. Specifically, the fading coefficients from node $m$ to node $t$, $m \in \{1,2,\mathsf{R}\}, t \in \{\mathsf{R},\mathsf{D}\}, m \neq t$, in the first band  are denoted by $H_{mt,i} \eqdef \sqrt{G_{mt,i}}e^{j \Theta_{mt,i}}$, while those in the second band by $\bar{H}_{mt,\ell} \eqdef \sqrt{\bar{G}_{mt,\ell}} e^{j \bar{\Theta}_{mt,\ell}}$, with $j \eqdef \sqrt{-1}$. Here,  $\Theta_{mt,i},\bar{\Theta}_{mt,\ell} \sim $ $\mathcal{U}[0, 2\pi)$ i.i.d., and $G_{mt,i}, \bar{G}_{mt,\ell}  \in \mathbb{R}_+$ are i.i.d. RVs that depend on the inter-node distance ${\sf d}_{mt}$, as well as the pathloss exponent $\beta_1$ (for the first band) and $\beta_2$ (for the second band). For example, when specializing to phase fading, we take  $G_{mt,i} \eqdef 1/{{\sf d}_{mt}^{\beta_1}}$ and $\bar{G}_{mt,\ell} \eqdef 1/{{\sf d}_{mt}^{\beta_2}}$ to be constant, and for Rayleigh fading, we take  $G_{mt,i} \sim \mathsf{exp}(1/{{\sf d}_{mt}^{\beta_1}})$ and  $\bar{G}_{mt,\ell} \sim \mathsf{exp}(1/{{\sf d}_{mt}^{\beta_2}})$ i.i.d., where $\mathsf{exp}(\mu)$ is an exponential distribution with mean $\mu$.

We also assume that  $(\mathrm{i})$ the long term parameters, \ied the distances and the pathloss exponents, are known at all nodes; $(\mathrm{ii})$ the instantaneous channel state information (CSI), \ied the phase and magnitude of the fading coefficients, are \emph{not available} to any transmitter; and $(\mathrm{iii})$ each receiver knows the CSI on all its incoming channels, but has no CSI of other channels. This models practical scenarios where CSI feedback to a transmitter is unavailable, while a receiver can reliably estimate the CSI. Also, this is less restrictive than \cite{lgr_mai_vu} where full or partial CSI is also available at a transmitter.

Note that given a BMF $\alpha$, for $n$ uses of the microwave band, the mm-wave band is used  $n_1(n) \eqdef \floor{\alpha n}$ times, while for $n_1$ uses in the mm-wave band, the microwave band is used $n(n_1) \eqdef \floor{n/\alpha}$ times. We  define a   $(2^{nR_1}, 2^{nR_2},n,\alpha)$ code for the DR-MARC that consists of ($\mathrm{i}$) two independent, uniformly distributed message sets $\mathcal{M}_k=\{ 1,\ldots,2^{nR_k}\}, k \in \{1,2\}$, one for each source; ($\mathrm{ii}$) two encoders $\phi_1$ and $\phi_2$ such that $ \phi_k: \mathcal{M}_k   \rightarrow   \mathbb{C}^{n} \times  \mathbb{C}^{n_1(n)} \times  \mathbb{C}^{n_1(n)}, k \in \{1,2\}$; ($\mathrm{iii}$) a set of relay encoding functions, $\{f_i \}_{i=1}^{n}$ and $\{\bar{f}_\ell \}_{\ell=1}^{n_1(n)}$, such that $x_{\mathsf{R},i} = f_i(y_{\mathsf{R}}^{i-1}, \{h_{k\mathsf{R}}^{i-1}, \bar{y}_{k\mathsf{R}}^{n_1(i-1)},  \bar{h}_{k\mathsf{R}}^{n_1(i-1)}\}_{k=1}^2)$ and $\bar{x}_{\mathsf{R},\ell} = \bar{f}_\ell(y_{\mathsf{R}}^{n(\ell-1)},  \{\bar{y}_{k\mathsf{R}}^{\ell-1}, \bar{h}_{k\mathsf{R}}^{\ell-1}, {h}_{k\mathsf{R}}^{n(\ell-1)}\}_{k=1}^2)$, $x_{\mathsf{R},i}, \bar{x}_{\mathsf{R},\ell} \in \mathbb{C}$; and ($\mathrm{iv}$) a decoder  $\psi$ at $\mathsf{D}$ such that $ \psi: \mathbb{C}^{n} \times \mathbb{C}^{3 n_1(n)} \times \mathbb{C}^{3 n}  \times  \mathbb{C}^{3 n_1(n)}   \rightarrow   \mathcal{M}_1 \times \mathcal{M}_2$. 

The relay helps by computing $\{x_{\mathsf{R},i}\}_{i=1}^n$ and $\{\bar{x}_{\mathsf{R},\ell}\}_{\ell=1}^{n_1(n)}$ causally by applying functions $\{f_i\}_{i=1}^{n}$ and $\{\bar{f}_\ell\}_{\ell=1}^{n_1(n)}$ on its past received signals and CSI as above and transmitting them to $\mathsf{D}$.	
A rate tuple $(R_1, R_2)$ is said to be achievable if there exists a sequence of  $(2^{nR_1}, 2^{nR_2},n,\alpha)$ codes such that the average probability of error  $ P_{e}^{(n)} \eqdef \text{Pr}[\psi({Y}_{\mathsf{D}}^{n}, \{\bar{Y}_{m\mathsf{D}}^{n_1}, H_{m\mathsf{D}}^n, \bar{H}_{m\mathsf{D}}^{n_1} \}_{m \in \{1,2,\mathsf{R} \}}) \neq (M_1,M_2) ] \rightarrow 0$  as $n   \rightarrow  \infty$  \cite[Chap.~15.3]{cover_thomas}. Finally, the system model of the R-MARC is defined from that of the DR-MARC by setting $\bar{{X}}_{k\mathsf{D},l} = \bar{{Y}}_{k\mathsf{D},l} = \emptyset, k=1,2$.

	\section{Decomposition Result on the DR-MARC}  T
We show that the capacity of the DR-MARC with BMF $\alpha$, denoted $\mathcal{C}_{DR}(\alpha)$, can be decomposed into the capacity of the underlying R-MARC, denoted $\mathcal{C}_{R}(\alpha)$, and the two $\mathsf{S}_k$-$\mathsf{D}$ direct links.
\begin{theorem} 
	\label{thm1}
	$C_{DR}(\alpha)$ is given by the set of all non-negative rate tuples $(R_1, R_2)$ that satisfy 
 \begin{align} \begin{split}
	C_{DR}(\alpha ) = \{(R_1,R_2): \; & R_1 \leq r_1  + \alpha \mathbb{E}[ \mathsf{C}(\bar{G}_{1\mathsf{D}} \bar{P}_1)], \notag \\ 
	& R_2 \leq r_2 + \alpha \mathbb{E}[ \mathsf{C}(\bar{G}_{2\mathsf{D}} \bar{P}_2)]\},\end{split} \notag \end{align} where $(r_1,r_2) \in \mathcal{C}_{R}(\alpha)$, and the expectations are taken over the corresponding RVs.
\end{theorem} 
The proof is relegated to Appendix \ref{app1}. For the special case of phase fading where $\bar{G}_{k\mathsf{D}} = 1/{\sf d}_{k \mathsf{D}}^{\beta_2}$ are constant, expectations in Theorem \ref{thm1} are not needed, while for Rayleigh fading expectations are  over $\bar{G}_{k\mathsf{D}} \sim \mathsf{exp}(1/{{\sf d}_{k\mathsf{D}}^{\beta_2}})$. Any $(R_1,R_2)$ in the DR-MARC can be achieved by achieving $(r_1,r_2)$ in the underlying R-MARC and supplementing it with the capacity of the direct links. Hence,  operating the direct links independently of the R-MARC is optimal, which simplifies the transmission. Since $\mathcal{C}_{DR}(\alpha)$ can be determined from  $\mathcal{C}_{R}(\alpha)$, it is sufficient to focus on  $\mathcal{C}_{R}(\alpha)$, considered next.

	\section{Capacity of a Class of R-MARC}  
Unlike the DR-MARC where separating the operation of the underlying R-MARC from the mm-wave direct links is optimal, in the R-MARC separating the underlying c-MARC  and the  mm-wave relay links is suboptimal in general. In fact,  capacity of the R-MARC is derived by operating the c-MARC jointly with the relay links. First, we characterize an {\em achievable rate region} for the R-MARC.  
\begin{theorem} 
	\label{thm2} 
	An achievable region of the R-MARC with BMF $\alpha$, denoted $\underline{\mathcal{C}}_{R}(\alpha)$, is given by the set of all non-negative rate tuples  $(R_1, R_2)$ that satisfy 
	\begin{align}
	R_1 &<  \mathbb{E}[\mathsf{C}(G_{1\mathsf{R}} P_1 )] + \alpha  \mathbb{E}[\mathsf{C}(\bar{G}_{1\mathsf{R}}\hat{P}_1 )], \label{r1_relay} \\ 
	R_2 &< \mathbb{E}[\mathsf{C}(G_{2\mathsf{R}} P_2 )] + \alpha \mathbb{E}[\mathsf{C}(\bar{G}_{2\mathsf{R}}\hat{P}_2 )], \label{r2_relay} \\ 
	R_1+R_2& < \mathbb{E}[\mathsf{C} (G_{1\mathsf{R}}P_1 + G_{2\mathsf{R}}P_2 )] +  \alpha \mathbb{E}[\mathsf{C} (\bar{G}_{1\mathsf{R}}\hat{P}_1)]  + \alpha \mathbb{E}[\mathsf{C} (\bar{G}_{2\mathsf{R}}\hat{P}_2)], \label{r1_r2_relay} \\ 
		R_1 &< \mathbb{E}[\mathsf{C}( G_{1\mathsf{D}}P_1 + G_{\mathsf{R}\mathsf{D}}P_{\mathsf{R}} )] + \alpha \mathbb{E}[\mathsf{C}(\bar{G}_{\mathsf{R}\mathsf{D}} \bar{P}_{\mathsf{R}} )], \label{r1_dest} \\ 
	R_2 & < \mathbb{E}[\mathsf{C}( G_{2\mathsf{D}}P_2 + G_{\mathsf{R}\mathsf{D}}P_{\mathsf{R}} )] + \alpha \mathbb{E}[\mathsf{C}(\bar{G}_{\mathsf{R}\mathsf{D}} \bar{P}_{\mathsf{R}} )], \label{r2_dest} \\ 
	R_1+R_2 &< \mathbb{E}[\mathsf{C}(G_{1\mathsf{D}}P_1 + G_{2\mathsf{D}}P_2 + G_{\mathsf{R}\mathsf{D}}P_{\mathsf{R}})]  + \alpha \mathbb{E}[\mathsf{C}(\bar{G}_{\mathsf{R}\mathsf{D}} \bar{P}_{\mathsf{R}} )], \label{r1_r2_dest}
	\end{align} where expectations are over the channel gains $G_{mt}$  and $\bar{G}_{mt}$, $m \neq t, m \in \{1,2,\mathsf{R}\}, t \in \{\mathsf{R},\mathsf{D}\}$.
\end{theorem}   The achievable region $\underline{\mathcal{C}}_{R}(\alpha)$ is obtained by performing block Markov encoding and backward decoding for the relay, as outlined in Appendix \ref{app2}. Moreover, the same message is \emph{jointly} encoded into codewords that are transmitted simultaneously in both bands. Interestingly, the bounds in  \eqref{r1_dest}-\eqref{r1_r2_dest} can be interpreted as that of the MAC from the sources to the destination aided by the relay.

In \cite{kramer_relay}, the capacity of the \emph{near} c-MARC, where the source-relay links can support higher rates than source-destination links, was characterized. In contrast, for R-MARCs with \emph{far} underlying c-MARC, if the following conditions hold, then the scheme of Theorem \ref{thm2} is also capacity achieving. 
\begin{theorem} \label{thm3}
	If the channel parameters of the Gaussian R-MARC with BMF $\alpha$ satisfy 
	\begin{align}
	  \mathbb{E}[\mathsf{C}( G_{1\mathsf{D}}P_1 + G_{\mathsf{R}\mathsf{D}}P_{\mathsf{R}})] + \alpha \mathbb{E}[\mathsf{C}(\bar{G}_{\mathsf{R}\mathsf{D}} \bar{P}_{\mathsf{R}})]  & \leq \mathbb{E}[\mathsf{C}(G_{1\mathsf{R}} P_1 )] + \alpha \mathbb{E}[\mathsf{C}(\bar{G}_{1\mathsf{R}}\hat{P}_1)], \label{r1_cond} \\ 
	\mathbb{E}[\mathsf{C}( G_{2\mathsf{D}}P_2 + G_{\mathsf{R}\mathsf{D}}P_{\mathsf{R}})] + \alpha \mathbb{E}[\mathsf{C}(\bar{G}_{\mathsf{R}\mathsf{D}} \bar{P}_{\mathsf{R}})] & \leq \mathbb{E}[\mathsf{C}(G_{2\mathsf{R}} P_2)] + \alpha \mathbb{E}[\mathsf{C}(\bar{G}_{2\mathsf{R}}\hat{P}_2)], \label{r2_cond}  \\ 
 \mathbb{E}[\mathsf{C}( G_{1\mathsf{D}}P_1 + G_{2\mathsf{D}}P_2 + G_{\rel\mathsf{D}}P_\rel)] + \alpha \mathbb{E}[ \mathsf{C}(\bar{G}_{\mathsf{R}\mathsf{D}} \bar{P}_{\mathsf{R}} )]  	&  \leq \mathbb{E}[\mathsf{C} (G_{1\mathsf{R}}P_1 + G_{2\mathsf{R}}P_2 )] + \alpha  \textstyle \sum\nolimits_{k=1}^{2}   \mathbb{E}[\mathsf{C} (\bar{G}_{k\mathsf{R}}\hat{P}_k)], \label{r1_r2_cond}
	\end{align} then its capacity is given by the set of all non-negative rate tuples $(R_1, R_2)$  that satisfy \eqref{r1_dest}-\eqref{r1_r2_dest}. Here, the expectations are  over channel gains $G_{mt}$, $\bar{G}_{mt}$, $m \neq t, m \in \{1,2,\mathsf{R}\}, t \in \{\mathsf{R},\mathsf{D}\}$.
\end{theorem} 

While the proof is relegated to Appendix \ref{app3}, we discuss the key steps here. First, in the proof of the outer bounds in steps (e)-(f) of \eqref{outer_bnd_phase}, the cross-correlation coefficients between the source and relay signals are set to zero. Since instantaneous CSI are not available to the transmitters and the phase of the fading coefficients $\sim \mathcal{U}[0, 2\pi)$, i.i.d., setting the cross-correlation  to zero proves optimal, resulting in outer bounds \eqref{r1_dest}-\eqref{r1_r2_dest}. Next,  in Theorem \ref{thm2}, if conditions \eqref{r1_cond}-\eqref{r1_r2_cond} hold, the achievable rates \eqref{r1_dest}-\eqref{r1_r2_dest} for the destination are smaller than those in \eqref{r1_relay}-\eqref{r1_r2_relay} for the relay. Hence, the relay can decode both messages without becoming a bottleneck to the rates. Thus, under \eqref{r1_cond}-\eqref{r1_r2_cond},  rates \eqref{r1_dest}-\eqref{r1_r2_dest} are achievable and they match the outer bounds.

Note that the rates in Theorem \ref{thm2} are achieved by encoding jointly over both bands. Hence, while capacity of the c-MARC is known only when the source-relay links are stronger in the microwave band (near case), in the R-MARC, they only need to be stronger \emph{jointly} over both bands. Thus, even if sources are not near the relay in the microwave band, for sufficiently \emph{strong} mm-wave relay links, they can become  ``\emph{jointly near}'' over both bands, where the scheme of Theorem \ref{thm2} achieves capacity.

The above result applies directly to phase and Rayleigh fading: for phase fading, $G_{mt}$ and $\bar{G}_{mt}$ are geometry determined constants, and thus the expectations in Theorem \ref{thm2} are not needed, while for Rayleigh fading, the expectations are over $G_{mt} \sim \mathsf{exp}(1/{{\sf d}_{mt}^{\beta_1}})$ and $\bar{G}_{mt} \sim \mathsf{exp}(1/{{\sf d}_{mt}^{\beta_2}})$.

\begin{figure}[t] 
	\captionsetup[subfloat]{captionskip=1mm}
	\centering    
	\subfloat[]{\includegraphics[width=8 cm,height=5cm]{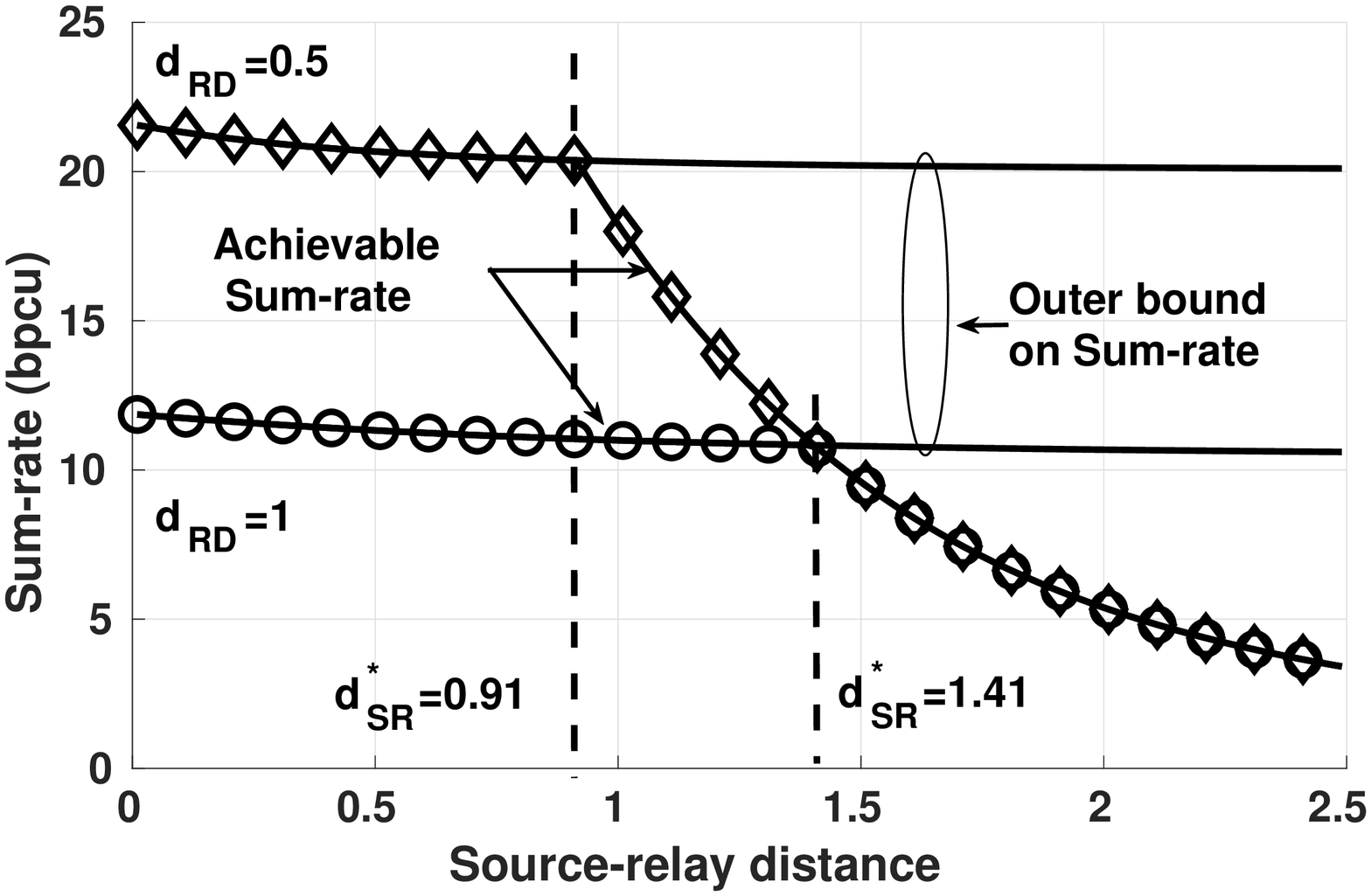}%
		\label{fig2:b}} \hspace*{10mm}
	\subfloat[]{\includegraphics[width=7.7cm,height=7cm]{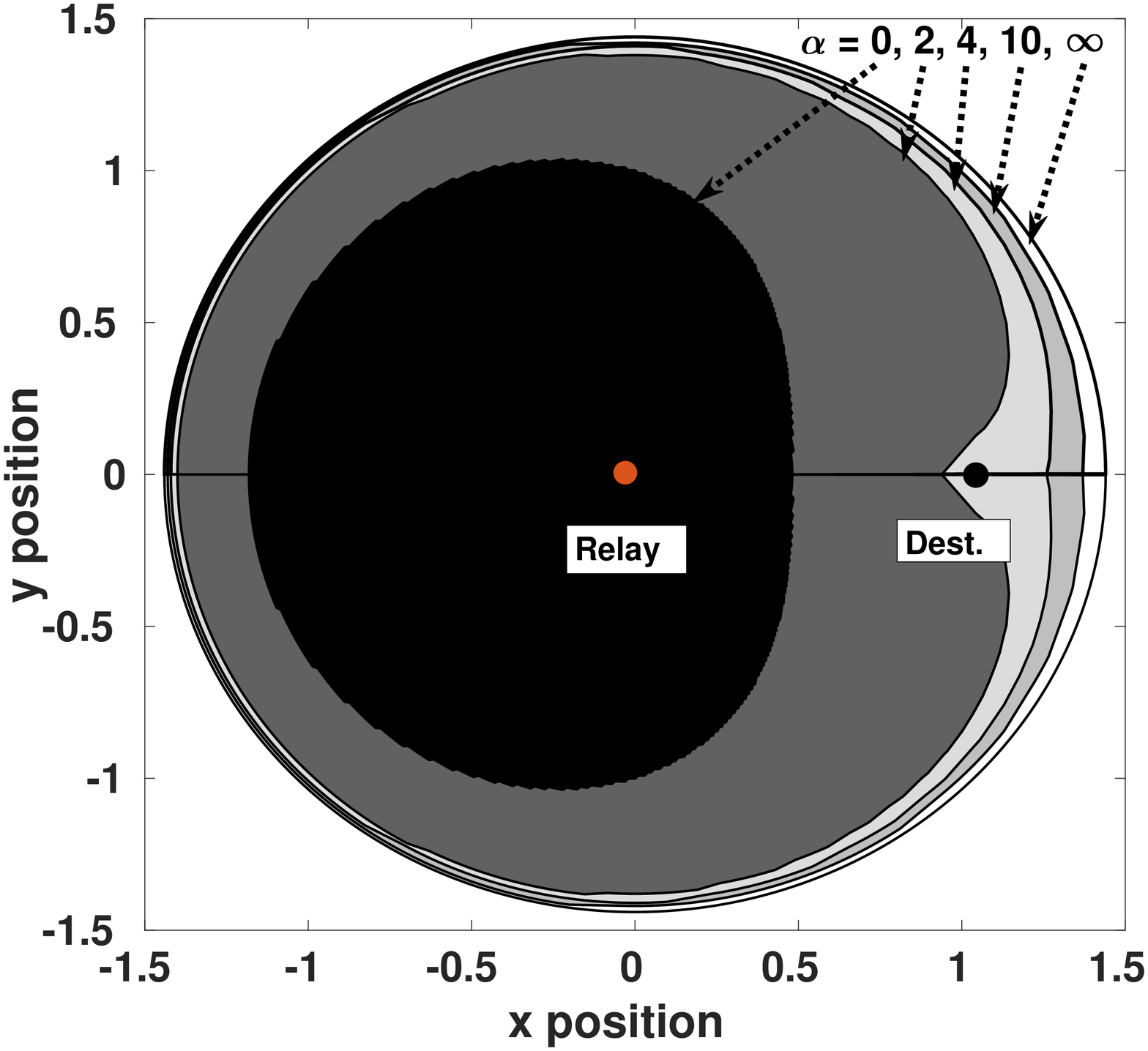}%
		\label{fig2:c}}
	\caption{ \small (a) The ASR matches the OB if ${\sf d}_{\mathsf{S}\mathsf{R}} \leq {\sf d}_{\mathsf{S}\mathsf{R}}^*$ for both cases of ${\sf d}_{\mathsf{R}\mathsf{D}}$. (b) The source locations for which the scheme of Theorem \ref{thm2} achieves the capacity of the R-MARC (\ied the locations at coordinates $(x,\pm y)$ in the shaded regions).} 
\end{figure}

\vspace*{2mm}
\textit{Numerical Examples:} To illustrate the impact of mm-wave links on the capacity of the R-MARC, we consider a two-dimensional topology as in Fig.~\ref{fig2:a} where $\mathsf{R}$ and $\mathsf{D}$ are located on the x-axis at $(0,0)$ and $(0, {\sf d}_{\mathsf{R}\mathsf{D}})$, and $\mathsf{S}_1$ and $\mathsf{S}_2$ are located symmetrically at $(-{\sf d}_{\mathsf{S}\mathsf{R}} \cos \phi, \pm {\sf d}_{\mathsf{S}\mathsf{R}} \sin \phi )$, with $\phi$ being the angle between a source and $\mathsf{R}$ and ${\sf d}_{\mathsf{S}\mathsf{D}} = ({\sf d}_{\mathsf{S}\mathsf{R}}^2 + {\sf d}_{\mathsf{R}\mathsf{D}}^2 + 2 {\sf d}_{\mathsf{S}\mathsf{R}} {\sf d}_{\mathsf{R}\mathsf{D}} \cos \phi)^{1/2}$ the resulting source-destination distance.  We take both bands in the R-MARC to be under phase fading as in \cite{kramer_relay}. Hence, expectations in conditions \eqref{r1_cond}-\eqref{r1_r2_cond} and Theorem~\ref{thm2} are not needed, and observations can be interpreted in terms of distances. Also, power constraints in the R-MARC are set to $10$ and $\beta_1 = 2, \beta_2 = 4$.

First, note that under condition \eqref{r1_r2_cond}, the sum-rate outer bound (OB), given by the r.h.s. of \eqref{r1_r2_dest}, matches the achievable sum-rate (ASR) in Theorem \ref{thm2}, given by the minimum of r.h.s. of \eqref{r1_r2_relay} and \eqref{r1_r2_dest}. For ease of exposition, we fix ${\sf d}_{\mathsf{R}\mathsf{D}}, \phi$ and BMF $\alpha$. Hence, condition \eqref{r1_r2_cond} is equivalent to ${\sf d}_{\mathsf{S}\mathsf{R}} \leq {\sf d}_{\mathsf{S}\mathsf{R}}^*({\sf d}_{\mathsf{R}\mathsf{D}}, \phi,\alpha)$ for some threshold source-destination distance ${\sf d}_{\mathsf{S}\mathsf{R}}^*({\sf d}_{\mathsf{R}\mathsf{D}}, \phi,\alpha)$. We verify this for fixed $\phi = \pi/4$ and $\alpha=2$ and two cases of ${\sf d}_{\mathsf{R}\mathsf{D}} \in \{1,0.5\}$ by plotting the ASR and the OB as functions of ${\sf d}_{\mathsf{S}\mathsf{R}}\in (0,2.5]$  in Fig.~\ref{fig2:b}. We observe that the ASR matches the OB if ${\sf d}_{\mathsf{S}\mathsf{R}} \leq {\sf d}_{\mathsf{S}\mathsf{R}}^*$ with ${\sf d}_{\mathsf{S}\mathsf{R}}^* \approx 1.41$ for ${\sf d}_{\mathsf{R}\mathsf{D}}=1$, and ${\sf d}_{\mathsf{S}\mathsf{R}}^* \approx 0.91$   for ${\sf d}_{\mathsf{R}\mathsf{D}}=0.5$, otherwise the ASR is strictly smaller. As ${\sf d}_{\mathsf{R}\mathsf{D}}$ reduces from $1$ to $0.5$, for condition \eqref{r1_r2_cond}  to hold, ${\sf d}_{\mathsf{S}\mathsf{R}}^*$ also {\em reduces} from ${\sf d}_{\mathsf{S}\mathsf{R}}^* \approx 1.41$ to $\approx 0.91$.

Next, to illustrate the impact of the mm-wave links, in Fig.~\ref{fig2:c} we depict the source locations relative to the relay and the destination for which all of conditions \eqref{r1_cond}-\eqref{r1_r2_cond} are satisfied and therefore the scheme of Theorem \ref{thm2} achieves capacity. As such, we fix ${\sf d}_{\mathsf{R}\mathsf{D}}=1$, vary $\phi \in (0, \pi)$ and ${\sf d}_{\mathsf{S}\mathsf{R}}\in (0,2)$ to vary source locations, and plot the resulting regions: we overlay the region for the case without mm-wave links ($\alpha=0$) on those with mm-wave links with  BMF $\alpha \in \{2,4,10\}$ as well as $\alpha   \rightarrow   \infty$. 

First, for the case without mm-wave links ($\alpha =0$), conditions \eqref{r1_cond}-\eqref{r1_r2_cond} hold only when sources are within the innermost black region in Fig.~\ref{fig2:c}. Noting that for each $\phi$, the resulting threshold distance ${\sf d}_{\mathsf{S}\mathsf{R}}^*(\phi)$ is at the boundary of this region,  as $\phi$ increases from $\phi = 0$ to $\phi  = \pi$, ${\sf d}_{\mathsf{S}\mathsf{R}}^*(\phi)$ decreases monotonically from $\approx 1.2$ to $\approx 0.47$. We thus observe that conditions \eqref{r1_cond}-\eqref{r1_r2_cond} hold for much larger threshold distance ${\sf d}_{\mathsf{S}\mathsf{R}}^*$ when sources are located far away from destination (\ied $\phi \approx 0$), and threshold distance ${\sf d}_{\mathsf{S}\mathsf{R}}^*$ reduces considerably when sources are closer to the destination (\ied $\phi \approx \pi$).

We note that the above trends continue to hold when mm-wave links are used ($\alpha >0$), however, the resulting region (union of the inner black and outer gray regions) now extends much closer to the destination. For example, for the region with  $\alpha = 2$, ${\sf d}_{\mathsf{S}\mathsf{R}}^*$ reduces to only $\approx 0.96$ near the destination, compared to $\approx 0.47$ with $\alpha=0$. Moreover, the resulting region grows with $\alpha$ but the growth saturates for higher values of $\alpha$, with $\alpha=10$  producing almost the same region as that for  $\alpha   \rightarrow   \infty$.

	\section{The Optimal Sum-Rate Problem}  \label{sum_rate_opt}
\label{sec5}
Since mm-wave links can have significantly larger bandwidth than the microwave links, they can significantly affect the performance limits of the DR-MARC. To understand this impact, we study how the sum-rate achievable on the DR-MARC (with the scheme of Theorem \ref{thm2}) is maximized by optimally allocating power to the mm-wave direct and relay links. We observe that the resulting scheme allocates power to the mm-wave links in different \emph{modes} depending on whether certain channel conditions hold. This characterization reveals insights into the nature of the scheme, and can serve as an effective resource allocation strategy for such dual-band networks in practice. 

For ease of exposition, the mm-wave band is assumed to be under \emph{phase} fading while the microwave band is assumed to be under the general fading of Section II.  Here, phase fading is a good model for mm-wave links such as those in \cite{directed_1},  as phase fading is a special case of the general fading model \cite{simple_phase_fading} when the diffuse component associated with the non-LoS propagation is not present. Furthermore, this simplification reveals useful insights into the optimal power allocation.

Under phase fading, the link gain in the $\mathsf{S}_k$-$\mathsf{D}$ direct link (referred to as $\mathtt{DL}_k$) is $\bar{G}_{k\mathsf{D}} = 1/{\sf d}_{k\mathsf{D}}^{\beta_2} > 0$, and that in the $\mathsf{S}_k$-$\mathsf{R}$ relay link (referred to as $\mathtt{RL}_k$) is $\bar{G}_{k\mathsf{R}} = 1/{\sf d}_{k\mathsf{R}}^{\beta_2} > 0$, which are constants. For convenience, we denote the link gains of $\mathtt{DL}_k$  and $\mathtt{RL}_k$ by $d_k \eqdef \bar{G}_{k\mathsf{D}}$ and $r_k \eqdef \bar{G}_{k\mathsf{R}}, k \in \{1,2\}$.  We assume that the transmit power in $\mathtt{DL}_k$ ($p_k$) and $\mathtt{RL}_k$ ($q_k$) from source $\mathsf{S}_k$ satisfy a total power budget    \begin{align} 
p_k + q_k = P, \;\; k \in \{1,2\}. \label{power_budget}
\end{align} 	

For a fixed power allocation $(p_1, q_1, p_2, q_2)$, $R$ is an achievable sum-rate of the DR-MARC iff  \begin{align} R \leq \min \{\Sigma_{\mathsf{R}}, \Sigma_{\mathsf{D}}\}. \end{align} Here, $\Sigma_{\mathsf{R}}$ and $\Sigma_{\mathsf{D}}$ denote the sum-rates achievable at the relay and  destination, and are given by  \begin{align}
& \Sigma_{\mathsf{R}} \eqdef \sigma_{\mathsf{R}} + \alpha  \textstyle\sum\nolimits_{k=1}^{2}  \log(1+r_k q_k) + \log(1+d_k p_k), \label{zr13}\\
& \Sigma_{\mathsf{D}} \eqdef \sigma_{\mathsf{D}} + \alpha \textstyle\sum\nolimits_{k=1}^{2} \log(1+d_k p_k), \label{zr23}
\end{align} where $ \sigma_{\mathsf{D}} \eqdef \mathbb{E}[\mathsf{C} (G_{1\mathsf{D}}P_1 + G_{2\mathsf{D}}P_2 + G_{\mathsf{R}\mathsf{D}}P_{\mathsf{R}} )] + \alpha \mathsf{C}(\bar{G}_{\mathsf{R}\mathsf{D}} \bar{P}_{\mathsf{R}} )$ and $\sigma_{\mathsf{R}} \eqdef \mathbb{E}[\mathsf{C} (G_{1\mathsf{R}}P_1 + G_{2\mathsf{R}}P_2 )]$, with the expectations taken over the RVs involved.  Note that $\Sigma_{\mathsf{R}}$ and $\Sigma_{\mathsf{D}}$ are obtained as follows. For direct link powers $(p_1,p_2)$, it follows from the decomposition result in Theorem~\ref{thm1} that the sum-rate of the DR-MARC is given by the sum of the sum-rate of the R-MARC and the total rate of the direct links, \ied $\alpha \textstyle\sum_{k=1}^2 \log(1+ d_k p_k)$. Now, for given relay link powers $(q_1,q_2)$, the sum-rate of the R-MARC is given by the minimum of r.h.s. of \eqref{r1_r2_relay} and \eqref{r1_r2_dest}. Hence,  $\Sigma_{\mathsf{R}}$ is given by the sum of the r.h.s. of \eqref{r1_r2_relay} and $\alpha \textstyle\sum_{k=1}^2 \log(1+ d_k p_k)$ as expressed in \eqref{zr13}, while $\Sigma_{\mathsf{D}}$ is obtained by the sum of the r.h.s. of \eqref{r1_r2_dest} and $\alpha \textstyle\sum_{k=1}^2 \log(1+ d_k p_k)$, as given in \eqref{zr23}.

The problem of maximizing $R$ over the transmit powers ($p_1, q_1, p_2, q_2$) is then
\begin{align}
[\mathcal{P}1] \quad \text{maximize} &\quad R  \notag \\ 
\text{subject to} &\quad R \leq \Sigma_{\mathsf{R}}, \label{conv_1} \\ 
&\quad R \leq \Sigma_{\mathsf{D}}, \label{conv_2} \\ 
&\quad  p_k +   q_k = P, \;\; k \in \{1,2\}\label{apc}\\ 
&\quad(p_1, q_1, p_2, q_2, R) \succeq \mathbf{0} \label{nonneg1}.
\end{align}

Note that $[\mathcal{P}1]$ is a convex optimization problem as the objective is linear, constraints in \eqref{apc} are affine, and those in \eqref{conv_1}--\eqref{conv_2} are convex. Hence, it can be solved by formulating the Lagrangian function of $[\mathcal{P}1]$ by associating a Lagrange multiplier to each constraint in \eqref{conv_1}-\eqref{nonneg1}, and then deriving and solving the KKT conditions \cite{Boyd}. See Appendix \ref{app4} for details.

\subsection{Link Gain Regimes and Optimal Power Allocation}  
To gain insights, we derive the optimal power allocation in closed form, and describe it in terms of \emph{link-gain regimes} (LGR) which are partitions of the set of all tuples of link gains and power budget $P$, found while solving the KKT conditions for $[\mathcal{P}1]$. Specifically, we derive the KKT conditions and solve for the optimal primal variables (\ied transmit powers) and the optimal Lagrange multipliers (OLM). To simplify the procedure, we consider the set of tuples of OLMs associated with \emph{inequality} constraints in \eqref{conv_1}, \eqref{conv_2} and \eqref{nonneg1}, and partition this set into a few subsets based on whether the OLMs in the set are \emph{positive} or \emph{zero}, \ied whether the associated primal constraints are \emph{tight} or \emph{not} (detailed in Appendix \ref{app4}). For each resulting partition of the set of OLM tuples, we first derive the expression for the optimal powers in closed form. However, the conditions that define these partitions are still characterized in terms of the OLMs. Therefore, to express the optimal power allocation explicitly in terms of link gains $(r_1, r_2, d_1, d_2)$ and power budget $P$, we express the conditions that partition the set of the OLM tuples in terms of link gains, $P$, and parameter $\gamma$, defined as   \begin{align} \gamma \eqdef 2^{(\sigma_{\mathsf{D}} - \sigma_{\mathsf{R}})/\alpha} \label{gamma_def} \end{align}  which models the effect of microwave band parameters, with  $\sigma_{\mathsf{D}} $ and $ \sigma_{\mathsf{R}}$ defined in \eqref{zr13}-\eqref{zr23}.  \begin{remark}
	The parameter $\gamma$ in \eqref{gamma_def} is used only to simplify the exposition. When interpreting the optimum transmit powers, we often compare $\Sigma_{\mathsf{R}} $ and $ \Sigma_{\mathsf{D}}$. Substituting their expressions  in \eqref{zr13} and \eqref{zr23},  the comparison between $\Sigma_{\mathsf{R}} $ and $ \Sigma_{\mathsf{D}}$ reduces to that between   $\sigma_{\mathsf{R}} + \alpha \textstyle \sum_{k=1}^2  \log(1+ r_k q_k)$ and $\sigma_{\mathsf{D}}$, \ied equivalently  between $(1+ r_1 q_1)(1+ r_2 q_2)  $ and $ 2^{(\sigma_{\mathsf{D}} - \sigma_{\mathsf{R}})/\alpha}$. We thus define $\gamma = 2^{(\sigma_{\mathsf{D}} - \sigma_{\mathsf{R}})/\alpha}$.
\end{remark} 

As a result, the set of  $ (r_1, r_2, d_1, d_2, \gamma, P)$-tuples is partitioned into a few subsets (LGRs), each corresponding to one and only one subset of OLM tuples. The conditions for each LGR is then simplified and expressed as upper and lower bounds (\emph{threshold powers}) on power budget $P$ where the threshold powers depend on $ (r_1, r_2, d_1, d_2, \gamma)$. This results in partitioning the power budget $P \geq 0$  into a few intervals,  each describing an LGR.  Specifically, we consider two cases $\sigma_{\mathsf{D}} \leq \sigma_{\mathsf{R}}$ and $\sigma_{\mathsf{D}} > \sigma_{\mathsf{R}}$, which are equivalent to $\gamma \leq 1$ and $\gamma > 1$ respectively. 

\begin{table*}[tbp]  
	\renewcommand{\arraystretch}{1.1}
	\tabcolsep = 3pt
	\caption{ \small LGRs and corresponding optimal power allocation for the case with $\gamma >1$.} 
	\label{table_thm_1} 
	\centering 
	\begin{tabular}{l|l}
		\hlinewd{1pt}
		Definition of LGR & Optimal power allocation \\  \hlinewd{1pt}
		$\mathcal{A}_{\sf d,d}   \eqdef \{\boldsymbol{c}: 0 \leq P \leq \min(\mathsf{P}_{\sf d,d},\hat{\mathsf{P}}_{\sf d,d}) \}$  & $p_1 = P,$ \quad\quad $  q_1 = 0, $ \quad\quad $  p_2 = P,$ \quad\quad  $  q_2 = 0$  \\ 
		$ \mathcal{A}_{\sf d,r}  \eqdef  \{\boldsymbol{c}: 0 \leq  P \leq \min(\hat{\mathsf{P}}^{\prime}_{\sf d,d},\mathsf{P}_{\sf d,d},\mathsf{P}_{\sf d,r}) \}$  & $p_1 = P,$ \quad\quad $  q_1 = 0, $ \quad\quad $  p_2 = 0,$ \quad\quad  $  q_2 = P$  \\ 
		$ \mathcal{A}_{\sf r,d} \eqdef  \{\boldsymbol{c}: 0 \leq  P \leq \min(\mathsf{P}^{\prime}_{\sf d,d},\hat{\mathsf{P}}_{\sf d,d},\mathsf{P}_{\sf r,d}) \}$ & $p_1 = 0,$ \quad\quad $  q_1 = P, $ \quad\quad $  p_2 = P,$ \quad\quad  $q_2 = 0$  \\ 
		$\mathcal{A}_{\sf r,r}  \eqdef  \{\boldsymbol{c}: 0 \leq  P \leq \min(\mathsf{P}^{\prime}_{\sf d,d},\hat{\mathsf{P}}^{\prime}_{\sf d,d},\mathsf{P}_{\sf r,r}) \} $  & $p_1 = 0,$ \quad\quad $  q_1 = P, $ \quad\quad $  p_2 = 0,$ \quad\quad  $  q_2 = P$  \\[1mm]  \hlinewd{.5pt} \\[-3mm] 
		$\mathcal{A}_{\sf rd,d}  \eqdef  \{\boldsymbol{c}: \max(\mathsf{P}_{\sf d,d},\mathsf{P}^{\prime}_{\sf d,d})   <  P \leq \min(\hat{\mathsf{P}}_{\sf d,d},\mathsf{P}_{\sf rd,d}) \}  $ & $p_1 = \dfrac{1}{2}\left(P+\dfrac{1}{r_1}-\dfrac{1}{d_1}\right), \quad\quad q_1 =\dfrac{1}{2}\left(P-\dfrac{1}{r_1}+\dfrac{1}{d_1}\right), \quad\quad p_2 = P, \quad\quad  q_2 = 0$\\[3mm] 
		$\mathcal{A}_{\sf d,rd} \eqdef  \{\boldsymbol{c}: \max(\hat{\mathsf{P}}_{\sf d,d},\hat{\mathsf{P}}^{\prime}_{\sf d,d})  <  P \leq \min(\mathsf{P}_{\sf d,d},\mathsf{P}_{\sf d,rd}) \}  $ & $p_1 = P, \quad\quad  q_1 = 0, \quad\quad p_2 = \dfrac{1}{2}\left(P+\dfrac{1}{r_2}-\dfrac{1}{d_2}\right), \quad\quad q_2 =\dfrac{1}{2}\left(P-\dfrac{1}{r_2}+\dfrac{1}{d_2}\right)$  \\[3mm]  \hlinewd{.5pt} \\[-4mm] 
		$ \mathcal{A}_{\sf r,rd}  \eqdef \{\boldsymbol{c}: \max(\hat{\mathsf{P}}_{\sf d,d},\hat{\mathsf{P}}^{\prime}_{\sf d,d})  <  P \leq \min(\mathsf{P}^{\prime}_{\sf d,d},\mathsf{P}_{\sf r,rd}) \}    $ & $  p_1 = 0, \quad\quad q_1 = P, \quad\quad p_2 = \dfrac{1}{2}\left(P+\dfrac{1}{r_2}-\dfrac{1}{d_2}\right), \quad\quad q_2 =\dfrac{1}{2}\left(P-\dfrac{1}{r_2}+\dfrac{1}{d_2}\right)$  \\[3mm]
		$ \mathcal{A}_{\sf rd,r}  \eqdef \{\boldsymbol{c}: \max(\mathsf{P}_{\sf d,d},\mathsf{P}^{\prime}_{\sf d,d})  <  P \leq \min(\hat{\mathsf{P}}^{\prime}_{\sf d,d},\mathsf{P}_{\sf rd,r}) \}    $ &  $p_1 = \dfrac{1}{2}\left(P+\dfrac{1}{r_1}-\dfrac{1}{d_1}\right), \quad\quad q_1 =\dfrac{1}{2}\left(P-\dfrac{1}{r_1}+\dfrac{1}{d_1}\right), \quad\quad p_2 = 0, \quad\quad  q_2 = P$ \\[2mm] \hlinewd{.5pt} \\[-2mm] 
		$\mathcal{A}_{\sf rd,rd}   \eqdef  \{\boldsymbol{c}:  \max(\mathsf{P}_{\sf d,d},\hat{\mathsf{P}}_{\sf d,d},\mathsf{P}^{\prime}_{\sf d,d},\hat{\mathsf{P}}^{\prime}_{\sf d,d})  <  P  \leq  \mathsf{P}_{\sf rd,rd}$ & $p_k =\dfrac{1}{2}\left(P+\dfrac{1}{r_k}-\dfrac{1}{d_k}\right), \quad\quad  q_k =  \dfrac{1}{2}\left(P-\dfrac{1}{r_k}+\dfrac{1}{d_k}\right), \quad\quad k \in \{1,2\}$ \\[3mm] \hlinewd{.5pt} \\[-4mm]  
		$ \mathcal{S}_{\sf r,rd} \eqdef \{\boldsymbol{c}: \max(\mathsf{P}_{\sf r,r},\mathsf{P}_{\sf r,rd})  <  P \leq \min(\overbar{\mathsf{P}}_{\sf r,rd},\mathsf{P}_{\sf r,d}) \} $ &  $  p_1 = 0, \quad\quad q_1 = P, \quad\quad p_2 = P - q_2, \quad\quad q_2 =  r_2^{-1}\left({\gamma}/({1+Pr_1})-1\right)$\\[1mm] 
		$\mathcal{S}_{\sf rd,r} \eqdef \{\boldsymbol{c}: \max(\mathsf{P}_{\sf r,r},\mathsf{P}_{\sf rd,r})  <  P \leq \min(\overbar{\mathsf{P}}_{\sf rd,r},\mathsf{P}_{\sf d,r}) \} $ & $  p_1 = P-q_1, \quad\quad q_1 =  r_1^{-1}\left({\gamma}/({1+Pr_2}\right)-1), \quad\quad p_2 = 0, \quad\quad q_2 =  P$ \\[2mm] \hlinewd{.5pt} \\[-4mm] 
		$\mathcal{S}_{\sf rd,d} \eqdef \{\boldsymbol{c}: \boldsymbol{r} \in \mathcal{R}_{S1},  \max(\mathsf{P}_{\sf r,d},\mathsf{P}_{\sf rd,d},\overbar{\mathsf{P}}_{\sf rd,d}) < P\}$ & $p_1 = P - (\gamma -1)r_1^{-1}, \quad\quad q_1 =  (\gamma -1)r_1^{-1}, \quad\quad p_2=P, \quad\quad q_2 = 0$  \\ 
		$ \cup \{ \boldsymbol{c}: \boldsymbol{r} \in (\mathcal{R}_{1} \cup \mathcal{R}_{2} \cup \mathcal{R}_{S2}), \max(\mathsf{P}_{\sf r,d},\mathsf{P}_{\sf rd,d}) < P < \overbar{\mathsf{P}}_{\sf rd,d} \}$ & \\
		$\mathcal{S}_{\sf d,rd} \eqdef \{\boldsymbol{c}: \boldsymbol{r} \in \mathcal{R}_{S2}, \max(\mathsf{P}_{\sf d,r},\mathsf{P}_{\sf d,rd},\overbar{\mathsf{P}}_{\sf d,rd}) < P \}$ & $p_1 = P, \quad\quad q_1 = 0, \quad\quad p_2 = P - (\gamma -1)r_2^{-1}, \quad\quad q_2 =  (\gamma -1)r_2^{-1}$  \\
		$ \cup \{ \boldsymbol{c}:\boldsymbol{r} \in (\mathcal{R}_{1} \cup \mathcal{R}_{2} \cup \mathcal{R}_{S1}), \max(\mathsf{P}_{\sf d,r},\mathsf{P}_{\sf d,rd}) < P < \overbar{\mathsf{P}}_{\sf d,rd} \}$ & \\[1 mm]   \hlinewd{.5pt} \\[-4mm] 
		$\mathcal{S}_{\sf rd,rd} \eqdef $ & $p_1 = P - q_1,$  \\
		$\{\boldsymbol{c}: \boldsymbol{r} \in (\mathcal{R}_1 \cup \mathcal{R}_2), \max(\overbar{\mathsf{P}}_{\sf rd,d}, \overbar{\mathsf{P}}_{\sf d,rd}, \overbar{\mathsf{P}}_{\sf rd,r},\overbar{\mathsf{P}}_{\sf r,rd},\mathsf{P}_{\sf rd,rd}) < P \}$ & $  q_1 =  r_1^{-1}\left(\gamma({ { P r_1 + r_1 d_1^{-1} + 1})/({P r_2 + r_2 d_2^{-1} + 1}})\right)^{1/2} - r_1^{-1},$ \\ 
		$ \cup \{ \boldsymbol{c}: \boldsymbol{r}  \in  \mathcal{R}_{S1}, \max(\overbar{\mathsf{P}}_{\sf d,rd}, \overbar{\mathsf{P}}_{\sf rd,r},\overbar{\mathsf{P}}_{\sf r,rd},\mathsf{P}_{\sf rd,rd})  <  P   \leq  \overbar{\mathsf{P}}_{\sf rd,d} \}     $ & $p_2 = P - q_2,$    \\ 
		$ \cup \{ \boldsymbol{c}:\boldsymbol{r}  \in  \mathcal{R}_{S2}, \max(\overbar{\mathsf{P}}_{\sf rd,d}, \overbar{\mathsf{P}}_{\sf rd,r},\overbar{\mathsf{P}}_{\sf r,rd},\mathsf{P}_{\sf rd,rd})  <  P  \leq  \overbar{\mathsf{P}}_{\sf d,rd} \}    $  & $  q_2 =  r_2^{-1}\left(\gamma({ { P r_2 + r_2 d_2^{-1} + 1})/({P r_1 + r_1 d_1^{-1} + 1}})\right)^{1/2} - r_2^{-1}$ \\[2mm] \hlinewd{1pt}
	\end{tabular} 
\end{table*}

\begin{table*}[!t] 
	\renewcommand{\arraystretch}{1}
	\tabcolsep = 10pt
	\caption{ \small Definition of the threshold powers in LGRs.} 
	\label{table_critical_power} 
	\centering  \begin{tabular}{l l}
		\hlinewd{1pt} \\[-4mm]
		$\mathsf{P}_{\sf d,d} \eqdef r_1^{-1}-d_1^{-1},$ & $\mathsf{P}^{\prime}_{\sf d,d} \eqdef -\mathsf{P}_{\sf d,d},$ \\
		$\hat{\mathsf{P}}_{\sf d,d} \eqdef  r_2^{-1}-d_2^{-1},$ & $\hat{\mathsf{P}}^{\prime}_{\sf d,d}  \eqdef  -\hat{\mathsf{P}}_{\sf d,d},$\\
		$\mathsf{P}_{\sf r,d} \eqdef (\gamma-1)r_1^{-1},$  & $\mathsf{P}_{\sf d,r} \eqdef (\gamma-1)r_2^{-1},$		\\ 
		$\mathsf{P}_{\sf r,r} \eqdef \varrho[(1+x r_1)(1+x r_2)-\gamma],$ &  \\
		$\mathsf{P}_{\sf rd,d}  \eqdef (2\gamma-1)r_1^{-1}-d_1^{-1},$ & $\mathsf{P}_{\sf d,rd}  \eqdef (2\gamma-1)r_2^{-1}-d_2^{-1},$ \\
		$\mathsf{P}_{\sf r,rd} \eqdef \varrho[(1+r_2 d_2^{-1}+xr_2)(1+x r_1)-2\gamma],$ &  $\mathsf{P}_{\sf rd,r} \eqdef \varrho[(1+r_1 d_1^{-1}+xr_1)(1+x r_2)-2\gamma],$  \\ 
		$\mathsf{P}_{\sf rd,rd}  \eqdef \varrho[(1+r_1 d_1^{-1}+xr_1)(1+r_2 d_2^{-1}+xr_2) -4\gamma],$ \\ 
		$\overbar{\mathsf{P}}_{\sf r,rd}  \eqdef \varrho[(1+r_2/d_2+xr_2)(1+x r_1)^2   -  \gamma (1+r_1/d_1+xr_1)],$ & $\overbar{\mathsf{P}}_{\sf rd,r}  \eqdef \varrho[(1+r_1/d_1+xr_1)(1+x r_2)^2  -  \gamma (1+r_2/d_2+xr_2)],$ \\ 
		$\overbar{\mathsf{P}}_{\sf rd,d} \eqdef   (\gamma -1 + \gamma r_2d_2^{-1} - r_1d_1^{-1})/(r_1 - \gamma r_2),$ & $\overbar{\mathsf{P}}_{\sf d,rd} \eqdef   (\gamma -1 + \gamma r_1 d_1^{-1} - r_2 d_2^{-1})/(r_2 - \gamma r_1).$   \\[1mm]
		\hlinewd{1pt}
	\end{tabular} 
\end{table*}

For $\gamma \leq 1$, the set of all  $ (r_1, r_2, d_1, d_2, \gamma, P)$-tuples turn out to belong to a \emph{single} LGR where the allocation $(p_1, q_1, p_2, q_2) = (P,0,P,0)$ is optimal for all $P \geq 0$. Since $\gamma \leq 1 $ implies $ \sigma_{\mathsf{D}} \leq \sigma_{\mathsf{R}}$ from \eqref{gamma_def}, any feasible allocation results in $R = \Sigma_{\mathsf{D}} \leq \Sigma_{\mathsf{R}}$, with $\Sigma_{\mathsf{D}} $ and $ \Sigma_{\mathsf{R}}$ in \eqref{zr13}-\eqref{zr23}. Since $R = \Sigma_{\mathsf{D}}$  only increases by increasing $p_1$ and $p_2$, $\Sigma_{\mathsf{D}}$ is maximized with $p_1=p_2 = P$. Thus, $P$ should always be entirely allocated to the direct links. 

For the case with $\gamma > 1$, the set of $\boldsymbol{c} \eqdef  (r_1, r_2, d_1, d_2,  P)$-tuples is partitioned into $14$ LGRs, and thus the optimal power allocation (referred to as {OA}) is more involved. In Table  \ref{table_thm_1}, we define the $14$ LGRs and present the optimal powers for each LGR. Here, $\boldsymbol{r} \eqdef (r_1, r_2)$, and the threshold powers  for the LGRs are defined in Table \ref{table_critical_power}, with $\varrho[{\sf f}(x)]$ denoting the positive root of polynomial ${\sf f}(x)$.

For LGRs $\mathcal{A}_{x,y}, x, y \in \{\sf d,r, rd\}$, $x$ and $y$ denote the transmission status in the mm-wave links of sources $\mathsf{S}_1$ and $\mathsf{S}_2$ respectively: for each source, $\sf d$, $\sf r$ and $\sf rd$ denotes that \oa transmits in the direct link only, in the relay link only and in both links, respectively. For example, in LGR $\mathcal{A}_{\sf rd,d}$ \oa transmits in both links of source $\mathsf{S}_1$ and only in the direct link of source $\mathsf{S}_2$. While LGRs $\mathcal{S}_{(.,.)}$ can be similarly interpreted, $\mathcal{S}_{(.,.)}$ and $\mathcal{A}_{(.,.)}$ are associated with two distinct properties of \oa discussed shortly.  Moreover, the threshold powers $\mathsf{P}_{(.,.)}, \hat{\mathsf{P}}_{(.,.)}, \mathsf{P}^{\prime}_{(.,.)}, \hat{\mathsf{P}}^{\prime}_{(.,.)}$ and $\overbar{\mathsf{P}}_{(.,.)}$ follow the same notation as the LGRs, with $\mathsf{P}^{\prime}_{(.,.)} \eqdef - \mathsf{P}_{(.,.)}$. Also, while $\overbar{\mathsf{P}}_{(.,.)}$ are used for LGRs  $\mathcal{S}_{(.,.)}$ only, all other threshold powers are used for both type of LGRs $\mathcal{A}_{(.,.)}$ and $\mathcal{S}_{(.,.)}$.

Note that all LGRs in Table \ref{table_thm_1} are mutually exclusive in that, for a given tuple $\boldsymbol{c} = (r_1, r_2, d_1, d_2, P)$, the condition for one and only one LGR holds. For example, suppose a tuple $\boldsymbol{c} \in \mathcal{A}_{\sf d,d}$, hence it satisfies $\min(\mathsf{P}_{\sf d,d},\hat{\mathsf{P}}_{\sf d,d})  \geq P  \geq 0$. From Table \ref{table_critical_power}, since $\mathsf{P}^{\prime}_{\sf d,d}  \eqdef  -\mathsf{P}_{\sf d,d}$, $\hat{\mathsf{P}}^{\prime}_{\sf d,d}  \eqdef  -\hat{\mathsf{P}}_{\sf d,d}$ the condition  $(\mathsf{P}_{\sf d,d},\hat{\mathsf{P}}_{\sf d,d}) \succeq \mathbf{0}$ for $\mathcal{A}_{\sf d,d}$ requires $(\mathsf{P}^{\prime}_{\sf d,d},\hat{\mathsf{P}}^{\prime}_{\sf d,d}) \preceq \mathbf{0}$, \ied $\mathcal{A}_{\sf r,r}  =  \mathcal{A}_{\sf r,d}  =  \mathcal{A}_{\sf d,r}  =  \mathcal{A}_{\sf r,rd}  =  \mathcal{A}_{\sf rd,r}  =  \emptyset$.   Next, $\boldsymbol{c} \not\in \mathcal{A}_{\sf rd,d}$ as condition $\mathsf{P}_{\sf d,d} < P$ for $\mathcal{A}_{\sf rd,d}$ violates condition $ \mathsf{P}_{\sf d,d} \geq P$ for $\mathcal{A}_{\sf d,d}$; similarly $\boldsymbol{c} \not\in \mathcal{A}_{\sf d,rd}$ and $\boldsymbol{c}  \not\in  \mathcal{A}_{\sf rd,rd}$. Also, $\boldsymbol{c} \not\in \mathcal{S}_{\sf rd,d}$ as condition $\mathsf{P}_{\sf rd,d} < P$ for $\mathcal{S}_{\sf rd,d}$ violates $\mathsf{P}_{\sf d,d}  >  P$ for $\mathcal{A}_{\sf d,d}$ since $\mathsf{P}_{\sf d,d}  <  \mathsf{P}_{\sf rd,d}$; similarly $\boldsymbol{c} \not\in \mathcal{S}_{\sf d,rd}$. We can also show that $\boldsymbol{c} \not\in \mathcal{S}_{\sf r,rd}$,  $\boldsymbol{c}  \not\in  \mathcal{S}_{\sf rd,r}$ and  $\boldsymbol{c}  \not\in  \mathcal{S}_{\sf rd,rd}$ via  simple algebraic manipulations. Similarly any other LGR-pair can be shown to be mutually exclusive.

\subsection{Properties of \oa} \label{prop_OA}
We observe that \oa has two underlying properties. First, there exists a certain \emph{saturation threshold} $\mathsf{P_{sat}}$ such that for power budget  $P < \mathsf{P_{sat}}$, \oa allocates powers as follows: \begin{itemize}
	\item if $P$ is \emph{sufficiently small} (\ied $P$ satisfies the condition of one of $\mathcal{A}_{x,y}, x, y \in \{ \sf d,r\}$), for each source \oa transmits only in the \emph{strongest} of the relay and direct links from that source.
	\item as $P$ increases, for at least one source, \oa transmits in \emph{both} the relay and direct links of that source, and \oa thus transmits in $3$ of the $4$ mm-wave links. As $P$ increases further, depending on link gains, \oa may eventually transmit in the only remaining link as well. Thus, for $P < \mathsf{P_{sat}}$, all link powers are either zero, or increase piecewise linearly with $P$. \end{itemize}

This property of \oa resembles the Waterfilling (WF) \cite[Chap.~10.4]{cover_thomas} property for parallel AWGN channels and thus is referred to as the \emph{WF-like} property. All LGRs satisfying this property are denoted by LGRs $\mathcal{A}_{x,y}, x, y \in \{\sf d,r, rd\}$. 	 Specifically, depending on the direct and relay link gains, \oa transmits in one of the following sets of links: $\mathrm{(i)}$ $\mathtt{DL}_1$ and $\mathtt{DL}_2$ if $d_1 \geq r_1$, $d_2 \geq r_2$, $\mathrm{(ii)}$ $\mathtt{RL}_1$ and $\mathtt{RL}_2$ if $r_1 > d_1$,  $r_2 > d_2$, $\mathrm{(iii)}$ $\mathtt{DL}_1$ and $\mathtt{RL}_2$ if $d_1 \geq r_1$,  $r_2 > d_2$, and  $\mathrm{(iv)}$ $\mathtt{DL}_2$ and $\mathtt{RL}_1$ if $d_2 \geq r_2$,  $r_1 > d_1$. Clearly, the corresponding LGRs are $\mathcal{A}_{\sf d,d}$, $\mathcal{A}_{\sf r,r}$,  $\mathcal{A}_{\sf d,r}$ and $\mathcal{A}_{\sf r,d}$.

Since the marginal return from transmitting only in the strongest link of each source diminishes as $P$ increases, for sufficiently large $P$ (that is below $\mathsf{P_{sat}}$) \oa transmits in one additional link. For example, consider a given $(r_1, r_2, d_1, d_2, P)$-tuple such that for $P  <  \min(\mathsf{P}_{\sf d,d},\hat{\mathsf{P}}_{\sf d,d})$, \oa transmits in links $\mathtt{DL}_1$ and $\mathtt{DL}_2$ only as in $\mathcal{A}_{\sf d,d}$. Now, if $\mathsf{P}_{\sf d,d}  <  \hat{\mathsf{P}}_{\sf d,d}$ holds, then for $\mathsf{P}_{\sf d,d}  \leq  P \leq \min(\hat{\mathsf{P}}_{\sf d,d}, \mathsf{P}_{\sf rd,d})$, \oa transmits in relay link $\mathtt{RL}_1$ for source $\mathsf{S}_1$ as well following the allocation in LGR $\mathcal{A}_{\sf rd,d}$. Note that through LGRs $\mathcal{A}_{\sf d,d}$ and $\mathcal{A}_{\sf rd,d}$, the  powers $p_1$ and $q_1$ increase piecewise linearly with $P$, while $p_2=P$ increasing linearly with $P$ and $q_2 = 0$, as per the WF-like property.  

Similar to $\mathcal{A}_{\sf rd,d}$, LGRs $\mathcal{A}_{\sf d,rd}$, $\mathcal{A}_{\sf r,rd}$ and $\mathcal{A}_{\sf rd,r}$ follow the WF-like property as well. Specifically, the intuition behind LGR $\mathcal{A}_{\sf d,rd}$ follows by swapping the roles of the sources as in $\mathcal{A}_{\sf rd,d}$, whereas the intuition behind  $\mathcal{A}_{\sf r,rd}$ and $\mathcal{A}_{\sf rd,r}$ follow from $\mathcal{A}_{\sf d,rd}$ and $\mathcal{A}_{\sf rd,d}$ respectively by exchanging the roles of the relay and direct links. Finally, in $\mathcal{A}_{\sf rd,rd}$ \oa transmits in all $4$ links as in WF.

While for $P <\mathsf{P_{sat}}$, \oa follows the WF-like property, for  $P  \geq  \mathsf{P_{sat}}$, \oa limits the relay link powers such that $(1+r_1 q_1)(1+r_2 q_2)=\gamma$, \ied the saturation condition, holds. Thus, as $P$ increases beyond $\mathsf{P_{sat}}$, $q_1$ and $q_2$ can no longer both increase with $P$. However, the direct link powers $p_k = P-q_k,$ increase unbounded with $P$. This property is referred to as the saturation property and is clearly unlike  WF.  The $5$ LGRs satisfying this property are denoted by $\mathcal{S}_{(.,.)}$ in Table \ref{table_thm_1}. 
Given a $(r_1, r_2, d_1, d_2, \gamma)$-tuple, saturation \emph{first} occurs in one of LGRs $\mathcal{S}_{(.,.)}$, called the \emph{saturation LGR}, which is determined by how the resulting threshold powers compare. In either case, $\mathsf{P_{sat}}$ is given by the lower bound on $P$ in the respective LGR $\mathcal{S}_{(.,.)}$ in Table \ref{table_thm_1}, \egd if the saturation LGR is $\mathcal{S}_{\sf r,rd}$, then $\mathsf{P_{sat}} =\max(\mathsf{P}_{\sf r,r},\mathsf{P}_{\sf r,rd})$.

To understand saturation, suppose that for a given link gain tuple, saturation occurs in some LGR $\mathcal{S}_{(.,.)}$ for $P $ larger than the corresponding $\mathsf{P_{sat}}$. Also, recall that the objective of \oa is to maximize $R = \min\{ \Sigma_{\mathsf{R}},\Sigma_{\mathsf{D}}\}$. Note that at $P=0$ the  resulting allocation $p_k=q_k=0$ achieves $\Sigma_{\mathsf{R}} = \sigma_{\mathsf{R}}$ and $\Sigma_{\mathsf{D}}=\sigma_{\mathsf{D}}$, and since $\gamma > 1$  implies $ \sigma_{\mathsf{R}} < \sigma_{\mathsf{D}}$ from \eqref{gamma_def}, at $P=0$ only $R =  \Sigma_{\mathsf{R}} < \Sigma_{\mathsf{D}}$ is achieved. 

As $P$ increases, and consequently $p_k$ and $q_k$ increase following the WF-like property, $\Sigma_{\mathsf{R}}$ and $\Sigma_{\mathsf{D}}$  in \eqref{zr13}-\eqref{zr23} increase differently. As $P$ increases, the resulting increase in $p_k$ increases $\Sigma_{\mathsf{R}}$ and $\Sigma_{\mathsf{D}}$ equally, and hence $R =  \Sigma_{\mathsf{R}} \leq \Sigma_{\mathsf{D}}$ is maintained and the sum-rate-gap $\Delta R \eqdef \Sigma_{\mathsf{D}} - \Sigma_{\mathsf{R}} \geq 0$ is not affected by the increase in $p_k$. However, as $P$ increases, the resulting increase in $q_k$  increases only $\Sigma_{\mathsf{R}}$, and thus  $\Delta R$ decreases gradually.	 Naturally, at some $P=\mathsf{P_{sat}}$, $q_1$ and $q_2$ are alloted enough power such that $R =  \Sigma_{\mathsf{R}} = \Sigma_{\mathsf{D}}$, \ied $\Delta R = 0$ or equivalently  $(1+r_1 q_1)(1+r_2 q_2) = \gamma$ is achieved. For all $P \geq \mathsf{P_{sat}}$, $q_1$ and $q_2$ are then constrained to maintain $R =  \Sigma_{\mathsf{R}} = \Sigma_{\mathsf{D}}$, and the rest of the budget, \ied $p_k = P - q_k$ are alloted to the direct links.  

As earlier noted, for a given $(r_1, r_2, d_1, d_2, \gamma)$-tuple, saturation first occurs  in one of $5$ LGRs $\mathcal{S}_{(.,.)}$, and in each case, the optimal powers vary differently. Specifically, in $\mathcal{S}_{\sf r,rd}$, as $P$ increases, $q_1=P$ increases \emph{linearly} with $P$, and thus $p_1=P-q_1=0$. However, due to saturation, $q_2 = (\gamma/(1+Pr_1) - 1)/r_2$ \emph{decreases non-linearly} with $P$, and thus $p_2 = P-q_2$ increases non-linearly. The same trend is found in $\mathcal{S}_{\sf rd,r}$ where the role of the two sources are swapped as compared to $\mathcal{S}_{\sf r,rd}$.  In $\mathcal{S}_{\sf rd,rd}$, as $P$ increases, if $r_1 \geq r_2$ (resp. $r_1 < r_2$), $q_1$ and $q_2$ (resp. $q_2$ and $q_1$) monotonically increase and decrease non-linearly with $P$, while both $p_1$ and $p_2$ increase non-linearly. Finally, in $\mathcal{S}_{\sf rd,d}$, as $P$ increases, $q_1 = \frac{\gamma-1}{r_1}$ and $q_2 = 0$ remain fixed, and all additional increments of $P$ are allotted entirely to the direct links, whereas in $\mathcal{S}_{\sf d,rd}$, the same trend is followed with roles of the sources swapped.

Moreover, for a given  $(r_1, r_2, d_1, d_2, \gamma)$-tuple, while saturation \emph{first} occurs in one of LGRs $\mathcal{S}_{(.,.)}$ for $P \geq \mathsf{P}_{\sf sat}$ associated with that LGR, as $P$ increases further, one or more other LGRs $\mathcal{S}_{(.,.)}$ may become optimal where saturation continues to hold. Specifically, there exists a threshold $\mathsf{P}_{\sf fin} \geq \mathsf{P}_{\sf sat}$ such that for all $P \geq \mathsf{P}_{\sf fin}$, a specific LGR $\mathcal{S}_{(.,.)}$, denoted the \emph{final} LGR, remain active. To be more precise, we partition the relay link gains $\boldsymbol{r} \eqdef (r_1, r_2)$ into subsets $\mathcal{R}_{S1} \eqdef \{\boldsymbol{r}: r_1 \geq \gamma r_2\}$, $\mathcal{R}_1 \eqdef \{\boldsymbol{r}:  \gamma r_2 > r_1 \geq r_2\}$, $\mathcal{R}_2 \eqdef \{\boldsymbol{r}: \gamma r_1 > r_2 > r_1\}$, and $\mathcal{R}_{S2} \eqdef \{\boldsymbol{r}: r_2 \geq \gamma r_1\}$. Intuitively, in $\mathcal{R}_{S2}$, relay link $\mathtt{RL}_2$ is \emph{significantly stronger} than $\mathtt{RL}_1$ (\ied $r_2 \geq \gamma r_1$) while in $\mathcal{R}_{2}$, it is only stronger (\ied $r_2 > r_1$) but not significantly stronger (\ied $r_2 < \gamma r_1$). The intuitions for $\mathcal{R}_{S1}$ and $\mathcal{R}_{1}$ follow similarly. We observe that for a given $(r_1, r_2, d_1, d_2, \gamma)$-tuple, if
\begin{itemize}
	\item $\boldsymbol{r}  \in \mathcal{R}_1$ or $\mathcal{R}_2$: $\mathsf{P}_{\sf fin}\! =\! \max(\overbar{\mathsf{P}}_{\sf rd,d},  \overbar{\mathsf{P}}_{\sf d,rd}, \overbar{\mathsf{P}}_{\sf rd,r},\overbar{\mathsf{P}}_{\sf r,rd},$ $\mathsf{P}_{\sf rd,rd})$, and the final LGR is $\mathcal{S}_{\sf rd,rd}$.
	\item $\boldsymbol{r} \in \mathcal{R}_{S1}$: $\mathsf{P}_{\sf fin} = \max(\mathsf{P}_{\sf r,d},\mathsf{P}_{\sf rd,d},\overbar{\mathsf{P}}_{\sf rd,d})$, and the final LGR is $\mathcal{S}_{\sf rd,d}$.
	\item $\boldsymbol{r} \in \mathcal{R}_{S2}$: $\mathsf{P}_{\sf fin} = \max(\mathsf{P}_{\sf d,r},\mathsf{P}_{\sf d,rd},\overbar{\mathsf{P}}_{\sf d,rd})$, and the final LGR is $\mathcal{S}_{\sf d,rd}$. 
\end{itemize} Naturally, for some link gain tuples, the saturation and the final LGRs are the same; thus $\mathsf{P}_{\sf fin} = \mathsf{P}_{\sf sat}$.

\section{Evolution of Link Gain Regimes with the Power Budget} \label{wf_like_allocation}
In Table \ref{table_thm_1}, the LGRs are defined as partitions of the set of the power budget $P$. Since the threshold powers in Table \ref{table_critical_power} are functions of link gains, for a given $(r_1, r_2, d_1, d_2, \gamma)$-tuple and $P$, it is easy to determine which LGR is \emph{active} (\ied  according to which LGR,  \oa allocates the link powers). It is evident that, as $P$ increases, the active LGR changes as well, and thus \oa follows a set of active LGRs, called a \emph{LGR-path}, which reveals useful insights on the optimal power allocation.

Given a link gain tuple, the saturation can occur in one of $\mathcal{S}_{\sf r,rd}, \mathcal{S}_{\sf rd,r}, \mathcal{S}_{\sf rd,rd}, \mathcal{S}_{\sf rd,d}$ and $\mathcal{S}_{\sf d,rd}$, which leads to a vast number of LGR-paths and makes it difficult to interpret interesting insights. To simplify the exposition, we now assume the direct links to be symmetric, \ied $d  \eqdef d_1  = d_2 $.  Although this causes some loss of generality, the resulting paths are simplified. For example, under this assumption, for $\boldsymbol{r} \in \mathcal{R}_{2}$, LGRs $\mathcal{A}_{\sf r,d} = \mathcal{A}_{\sf rd,d} = \mathcal{A}_{\sf r,rd} = \mathcal{S}_{\sf r,rd} = \mathcal{S}_{\sf rd,d} = \mathcal{S}_{\sf d,rd} = \emptyset$, and  saturation can occur in either $\mathcal{S}_{\sf rd,r}$ or $\mathcal{S}_{\sf rd,rd}$ only. Nonetheless, the paths for the case with $ d_1  \neq d_2 $ can be similarly derived.

In this section, we  discuss the paths for $\boldsymbol{r} \in \mathcal{R}_2$ and $\boldsymbol{r} \in \mathcal{R}_{S2}$ only, as the paths for $\boldsymbol{r} \in \mathcal{R}_1$ and $\boldsymbol{r} \in \mathcal{R}_{S1}$ can be derived from those of $\boldsymbol{r} \in \mathcal{R}_2$ and $\boldsymbol{r} \in \mathcal{R}_{S2}$, by exchanging the roles of relay links $\mathtt{RL}_2$ and $\mathtt{RL}_1$ as well as direct links $\mathtt{DL}_2$ and $\mathtt{DL}_1$. 

\subsection{Case $\boldsymbol{r} \in \mathcal{R}_2$}  In this case, we have $7$  LGR-paths denoted $[S1],\ldots,[S7]$ and presented in Table \ref{table_LGR_sequence} with their underlying conditions, and the interval of $P$ for each LGR in the path.

\textit{{Initial LGR}:} While $ [S1],$ $ [S2],$ $ [S3]$ originate from the \emph{initial} LGR $\mathcal{A}_{\sf r,r}$, $[S4]$ originates from  $\mathcal{A}_{\sf d,d}$, and $[S5], [S6], [S7]$  from $\mathcal{A}_{\sf d,r}$. The initial LGRs vary based on how $d$ compares to $r_1$ and $r_2$. For example, if $d \geq r_2 \geq r_1 \iff 0 \leq \hat{\mathsf{P}}_{\sf d,d} \leq \mathsf{P}_{\sf d,d}$ (\ied each $\mathtt{DL}_k$ is stronger than $\mathtt{RL}_k$), following the WF-like property, \oa transmits only in the direct links as in LGR $\mathcal{A}_{\sf d,d}$. On the other hand, if $r_2 \geq r_1 > d \iff 0 \leq \mathsf{P}^{\prime}_{\sf d,d} \leq \hat{\mathsf{P}}^{\prime}_{\sf d,d} $ (\ied each $\mathtt{RL}_k$ is stronger than $\mathtt{DL}_k$), following the WF-like property, \oa transmits only in the relay links as in $\mathcal{A}_{\sf r,r}$. Furthermore, depending on how $ \mathsf{P}^{\prime}_{\sf d,d}, \hat{\mathsf{P}}^{\prime}_{\sf d,d}$ and $\mathsf{P}_{\sf r,r}$ compare, \oa follows one of the paths $[S1], [S2],[S3]$, as  in Table \ref{table_LGR_sequence}.

Similarly, for the  case of $r_2 > d > r_1\iff (\mathsf{P}_{\sf d,d}, \hat{\mathsf{P}}^{\prime}_{\sf d,d}) \succeq \mathbf{0}$, \oa transmits in the two stronger links $\mathtt{RL}_2$ and $\mathtt{DL}_1$ as in  $\mathcal{A}_{\sf d,r}$. Also, based on how $\mathsf{P}_{\sf d,d}, \hat{\mathsf{P}}^{\prime}_{\sf d,d}$ and $\mathsf{P}_{\sf rd,r}$ compare, one of paths $[S5], [S6],[S7]$ is followed. Nevertheless, the conditions in Table \ref{table_LGR_sequence} are indeed mutually exclusive and exhaustive for $\boldsymbol{r} \in \mathcal{R}_2$.

\textit{{Saturation cases}:}  In this case, saturation first occurs in either LGR $\mathcal{S}_{\sf rd,rd}$  or LGR $\mathcal{S}_{\sf rd,r}$   as follows. 

Saturation occurs in  $\mathcal{S}_{\sf rd,rd}$ if the condition of one of the paths $[S1],[S4],[S5]$ or $[S7]$ is met. Here, $\sf P_{ sat} = \mathsf{P}_{\sf rd,rd}$, and for all $P \geq \mathsf{P}_{\sf rd,rd}$, as $P$ increases, $q_2$ increases and $q_1$ decreases and approach constants  $q_k   \rightarrow  \bar{q}_k  \eqdef  \sqrt{ {\gamma}/{{r_l r_k}}}-r_k^{-1} >0$, as $P   \rightarrow   \infty$.  Intuitively, since in $\mathcal{S}_{\sf rd,rd}$, $(1+r_1 q_1)(1+r_2 q_2) = \gamma$ must hold, as $P$ increases, $q_1$ and $q_2$ both cannot increase. Since $\mathtt{RL}_2$ is stronger than $\mathtt{RL}_1$, as $P$ increases, \oa achieves the best rate by increasing $q_2$ and \emph{decreasing} $q_1$. However, since $\mathtt{RL}_2$ is not significantly stronger than $\mathtt{RL}_1$, \oa should transmit in both relay links for all $P \geq \mathsf{P}_{\sf rd,rd}$. Thus, $q_1$ and $q_2$ both remain non-zero and approach constant levels as $P   \rightarrow   \infty$. 

On the other hand, saturation first occurs in LGR $\mathcal{S}_{\sf rd,r}$ if the condition of one of the paths $[S2],[S3]$ or $[S6]$ holds. Here, $\sf P_{sat}  =  \max(\mathsf{P}_{\sf rd,r}, \mathsf{P}_{\sf r,r})$, and   $\mathcal{S}_{\sf rd,r}$ is active for only $\max(\mathsf{P}_{\sf rd,r}, \mathsf{P}_{\sf r,r}) \leq P \leq \overbar{\mathsf{P}}_{\sf rd,r}$. In $\mathcal{S}_{\sf rd,r}$, for source $\mathsf{S}_2$, \oa allocates  $(p_2, q_2) = (0,P)$. It shows that $\mathtt{RL}_2$ is significantly stronger than $\mathtt{DL}_2$ in the sense that transmitting only in $\mathtt{RL}_2$, as opposed to both in $\mathtt{RL}_2$ and $\mathtt{DL}_2$, provides the best rate. For source $\mathsf{S}_1$, \oa allocates $(p_1,q_1)  =  (P - q_1, \frac{1}{r_1}(\frac{\gamma}{1+Pr_2} -1 ))$. This indicates that neither of $\mathtt{RL}_1$ and $\mathtt{DL}_1$ is significantly stronger than the other in that transmitting in both links  results in the best rate. Clearly, as $P$ increases, $q_2 = P$ increases and $q_1 = \frac{1}{r_1}(\frac{\gamma}{1+Pr_2} -1 )$ \emph{decreases}, and hence \oa follows the same trend as in $\mathcal{S}_{\sf rd,rd}$. 

\textit{{Final LGR}:} For $P \geq \mathsf{P_{fin}} = \max(\mathsf{P}_{\sf rd,rd},\overbar{\mathsf{P}}_{\sf rd,r})$, all paths terminate at the  final LGR $\mathcal{S}_{\sf rd,rd}$.

\begin{table*}[t]  
	\renewcommand{\arraystretch}{1.1}\vspace*{0mm}
	\tabcolsep = 10pt
	\caption{  \small LGR paths for $\boldsymbol{r} \in \mathcal{R}_2$.  Table \ref{table_critical_power} provides the threshold powers in terms of link gains and $\gamma$. Each path originates from one of three initial LGRs $\mathcal{A}_{\sf r,r},\mathcal{A}_{\sf d,d}$  or $\mathcal{A}_{\sf d,r}$, and they terminate at the final LGR $ \mathcal{S}_{\sf rd,rd}$.}	\label{table_LGR_sequence}  
	\begin{center} \hspace*{0mm} 
		\begin{tabular}{ l   l   l  } \hlinewd{1pt}
			LGR path  & Condition  & Interval of $P$ in the respective LGRs in the path\\ \hlinewd{1pt} \\[-5mm]
			$[S1] : \mathcal{A}_{\sf r,r}    \rightarrow    \mathcal{A}_{\sf rd,r}   \rightarrow    \mathcal{A}_{\sf rd,rd}     \rightarrow     \mathcal{S}_{\sf rd,rd}$  & $0 \leq \mathsf{P}^{\prime}_{\sf d,d} \leq \hat{\mathsf{P}}^{\prime}_{\sf d,d}  \leq  \mathsf{P}_{\sf r,r}$ & $  [0, \mathsf{P}^{\prime}_{\sf d,d}) \; , \; [\mathsf{P}^{\prime}_{\sf d,d},\hat{\mathsf{P}}^{\prime}_{\sf d,d}) \; , \;  [\hat{\mathsf{P}}^{\prime}_{\sf d,d},\mathsf{P}_{\sf rd,rd})   \; , \;   [\mathsf{P}_{\sf rd,rd}, \infty)  $  \\  
			$[S2] : \mathcal{A}_{\sf r,r}    \rightarrow    \mathcal{A}_{\sf rd,r}   \rightarrow    \mathcal{S}_{\sf rd,r}     \rightarrow     \mathcal{S}_{\sf rd,rd}$  & $0 \leq \mathsf{P}^{\prime}_{\sf d,d} \leq  \mathsf{P}_{\sf r,r} \leq \hat{\mathsf{P}}^{\prime}_{\sf d,d}  $ & $  [0, \mathsf{P}^{\prime}_{\sf d,d})  \; , \;  [\mathsf{P}^{\prime}_{\sf d,d},\mathsf{P}_{\sf rd,r}) \; , \;   [\mathsf{P}_{\sf rd,r},\overbar{\mathsf{P}}_{\sf rd,r})   \; , \;   [\overbar{\mathsf{P}}_{\sf rd,r}, \infty)  $ \\  
			$[S3] : \mathcal{A}_{\sf r,r}    \rightarrow    \mathcal{S}_{\sf rd,r}    \rightarrow     \mathcal{S}_{\sf rd,rd}$ & $0 \leq  \mathsf{P}_{\sf r,r} \leq \mathsf{P}^{\prime}_{\sf d,d}  \leq \hat{\mathsf{P}}^{\prime}_{\sf d,d}  $ & $   [0, \mathsf{P}_{\sf r,r})  \; , \;    [\mathsf{P}_{\sf r,r},\overbar{\mathsf{P}}_{\sf rd,r})   \; , \;  [\overbar{\mathsf{P}}_{\sf rd,r}, \infty)  $ \\[2mm]  \hlinewd{.5pt} \\[-5mm]
			$[S4] : \mathcal{A}_{\sf d,d}$ $  \rightarrow   \mathcal{A}_{\sf d,rd}   \rightarrow    \mathcal{A}_{\sf rd,rd}    \rightarrow    \mathcal{S}_{\sf rd,rd}$  & $0 \leq \hat{\mathsf{P}}_{\sf d,d} \leq \mathsf{P}_{\sf d,d}$ & $   [0, \hat{\mathsf{P}}_{\sf d,d})  \; , \;  [\hat{\mathsf{P}}_{\sf d,d},\mathsf{P}_{\sf d,d}) \; , \;  [\mathsf{P}_{\sf d,d},\mathsf{P}_{\sf rd,rd})   \; , \;   [\mathsf{P}_{\sf rd,rd}, \infty)  $ \\[2mm] \hlinewd{.5pt}  \\[-5mm]
			$[S5] : \mathcal{A}_{\sf d,r}   \rightarrow   \mathcal{A}_{\sf rd,r}   \rightarrow    \mathcal{A}_{\sf rd,rd}    \rightarrow    \mathcal{S}_{\sf rd,rd}$  & $0 \leq \mathsf{P}_{\sf d,d} \leq \hat{\mathsf{P}}^{\prime}_{\sf d,d} \leq \mathsf{P}_{\sf rd,r}$ & $  [0, \mathsf{P}_{\sf d,d})  \; , \;  [\mathsf{P}_{\sf d,d},\hat{\mathsf{P}}^{\prime}_{\sf d,d}) \; , \;  [\hat{\mathsf{P}}^{\prime}_{\sf d,d},\mathsf{P}_{\sf rd,rd})   \; , \;   [\mathsf{P}_{\sf rd,rd}, \infty)  $ \\ 
			$[S6] : \mathcal{A}_{\sf d,r}     \rightarrow    \mathcal{A}_{\sf rd,r}   \rightarrow    \mathcal{S}_{\sf rd,r}     \rightarrow     \mathcal{S}_{\sf rd,rd}$  & $0 \leq \mathsf{P}_{\sf d,d} \leq \mathsf{P}_{\sf rd,r} \leq \hat{\mathsf{P}}^{\prime}_{\sf d,d} $ & $   [0, \mathsf{P}_{\sf d,d})  \; , \;   [\mathsf{P}_{\sf d,d},\mathsf{P}_{\sf rd,r}) \; , \;   [\mathsf{P}_{\sf rd,r},\overbar{\mathsf{P}}_{\sf rd,r})   \; , \;    [\overbar{\mathsf{P}}_{\sf rd,r}, \infty)  $ \\ 
			$[S7] : \mathcal{A}_{\sf d,r}     \rightarrow    \mathcal{A}_{\sf d,rd}   \rightarrow    \mathcal{A}_{\sf rd,rd}     \rightarrow     \mathcal{S}_{\sf rd,rd}$  & $ 0 \leq  \hat{\mathsf{P}}^{\prime}_{\sf d,d} \leq  \mathsf{P}_{\sf d,d} $  & $   [0, \hat{\mathsf{P}}^{\prime}_{\sf d,d})  \; , \;   [\hat{\mathsf{P}}^{\prime}_{\sf d,d},\mathsf{P}_{\sf d,d}) \; , \;   [\mathsf{P}_{\sf d,d},\mathsf{P}_{\sf rd,rd})   \; , \;    [\mathsf{P}_{\sf rd,rd}, \infty)  $\\[2mm] \hlinewd{1pt} 
		\end{tabular}  
	\end{center}  
\end{table*}

\textit{{LGR-paths}:} We discuss path $[S2] $ in detail and use the obtained insights to interpret the other paths. Note that path $[S2]$  is followed if $0 \leq \mathsf{P}^{\prime}_{\sf d,d} \leq  \mathsf{P}_{\sf r,r} \leq \hat{\mathsf{P}}^{\prime}_{\sf d,d}$,  which can be interpreted as follows: 

${\sf (i)}$ Since $\mathtt{RL}_k$ is stronger than $\mathtt{DL}_k$, \ied $0 \leq \mathsf{P}^{\prime}_{\sf d,d} \leq \hat{\mathsf{P}}^{\prime}_{\sf d,d}$, for $P \in [0, \mathsf{P}^{\prime}_{\sf d,d})$, \oa allocates $P$ entirely to $\mathtt{RL}_1$ and $\mathtt{RL}_2$ as in $\mathcal{A}_{\sf r,r}$ (WF). Thus, $q_1=q_2=P$ increase with $P$, while $p_1=p_2=0$.

${\sf (ii)}$ As $P$ increases, the return from transmitting only in the relay links decreases. Here, $\mathtt{DL}_1$ is stronger than  $\mathtt{RL}_1$ in that $0  \leq  \mathsf{P}^{\prime}_{\sf d,d} \leq  \mathsf{P}_{\sf r,r}$. Hence, for  $P  \in  [\mathsf{P}^{\prime}_{\sf d,d},\mathsf{P}_{\sf rd,r})$, \oa achieves the best rate by transmitting in both $\mathtt{DL}_1$ and $\mathtt{RL}_1$ as in LGR $\mathcal{A}_{\sf rd,r}$, as opposed to only in $\mathtt{RL}_1$. Hence, for $P  \in  [\mathsf{P}^{\prime}_{\sf d,d},\mathsf{P}_{\sf rd,r})$,  \oa allocates power as in $\mathcal{A}_{\sf rd,r}$ where  $p_1, q_1$ and $q_2$ increase with $P$, and $p_2=0$.

${\sf (iii)}$ On the other hand, here $\mathtt{DL}_2$ is weak enough compared to $\mathtt{RL}_2$ in the sense of $0 \leq   \mathsf{P}_{\sf r,r} \leq \hat{\mathsf{P}}^{\prime}_{\sf d,d}$. Hence, for $P \in [\mathsf{P}_{\sf rd,r},\overbar{\mathsf{P}}_{\sf rd,r})$, the best rate is achieved by transmitting only in $\mathtt{RL}_2$, as opposed to sharing $P$ with $\mathtt{DL}_2$. Meanwhile, saturation occurs at $P = \mathsf{P_{sat}} = \mathsf{P}_{\sf rd,r}$ and LGR $\mathcal{S}_{\sf rd,r}$ becomes active.  Then, for $P \in [\mathsf{P}_{\sf rd,r},\overbar{\mathsf{P}}_{\sf rd,r})$, $p_1$ and $q_2$ increase with $P$, while $q_1$ \emph{decreases} and $p_2$ is  $p_2=0$.

${\sf (iv)}$ Finally, for $P \geq \overbar{\mathsf{P}}_{\sf rd,r}$, LGR $\mathcal{S}_{\sf rd,rd}$ becomes active.

Path $[S1]$  is similar to $[S2]$ except that $\mathtt{DL}_2$ is now strong enough compared to $\mathtt{RL}_2$ in the sense of $0 \leq \hat{\mathsf{P}}^{\prime}_{\sf d,d}  \leq   \mathsf{P}_{\sf r,r}$, which is opposite to that in $[S2]$. Hence, instead of transmitting only in $\mathtt{RL}_2$ as in $[S2]$, \oa now achieves the best rate by transmitting in both $\mathtt{DL}_2$ and $\mathtt{RL}_2$ as in LGR $\mathcal{A}_{\sf rd,rd}$. Finally, as $P$ increases,  saturation occurs in LGR $\mathcal{S}_{\sf rd,rd}$, which remains active for $P \geq \mathsf{P_{sat}} = \mathsf{P}_{\sf rd,rd}$. 

Path $[S3]$  is similar to $[S2]$ except that both direct links are weaker than the relay links in  that $\mathsf{P}_{\sf r,r} \leq \mathsf{P}^{\prime}_{\sf d,d}  \leq \hat{\mathsf{P}}^{\prime}_{\sf d,d}$. Hence, as $P$ increases, instead of transmitting in $\mathtt{DL}_1$ as in $[S2]$, the best rate is achieved by transmitting only in the relay links. Thus, $\mathcal{A}_{\sf rd,r}$ is skipped as compared to $[S2]$. As $P$ increases further,  saturation occurs in  $\mathcal{S}_{\sf rd,r}$ at $P = \mathsf{P_{sat}} = \mathsf{P}_{\sf r,r}$, and for $P \geq \overbar{\mathsf{P}}_{\sf rd,r}$ LGR $\mathcal{S}_{\sf rd,rd}$ is active.

Path $[S4]$ is complementary to $[S2]$ in that each $\mathtt{DL}_k$ is now stronger than $\mathtt{RL}_k$, \ied  $0  \leq  \hat{\mathsf{P}}_{\sf d,d} \leq \mathsf{P}_{\sf d,d}$. Here, \oa follows $\mathcal{A}_{\sf d,d}$, $\mathcal{A}_{\sf d,rd}$, and $\mathcal{A}_{\sf rd,rd}$ according to the WF-like property, and then follows $\mathcal{S}_{\sf rd,rd}$ according to the saturation property as in Table \ref{table_LGR_sequence}, and thus the details are omitted.

Finally, for the  case of $r_2 > d > r_1$, where $\mathtt{DL}_1$ is stronger than $\mathtt{RL}_1$ and $\mathtt{DL}_2$ is weaker than $\mathtt{RL}_2$, \oa follows $[S5],[S6]$ and $[S7]$ similarly to $[S1],[S2]$ and $[S4]$ respectively. For $P \in [0,\min(\mathsf{P}_{\sf d,d},\hat{\mathsf{P}}^{\prime}_{\sf d,d}))$, \oa transmits only in $\mathtt{DL}_1$ and $\mathtt{RL}_2$ as in LGR $\mathcal{A}_{\sf d,r}$, and then transmits in another link following $\mathcal{A}_{\sf d,rd}$ or $\mathcal{A}_{\sf rd,r}$. Then, for large enough $P$, depending on whichever achieves the best rate, either $\mathcal{S}_{\sf rd,r}$ (saturation) or $\mathcal{A}_{\sf rd,rd}$ (WF fashion) becomes active as in path $[S6]$ or $[S5],[S7]$. Eventually, for large enough $P$, $\mathcal{S}_{\sf rd,rd}$ is active. The details are omitted to avoid repetition.

\begin{table*}[t] 
	\renewcommand{\arraystretch}{1.1}
	\tabcolsep = 10pt
	\caption{ \small LGR paths for $\boldsymbol{r} \in \mathcal{R}_{S2}$.  Table \ref{table_critical_power} provides the threshold powers in terms of link gains and $\gamma$. Each path originates from one of three different LGRs $\mathcal{A}_{\sf r,r},\mathcal{A}_{\sf d,d}$  or $\mathcal{A}_{\sf d,r}$, and they terminate at the final  LGR $ \mathcal{S}_{\sf d,rd}$.}
	\label{table_LGR_sequence_21}  
	\centering 
	\begin{tabular}{ l   l   l } 
		\hlinewd{1pt}
		LGR path  & Condition  & Interval of $P$ in each LGR respectively \\ \hlinewd{1pt}   \\[-5mm]
		$[T3]: \mathcal{A}_{\sf r,r}    \rightarrow    \mathcal{S}_{\sf rd,r}    \rightarrow     \mathcal{S}_{\sf rd,rd}    \rightarrow     \mathcal{S}_{\sf d,rd} $ & $0 \leq  \mathsf{P}_{\sf r,r} \leq \mathsf{P}^{\prime}_{\sf d,d}  \leq \hat{\mathsf{P}}^{\prime}_{\sf d,d},  \overbar{\mathsf{P}}_{\sf rd,r} \leq  \mathsf{P}_{\sf d,r} $ & $  [0, \mathsf{P}_{\sf r,r})    , \;     [\mathsf{P}_{\sf r,r},\overbar{\mathsf{P}}_{\sf rd,r})     , \;    [\overbar{\mathsf{P}}_{\sf rd,r}, \overbar{\mathsf{P}}_{\sf d,rd}), [\overbar{\mathsf{P}}_{\sf d,rd}, \infty)$  \\
		$[N1]: \mathcal{A}_{\sf r,r}    \rightarrow    \mathcal{S}_{\sf rd,r}     \rightarrow     \mathcal{S}_{\sf d,rd}$ & $0 \leq  \mathsf{P}_{\sf r,r} \leq \mathsf{P}^{\prime}_{\sf d,d}  \leq \hat{\mathsf{P}}^{\prime}_{\sf d,d},  \overbar{\mathsf{P}}_{\sf rd,r} >  \mathsf{P}_{\sf d,r}  $ & $   [0, \mathsf{P}_{\sf r,r})    , \;    [\mathsf{P}_{\sf r,r},\mathsf{P}_{\sf d,r})     , \;    [\mathsf{P}_{\sf d,r}, \infty)  $\\[2mm] \hlinewd{.5pt} \\[-5mm]
		$[T4]: \mathcal{A}_{\sf d,d}$ $  \rightarrow   \mathcal{A}_{\sf d,rd}   \rightarrow    \mathcal{A}_{\sf rd,rd}    \rightarrow  $  & $0 \leq \hat{\mathsf{P}}_{\sf d,d} \leq \mathsf{P}_{\sf d,d} \leq \mathsf{P}_{\sf d,rd} $ &  $   [0, \hat{\mathsf{P}}_{\sf d,d})   , \;   [\hat{\mathsf{P}}_{\sf d,d},\mathsf{P}_{\sf d,d})   , \;   [\mathsf{P}_{\sf d,d},\mathsf{P}_{\sf rd,rd}),$\\ 
		\quad \quad \;\;\;$\mathcal{S}_{\sf rd,rd}   \rightarrow     \mathcal{S}_{\sf d,rd}$ &&  $[\mathsf{P}_{\sf rd,rd}, \overbar{\mathsf{P}}_{\sf d,rd}), \; [\overbar{\mathsf{P}}_{\sf d,rd}, \infty) $ \\ 
		$[N2]: \mathcal{A}_{\sf d,d}$ $  \rightarrow   \mathcal{A}_{\sf d,rd}   \rightarrow    \mathcal{S}_{\sf d,rd}$  & $0 \leq \hat{\mathsf{P}}_{\sf d,d} \leq \mathsf{P}_{\sf d,rd}  \leq \mathsf{P}_{\sf d,d}$  &  $  [0, \hat{\mathsf{P}}_{\sf d,d})    , \;    [\hat{\mathsf{P}}_{\sf d,d},\mathsf{P}_{\sf d,rd})   , \;    [\mathsf{P}_{\sf d,rd}, \infty)  $ \\[2mm] \hlinewd{.5pt} \\[-5mm]
		$[T5]: \mathcal{A}_{\sf d,r}     \rightarrow    \mathcal{A}_{\sf rd,r}   \rightarrow    \mathcal{A}_{\sf rd,rd}     \rightarrow     $  & $ 0 \leq  \mathsf{P}_{\sf d,d} \leq  \hat{\mathsf{P}}^{\prime}_{\sf d,d}\leq  \mathsf{P}_{\sf rd,r}  \leq \mathsf{P}_{\sf d,r}$ & $  [0, \mathsf{P}_{\sf d,d})    , \;    [\mathsf{P}_{\sf d,d},\hat{\mathsf{P}}^{\prime}_{\sf d,d})   , \;   [\hat{\mathsf{P}}^{\prime}_{\sf d,d},\mathsf{P}_{\sf rd,rd}),$ \\
		\quad \quad \;\;\;$\mathcal{S}_{\sf rd,rd}   \rightarrow     \mathcal{S}_{\sf d,rd} $ & & $ [\mathsf{P}_{\sf rd,rd}, \overbar{\mathsf{P}}_{\sf d,rd}), \; [\overbar{\mathsf{P}}_{\sf d,rd}, \infty)  $ \\ 
		$[T6]:  \mathcal{A}_{\sf d,r}     \rightarrow    \mathcal{A}_{\sf rd,r}   \rightarrow    \mathcal{S}_{\sf rd,r}     \rightarrow    $ & $ 0 \leq  \mathsf{P}_{\sf d,d}  \leq \mathsf{P}_{\sf rd,r}  \leq  \hat{\mathsf{P}}^{\prime}_{\sf d,d} \leq \mathsf{P}_{\sf d,r}$, or &  $  [0, \mathsf{P}_{\sf d,d})    , \;    [\mathsf{P}_{\sf d,d},\mathsf{P}_{\sf rd,r})   , \;   [\mathsf{P}_{\sf rd,r},\overbar{\mathsf{P}}_{\sf rd,r})$ \\ 
		\quad \quad \;\;\;$ \mathcal{S}_{\sf rd,rd}   \rightarrow     \mathcal{S}_{\sf d,rd} $& $ 0 \leq  \mathsf{P}_{\sf d,d}  \leq \overbar{\mathsf{P}}_{\sf rd,r}  \leq  \mathsf{P}_{\sf d,r} \leq \hat{\mathsf{P}}^{\prime}_{\sf d,d} $ & $ [\overbar{\mathsf{P}}_{\sf rd,r}, \overbar{\mathsf{P}}_{\sf d,rd}), \; [\overbar{\mathsf{P}}_{\sf d,rd}, \infty) $ \\ 
		$[T7]: \mathcal{A}_{\sf d,r}     \rightarrow    \mathcal{A}_{\sf d,rd}   \rightarrow    \mathcal{A}_{\sf rd,rd}     \rightarrow    $ & $ 0 \leq  \hat{\mathsf{P}}^{\prime}_{\sf d,d} \leq \mathsf{P}_{\sf d,d}  \leq \mathsf{P}_{\sf d,r}$, or & $  [0, \hat{\mathsf{P}}^{\prime}_{\sf d,d})    , \;   [\hat{\mathsf{P}}^{\prime}_{\sf d,d},\mathsf{P}_{\sf d,d})   , \;   [\mathsf{P}_{\sf d,d},\mathsf{P}_{\sf rd,rd})$ \\
		\quad \quad \;\;\;$ \mathcal{S}_{\sf rd,rd}   \rightarrow     \mathcal{S}_{\sf d,rd} $&  $ 0 \leq  \hat{\mathsf{P}}^{\prime}_{\sf d,d} \leq \mathsf{P}_{\sf d,r}  \leq \mathsf{P}_{\sf d,d} \leq \mathsf{P}_{\sf d,rd}$ & $ [\mathsf{P}_{\sf rd,rd},\overbar{\mathsf{P}}_{\sf d,rd}),, \; [\overbar{\mathsf{P}}_{\sf d,rd}, \infty)   $  \\ 
		$[N3]: \mathcal{A}_{\sf d,r}     \rightarrow    \mathcal{A}_{\sf rd,r}     \rightarrow     \mathcal{S}_{\sf rd,r}   \rightarrow     \mathcal{S}_{\sf d,rd} $  & $ 0 \leq \mathsf{P}_{\sf d,d} \leq  \mathsf{P}_{\sf d,r}  \leq  \overbar{\mathsf{P}}_{\sf rd,r}  \leq  \hat{\mathsf{P}}^{\prime}_{\sf d,d}$ & $  [0, \mathsf{P}_{\sf d,d})    , \;   [\mathsf{P}_{\sf d,d},\mathsf{P}_{\sf rd,r})   , \;   [\mathsf{P}_{\sf rd,r},\mathsf{P}_{\sf d,r}), \; [\mathsf{P}_{\sf d,r}, \infty) $ \\ 
		$[N4]: \mathcal{A}_{\sf d,r}     \rightarrow    \mathcal{A}_{\sf d,rd}   \rightarrow     \mathcal{S}_{\sf d,rd} $  & $ 0 \leq  \hat{\mathsf{P}}^{\prime}_{\sf d,d} \leq  \mathsf{P}_{\sf d,r} \leq  \mathsf{P}_{\sf d,rd}  \leq \mathsf{P}_{\sf d,d}$ & $  [0, \hat{\mathsf{P}}^{\prime}_{\sf d,d})    , \;   [\hat{\mathsf{P}}^{\prime}_{\sf d,d},\mathsf{P}_{\sf d,rd})    , \;    [\mathsf{P}_{\sf d,rd}, \infty)  $ \\ 
		$[N5]: \mathcal{A}_{\sf d,r}      \rightarrow     \mathcal{S}_{\sf d,rd} $  & $ 0 \leq \mathsf{P}_{\sf d,r} \leq  \min( \hat{\mathsf{P}}^{\prime}_{\sf d,d},  \mathsf{P}_{\sf d,d})$ & $  [0, \mathsf{P}_{\sf d,r})   , \;    [\mathsf{P}_{\sf d,r}, \infty)  $ \\[2mm] \hlinewd{1pt}
	\end{tabular} 
\end{table*}

\subsection{Case $\boldsymbol{r} \in \mathcal{R}_{S2}$}  In this case,  we have $10$ paths, denoted $[T3],\ldots, [T7]$ and $[N1],\ldots, [N5]$ and given in Table \ref{table_LGR_sequence_21}. Paths $[T3],\ldots, [T7]$ are the counterparts of paths $[S3],\ldots, [S7]$ in Table \ref{table_LGR_sequence} with $\mathcal{S}_{\sf d,rd}$ appended as the final LGR, and thus are denoted in this manner. Also, paths $[S1]$ and $[S2]$ do not have any counterparts here, and thus $[T1]$ and $[T2]$ are not defined. Moreover, paths $[N1],\ldots, [N5]$ are valid exclusively for $\boldsymbol{r} \in \mathcal{R}_{S2}$.

\textit{Initial LGR:} While $[T3]$ and $[N1]$ originate from the initial LGR $\mathcal{A}_{\sf r,r}$, $[T4]$ and $[N2]$ originate from LGR $\mathcal{A}_{\sf d,d}$, and $[T5],\ldots [N5]$ originate from LGR $\mathcal{A}_{\sf d,r}$.  The initial LGRs vary depending on how $d$ compares to $r_1$  and $r_2$ as in the case of $\boldsymbol{r} \in \mathcal{R}_2$, hence is not repeated here. 

\textit{Saturation cases:} Saturation first occurs in one of LGRs $\mathcal{S}_{\sf rd,rd}$, $\mathcal{S}_{\sf d,rd}$ and $\mathcal{S}_{\sf rd,r}$.

Saturation first occurs in $\mathcal{S}_{\sf rd,rd}$ if the condition of one of the paths $[T4], [T5]$ or $[T7]$ is met. Here, $ \mathsf{P}_{\sf sat} = \max(\overbar{\mathsf{P}}_{\sf rd,r}, \mathsf{P}_{\sf rd,rd})$. Unlike in case $\boldsymbol{r} \in \mathcal{R}_2$, LGR $\mathcal{S}_{\sf rd,rd}$ is now active only for the finite range $\max(\overbar{\mathsf{P}}_{\sf rd,r}, \mathsf{P}_{\sf rd,rd}) \leq P \leq \overbar{\mathsf{P}}_{\sf d,rd}$. Intuitively, $\mathtt{RL}_2$ is now significantly stronger than $\mathtt{RL}_1$ (\ied $r_2 > \gamma r_1$), hence transmitting in both relay links as in $\mathcal{S}_{\sf rd,rd}$ is optimal only for this finite range.

Saturation first  occurs in $\mathcal{S}_{\sf rd,r}$ if the condition of one of the paths $[T3], [N1], [T6]$ or $[N3]$ hold. Here, $ \mathsf{P}_{\sf sat}  =  \max(\mathsf{P}_{\sf rd,r}, \mathsf{P}_{\sf r,r})$, and  $\mathcal{S}_{\sf rd,r}$ is active for the range  $\max(\mathsf{P}_{\sf rd,r}, \mathsf{P}_{\sf r,r})  \leq  P  \leq  \min(\overbar{\mathsf{P}}_{\sf rd,r}, \mathsf{P}_{\sf d,r})$.

Finally, saturation first occurs in $\mathcal{S}_{\sf d,rd}$ when the condition of one of the paths $[N2], [N4]$ or $[N5]$ hold. Here, for all $P \geq \mathsf{P}_{\sf sat} = \max(\mathsf{P}_{\sf d,r}, \mathsf{P}_{\sf d,rd}, \overbar{\mathsf{P}}_{\sf d,rd})$, LGR $\mathcal{S}_{\sf d,rd}$ is active. In $\mathcal{S}_{\sf d,rd}$, as $P$ increases, $q_2 = (\gamma-1)/r_2 > 0$ and $ q_1=0$ are fixed, and all additional increments of $P$ are allotted to the direct links only. Intuitively, since $\mathtt{RL}_2$ is significantly stronger than $\mathtt{RL}_1$, for all $P \geq \sf P_{ sat}$, the best rate is achieved by transmitting only in $\mathtt{RL}_2$.

\textit{Final LGR:}  For $P \geq \mathsf{P_{fin}} = \max(\overbar{\mathsf{P}}_{\sf d,rd}, \mathsf{P}_{\sf d,rd}, \mathsf{P}_{\sf d,r})$, all paths terminate at the final LGR  $\mathcal{S}_{\sf d,rd}$.

\textit{LGR-paths:} Since paths $[T3],\ldots, [T7]$ can be interpreted similarly to paths $[S3],\ldots, [S7]$, they  are not detailed here. Hence, we only discuss paths $[N1],\ldots, [N5]$ briefly.

Path $[N1]$ is similar to $[T3]$ with $\mathcal{S}_{\sf rd,rd}$ skipped. Compared  to $[T3]$, here $\mathtt{RL}_2$ is sufficiently stronger than $\mathtt{RL}_1$ in that $\mathsf{P}_{\sf d,r}  <  \overbar{\mathsf{P}}_{\sf rd,r}$. Hence,  and for $P  >  \mathsf{P}_{\sf d,r}$, the best rate is achieved by transmitting only in $\mathtt{RL}_2$ as in  $\mathcal{S}_{\sf d,rd}$ as compared to transmitting in both $\mathtt{RL}_1$ and $\mathtt{RL}_2$ as in  $\mathcal{S}_{\sf rd,rd}$. Hence, $\mathcal{S}_{\sf rd,rd}$ is skipped.

Path $[N2]$ is similar to $[T4]$ with $\mathcal{A}_{\sf rd,rd}$ and $\mathcal{S}_{\sf rd,rd}$ skipped. The conditions for $[N2]$ simplifies to $r_2 \geq r_1(2 \gamma -1)$. It shows that $\mathtt{RL}_2$ is so much stronger than $\mathtt{RL}_1$ that, for all $P \geq 0$,   the best rate is achieved by transmitting solely in $\mathtt{RL}_2$  and not transmitting in $\mathtt{RL}_1$ at all. Thus, compared to $[T4]$ where non-zero power is allocated to $\mathtt{RL}_1$ in LGRs $\mathcal{A}_{\sf rd,rd}$ and $\mathcal{S}_{\sf rd,rd}$, these LGRs are skipped here.

Likewise, $[N3]$ is similar to $[T6]$ with $\mathcal{S}_{\sf rd,rd}$ skipped, $[N4]$ to $[T7]$ with $\mathcal{A}_{\sf rd,rd}$ and $\mathcal{S}_{\sf rd,rd}$ skipped, and $[N5]$ to $[N4]$ with $\mathcal{A}_{\sf d,rd}$ skipped. The conditions for these paths can be  interpreted as $\mathtt{RL}_2$ being sufficiently stronger than $\mathtt{RL}_1$ in a sense similar to paths  $[N1]$ and  $[N2]$, so that for large enough $P$ \oa skips LGRs that allocate non-zero power to $\mathtt{RL}_1$ (\egd $\mathcal{S}_{\sf rd,rd}$, $\mathcal{A}_{\sf rd,rd}$ or $\mathcal{A}_{\sf d,rd}$).

\begin{figure}[] 
	\captionsetup[subfloat]{captionskip=2mm}
	\centering     
	\subfloat[{Path $[S5]$ with $(r_1, r_2, d, \gamma) = (1, 2.9, 1.3, 3)$.}]{\includegraphics[width=8 cm,height= 5.4 cm]{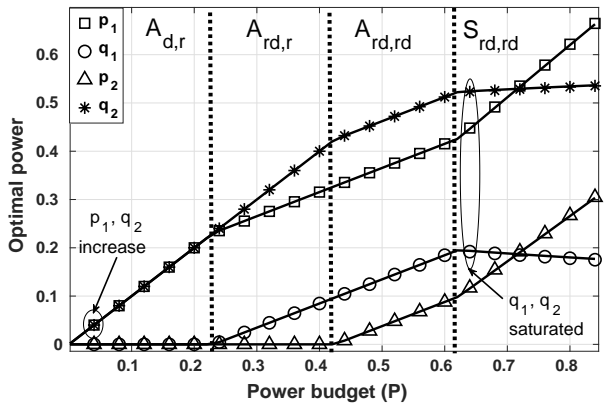}%
	\label{fig6:a}}  \hspace*{15mm}
	\subfloat[{Path $[T5]$ with $(r_1, r_2, d, \gamma) = (1, 4, 1.52, 3)$.}]{\includegraphics[width=8cm,height=5.4cm]{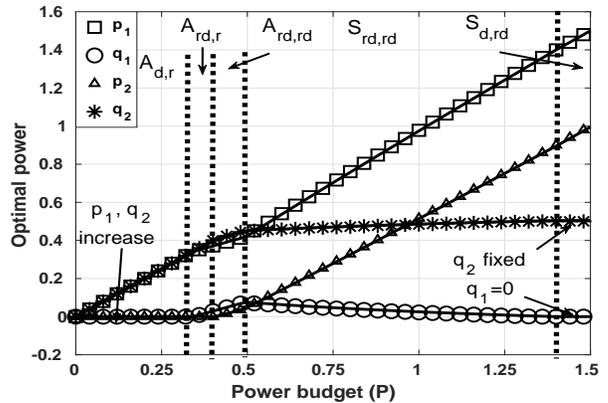}%
	\label{fig6:b}} 
	\caption{\small  (a) Path $[S5]$: for $P < \mathsf{P}_{\sf sat} = 0.62$, all link powers follow the WF-like property. At $P = \mathsf{P}_{\sf sat}$, saturation occurs in LGR $\mathcal{S}_{\sf rd,rd}$ and it remains active for all $P \geq \mathsf{P}_{\sf sat}$. (b) Path $[T5]$: saturation first occurs at $P = \mathsf{P}_{\sf sat}=0.49$ and LGR $\mathcal{S}_{\sf rd,rd}$ becomes active.   Then, for $P \geq \mathsf{P}_{\sf fin} = 1.34$, LGR $\mathcal{S}_{\sf d,rd}$ is active where $q_2=0.5$ and $q_1=0$ remain fixed. } 
\end{figure}

\vspace*{2mm} 
\textit{Numerical Examples:} We now illustrate examples of paths $[S5]$ and $[T5]$ in Fig.~\ref{fig6:a} and Fig.~\ref{fig6:b} respectively  by plotting the optimal link powers against budget $P$ for parameters $(r_1,r_2,d,\gamma)$ as noted in the respective figures.  In each example, the analytical expression of powers (marker-line) indeed match their numerically computed counterparts (solid line) using \texttt{CVX}  \cite{cvx}. We also verify that \oa follows the respective paths by labeling the active LGRs in the relevant intervals.

In Fig.~\ref{fig6:a}, we verify path $[S5]$ where  $\mathsf{P}_{\sf sat} =\mathsf{P}_{\sf fin}  =\mathsf{P}_{\sf rd,dr}=0.62$. Here, LGR $\mathcal{A}_{\sf d,r}$ is first active for $0 \leq P < \mathsf{P}_{\sf d,d}$, where $p_1 = q_2 = P$, while $q_1 =p_2=0$. Then, for  $\mathsf{P}_{\sf d,d} \leq P < \hat{\mathsf{P}}_{\sf d,d}$, LGR $\mathcal{A}_{\sf rd,r}$ becomes active where, in addition to $p_1$ and $q_2$, $q_1$ increases with $P$ as well. As $P$ increases, for $\hat{\mathsf{P}}^{\prime}_{\sf d,d} \leq P < \mathsf{P}_{\sf sat}$, LGR $\mathcal{A}_{\sf rd,rd}$ is active where all 4 powers increase with $P$. Finally, for $P \geq \mathsf{P}_{\sf sat}$, saturation occurs in $\mathcal{S}_{\sf rd,rd}$  where $q_2$ increases and $q_1$ decreases towards limits $\bar{q}_2 =0.67$ and $\bar{q}_1 =0.02$ (not shown in Fig.~\ref{fig6:a}), while $p_1,p_2$ grow unbounded with $P$.

We similarly verify $[T5]$ in Fig.~\ref{fig6:b} and omit the details since in $[T5]$, the first $4$ LGRs are the same as those of $[S5]$ in Fig.~\ref{fig6:a}. Nevertheless, for $[T5]$ while  saturation occurs at $\mathsf{P}_{\sf sat} = \mathsf{P}_{\sf rd,rd}=0.49$ in $\mathcal{S}_{\sf rd,rd}$, unlike  in $[S5]$, the final LGR is $\mathcal{S}_{\sf d,rd}$ where $q_2=0.5,q_1=0$ are fixed for all $P \geq \mathsf{P}_{\sf fin} = 1.34$.

\subsection{Special Cases and Further Insights} \label{special_case} 
\subsubsection{Symmetric case} For the symmetric case with $d = d_1=d_2$ and $r = r_1=r_2$, the symmetric power allocation $(p,q,p,q)$ is sum-rate optimal. Here, \oa follows one of the $3$ LGR-paths:

$({\sf i})$ if $d \geq r$ (\ied direct links are stronger than relay links): for $P \in [0,\frac{1}{r}-\frac{1}{d})$, \oa transmits only in the direct links as in  $\mathcal{A}_{\sf d,d}$, then for $P \in [\frac{1}{r}-\frac{1}{d}, \frac{2\gamma^{1/2}-1}{r} - \frac{1}{d})$ \oa transmits in all $4$ links as in $\mathcal{A}_{\sf rd,rd}$, and finally for $P \geq \frac{2\gamma^{1/2}-1}{r} - \frac{1}{d}$, saturation occurs in $\mathcal{S}_{\sf rd,rd}$ where $q = \frac{\gamma^{1/2}-1}{r}$ is fixed. 

$({\sf ii})$  if $d < r \leq d \gamma^{1/2}$ (\ied relay links are stronger but not significantly stronger): as opposed to  $\mathcal{A}_{\sf d,d}$ above, now $\mathcal{A}_{\sf r,r}$ is active for $P \in [0,\frac{1}{d}-\frac{1}{r})$, and then $ \mathcal{A}_{\sf rd,rd} $ and  $\mathcal{S}_{\sf rd,rd}$ become active as above.

$({\sf iii})$  if $r > d \gamma^{1/2}$ (relay links are significantly stronger): for $P \in [0,\frac{\gamma^{1/2}-1}{r})$ \oa transmits only in relay links as in $\mathcal{A}_{\sf r,r}$ until they saturate, and then for $P \geq \frac{\gamma^{1/2}-1}{r}$ $\mathcal{S}_{\sf rd,rd}$ becomes active.

\subsubsection{Large  mm-wave bandwidth} In this regime (\ied $\alpha   \rightarrow   \infty$),  $\gamma    \rightarrow    (1+\bar{G}_{\mathsf{R}\mathsf{D}} \bar{P}_{\mathsf{R}} )^2$, hence the saturation threshold is now a function of the mm-wave parameters only. We now examine how the optimal power allocation simplifies in two extreme scenarios. 	If $\bar{G}_{\mathsf{R}\mathsf{D}} \bar{P}_{\mathsf{R}} \gg 1$ (\ied $\gamma \gg 1$), saturation occurs for very large values of the power budget $P$. Hence, for practical finite $P$, when $P \geq \max(\mathsf{P}_{\sf d,d},\mathsf{P}^{\prime}_{\sf d,d},\hat{\mathsf{P}}_{\sf d,d},\hat{\mathsf{P}}^{\prime}_{\sf d,d})$, transmitting in all $4$ links as in LGR $\mathcal{A}_{\sf rd,rd}$ based on the WF-like property is optimal. 

Alternatively, if $\bar{G}_{\mathsf{R}\mathsf{D}} \bar{P}_{\mathsf{R}} \ll 1$  (\ied $\gamma \approx 1$),  saturation occurs for small values of $P$. Since allocating only a small proportion of $P$ to the relay links achieves saturation, as $P$ increases the remaining power (\ied almost all of $P$) is allotted to the direct links, resulting in an allocation similar to $\mathcal{A}_{\sf d,d}$.

\subsubsection{Optimum power allocation in a 2-D topology} \label{sub_sub3}
We now illustrate how the mode of optimal link powers varies as the source locations vary according to the 2-D topology of Fig.~\ref{fig2:a}, where  $\mathsf{R}$ and $\mathsf{D}$ are located on the x-axis at $(0,0)$ and $(0, {\sf d}_{\mathsf{R}\mathsf{D}})$, while the sources are located at $(-{\sf d}_{\mathsf{S}\mathsf{R}} \cos \phi, \pm {\sf d}_{\mathsf{S}\mathsf{R}} \sin \phi )$ with $\phi$ being the angle between the sources and the relay. Due to symmetric source placement,  the resulting link gains are symmetric, \ied $d = d_1=d_2$ and $r = r_1=r_2$, which simplifies the power allocation. Moreover, like the numerical section in Section IV, we assume that both bands are under phase fading. Thus, the  channel gains from node $s$ to $t$ in the microwave band are $G_{st} = 1/\sf{d}_{st}^{\beta_1}$ and  the mm-wave relay and direct link gains are $r = 1/\sf{d}_{\sf SR}^{\beta_2}$ and  $d = 1/\sf{d}_{\sf SD}^{\beta_2}$. 

\begin{figure}[t] 
	\centering    
	\includegraphics[width=7.5cm, height=7cm]{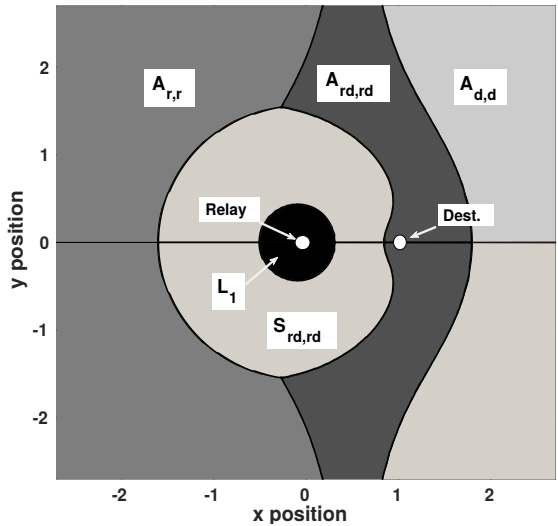} 
	\caption{\small For the 2-D network topology of the DR-MARC of Section \ref{sub_sub3}, the source locations are at coordinates $(x,\pm y)$. The set of source locations depicted here are partitioned into several regions, and for each regions the optimal transmission mode for the mm-wave links are labeled.}
	\label{LGR_symmetric}  
\end{figure}
For illustration, we take the following parameters $P_k=10, k \in \{1,2, \mathsf{R}\}, \bar{P}_{ \mathsf{R}}=1,$ $ \beta_1=2, \beta_2=4,$ $ \alpha=2$, while the power budget is $P=10$. We then plot the source locations in Fig.~\ref{LGR_symmetric} by varying $\phi \in (0,\pi)$ and $\sf{d}_{SR} \in (0,4)$  for fixed $\sf{d}_{RD}=1$ unit, and partition this space based on which mode of mm-wave transmission is optimal. First, in region $\sf L_1$, sources are much closer to the relay than the destination in that $\sigma_{\mathsf{R}} \geq \sigma_{\mathsf{D}}$ (\ied $\gamma \leq 1$), with $\sigma_{\mathsf{R}} $, $ \sigma_{\mathsf{D}}$ and $\gamma$ defined in \eqref{zr13}, \eqref{zr23} and \eqref{gamma_def}. Therefore, for sources located in  $\sf L_1$, it is optimal to transmit only in the direct links  for all $P \geq 0$.

All regions except $\sf L_1$, correspond to the case of $ \gamma > 1$, and depending on the budget $P$ and source locations (\ied the resulting direct and relay links gains), the optimal transmission mode in different regions vary. For example, the sources in the region labeled $\mathcal{A}_{\sf r,r}$ are not as close to the relay as in $\sf L_1$ but are sufficiently close  to the relay  such that $0 < P \leq {\sf d_{SD}^{\beta_2}} - {\sf d_{SR}^{\beta_2}}$ holds. Hence, for these source locations, allocating the budget $P$ entirely to the relay links is optimal. On the other hand, the sources in the region labeled $\mathcal{A}_{\sf d,d}$ are sufficiently close to the destination in that $0 < P \leq {\sf d_{SR}^{\beta_2}} - {\sf d_{SD}^{\beta_2}}$ holds. Hence, it is optimal to allocate the budget $P$ entirely to the direct links.  As opposed to these two regions, the sources in the region labeled $\mathcal{A}_{\sf rd,rd}$ are at an intermediate distance from the relay and the destination in that $P < ({2\gamma^{1/2}-1}){\sf d_{SR}^{\beta_2}} - {\sf d_{SD}^{\beta_2}}$ holds. Here, transmitting in all $4$ links as in $\mathcal{A}_{\sf rd,rd}$ is optimal. Finally, sources in the region $\mathcal{S}_{\sf rd,rd}$ are such that $P \geq ({2\gamma^{1/2}-1}){\sf d_{SR}^{\beta_2}} - {\sf d_{SD}^{\beta_2}}$ hold. Here, saturation occurs, and allocating power as in $\mathcal{S}_{\sf rd,rd}$ is optimal.
Clearly, for fixed $({\sf d_{SR}}, {\sf d_{SD}},\gamma)$, as $P$ increases, the region $\mathcal{S}_{\sf rd,rd}$  grows.

\begin{figure}[!tbp] 
	\captionsetup[subfloat]{captionskip=2mm}
	\centering  
	\subfloat[{Problem $[\mathcal{P}1]$}]{\includegraphics[width=8.4 cm, height=6  cm]{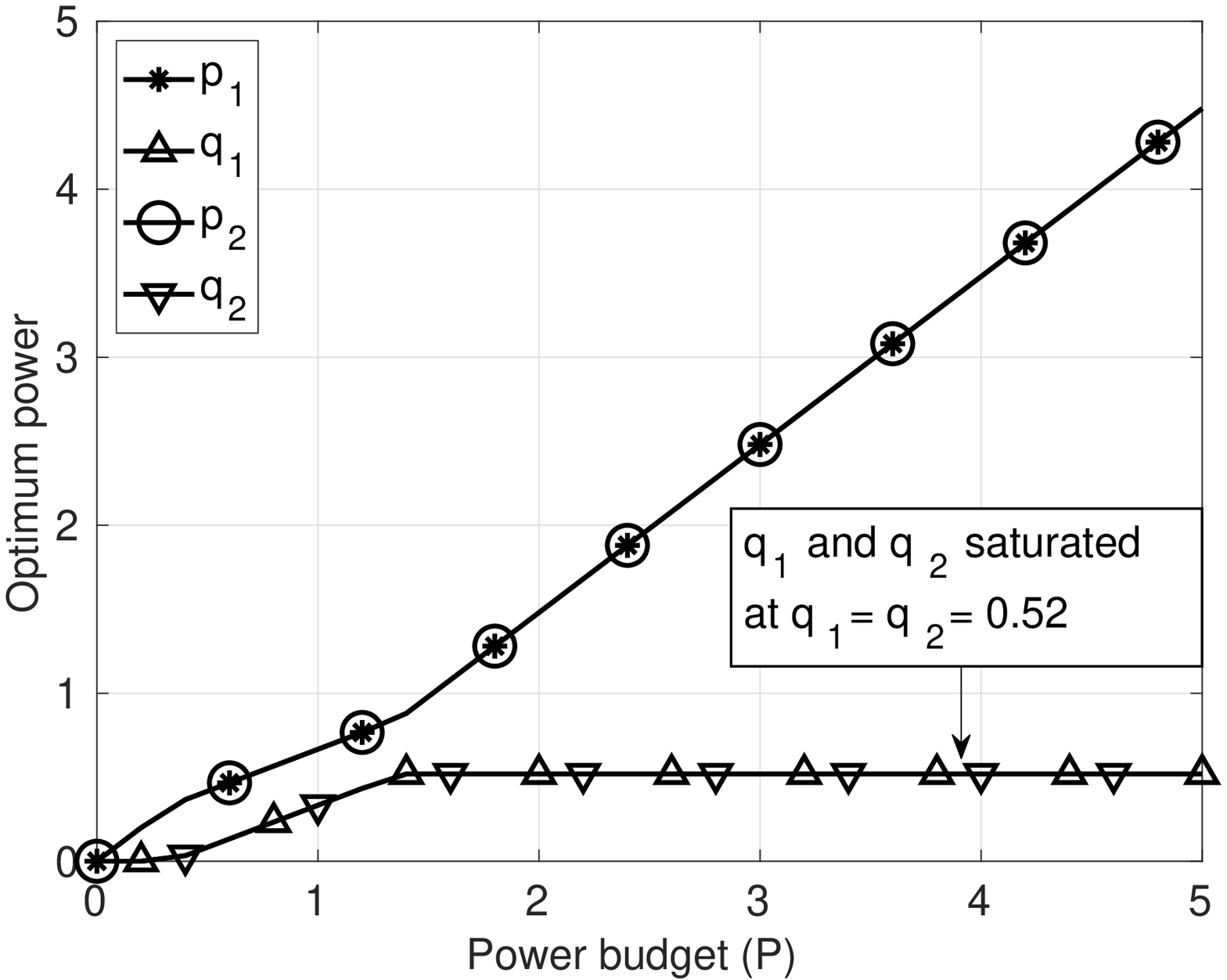} \hspace*{10mm} 
	\label{fig:02}}   
	\subfloat[{Problem $[\mathcal{P}2]$}]{\includegraphics[width=8.4 cm, height= 6  cm]{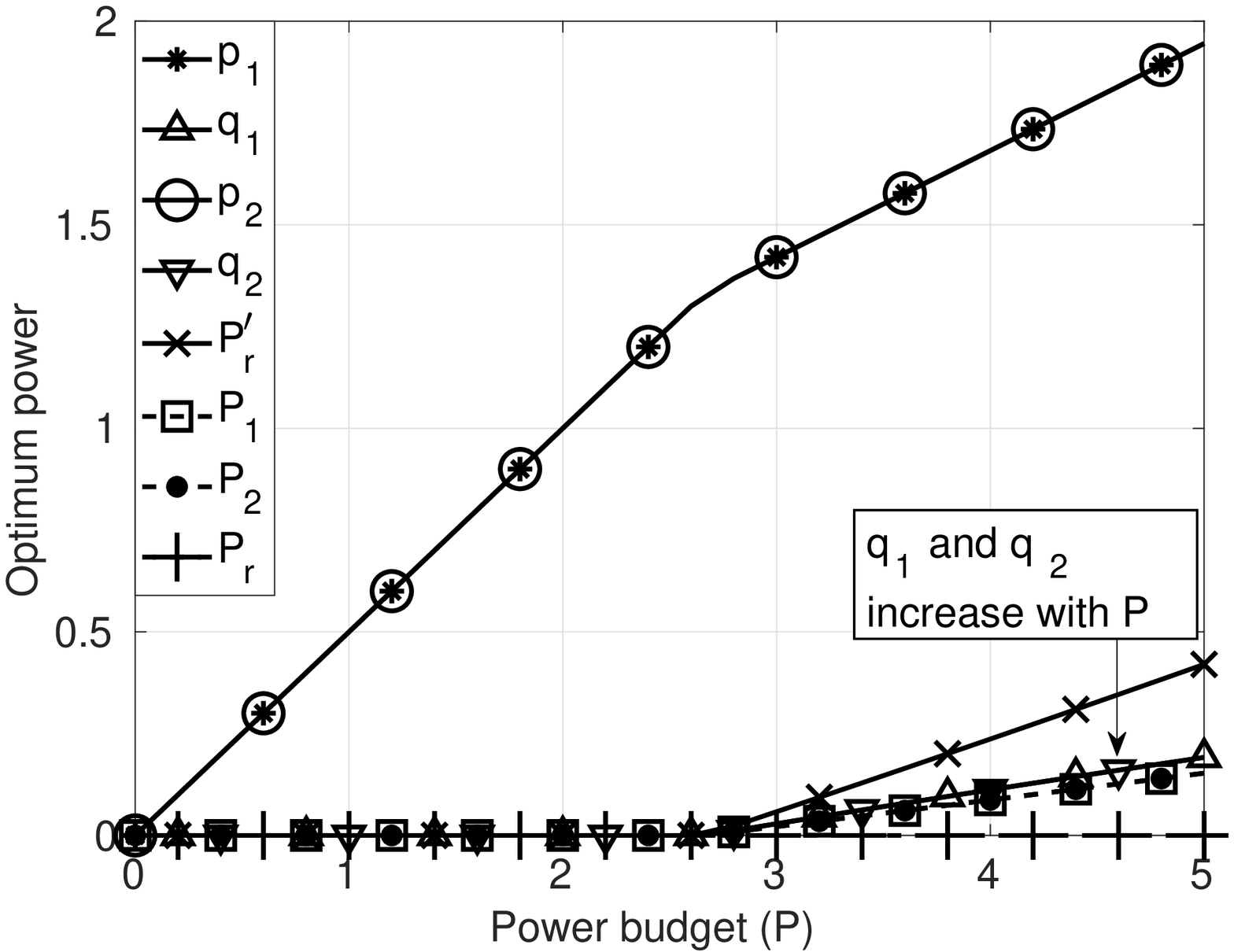}%
	\label{fig:04}} 
	\caption{\small Optimum transmit powers versus power budget $P$ for problems $[\mathcal{P}1]$ and $[\mathcal{P}2]$.}  
\end{figure}

\subsubsection{A joint optimum sum-rate problem over the integrated microwave and mm-wave dual-bands} \label{problem_P0}
As opposed to $[\mathcal{P}1]$ where the microwave link powers are fixed, it may also be interesting to study the optimum sum-rate problem when the total transmission power is to be shared by all mm-wave and microwave links to see whether the transmission powers have the same structure as in $[\mathcal{P}1]$. Nevertheless, sharing the power budgets for the microwave band and the mm-wave band may not be viable from practical and regulatory perspectives. Regulatory guidelines typically designate specific transmit power limits for each frequency band, and a transmit power scheme resulting from such a joint optimization may fail to comply with these limits. Moreover, the radio frequency chain of each frequency band is typically deployed separately and driven by dedicated power amplifiers, each with its own maximum power limit. A brief numerical study is presented below which demonstrates that the structure found in $[\mathcal{P}1]$ is not present when the problem is formulated with a sum-power constraint over mm-wave link and microwave link powers.

For a given total power budget $P$, the problem of jointly optimizing the sum-rate is formulated as 
\begin{align}
[\mathcal{P}2] \;\; \text{maximize} &\quad R  \notag \\ 
\text{subject to} &\quad R \leq \Sigma_{\mathsf{R}}, \label{conv_1_2}   \\ 
&\quad R \leq \Sigma_{\mathsf{D}},  \label{conv_2_2} \\  
&\textstyle  \quad   \sum\limits_{k \in \{1,2\}}  p_k + q_k +  \sum\limits_{k \in \{1,2,\rel\}} P_k + \bar{P}_{\sf R}   = P,   \label{apc_2} \\
&\quad(p_1, q_1, p_2, q_2, \bar{P}_{\sf R}, P_1, P_2, P_{\sf R},  R) \succeq \mathbf{0}   .\label{nonneg1_2}
\end{align} where $\Sigma_{\sf R}$ and $\Sigma_{\sf D}$ are defines in \eqref{zr13} and \eqref{zr23} respectively.  Problem $[\mathcal{P}2] $ turns out to be a convex problem \cite{Boyd}, and hence we are able to solve it numerically using the \texttt{CVX} package \cite{cvx}.

To understand the general behavior of the optimal powers of $[\mathcal{P}2]$, we numerically solve $[\mathcal{P}2]$ for a simplified setting where both bands are subject to phase fading, the mm-wave parameters are taken to be $d_1=d_2 = 1.5,  r_1=r_2 = 1, \bar{G}_{\mathsf{RD}} = 1$ and $\alpha=2$, and the microwave band parameters are $G_{1 \mathsf{D}} = G_{2 \mathsf{D}} = G_{1 \mathsf{R}} = G_{2 \mathsf{R}} = G_{\mathsf{RD}} = 1$. For reference, we also solve $[\mathcal{P}1]$ for the same setting as that for $[\mathcal{P}2]$, with the fixed transmission powers $P_1=P_2=P_{\sf R}=\bar{P}_{\sf R}=1$. 

The resulting optimal powers for $[\mathcal{P}1]$ and $[\mathcal{P}2]$ are plotted against the power budget $0 \leq P \leq 5$ in Fig.~\ref{fig:02} and Fig.~\ref{fig:04} respectively. As expected, the transmit powers for $[\mathcal{P}1]$ follow the Waterfilling (WF) property for $P \leq 1.3$, and for  $P > 1.3$  the relay link powers are saturated to a constant value  $q_1=q_2 \approx 0.52$. In contrast, the transmit powers for  $[\mathcal{P}2]$, depicted in Fig.~\ref{fig:04}, follow only the WF property: for  $P \leq 2.6$, the entire budget is shared between the direct links only, whereas  for $P > 2.6$ power is allocated to all other mm-wave links and the microwave links from both sources. Notably, unlike in $[\mathcal{P}1]$, the relay link powers in $[\mathcal{P}2]$ are \emph{not saturated}. Moreover,   solving $[\mathcal{P}2]$ for a larger range of  $0 \leq P \leq 100$ shows that none of the link transmit powers saturate. This indicates that the optimum power allocation in $[\mathcal{P}2]$ does not follow the saturation property in general.

\section{Conclusion}
 We considered the fading MARC over dual microwave and mm-wave bands where the mm-wave links to the relay and the destination are modeled as non-interfering AWGN links. We showed that the capacity of the DR-MARC can be decomposed into the capacity of the underlying R-MARC and the two mm-wave direct links, hence the direct links can be operated independently of the R-MARC without compromising optimal rates. Then, we characterized an achievable region for the R-MARC. Focusing on R-MARCs with  underlying far c-MARC, sufficient conditions were found under which the aforementioned achievable scheme is capacity achieving. This shows that even if the sources are not near in the  underlying c-MARC in the microwave band, for sufficiently strong source-relay mm-wave links, they become jointly near  over both bands such that capacity is achieved.

Next, the optimal power allocation over the phase faded mm-wave links was found that maximizes the achievable sum-rate. The resulting scheme allocates power in different modes depending on the power budget $P$ and the link gains (\ied active LGR), and all such modes were characterized. When the budget $P$ is sufficiently small, it is entirely allocated only to the strongest of the relay and direct links, and as $P$ increases but remains below the saturation threshold, power is allocated to other links as in WF solution.  However, for $P$ above the saturation threshold, if one relay link is stronger but not significantly stronger than the other, power in the two links respectively increases and decreases with $P$  and approach non-zero levels as $P   \rightarrow   \infty$. Otherwise, power in the significantly stronger relay link is fixed at a constant while that in the other is zero. Moreover, for large mm-wave bandwidth, the saturation threshold depends only on mm-wave parameters, and in addition, if the received power at the destination from the relay via the mm-wave band is  large, the saturation threshold becomes large, and therefore allocating powers as in WF is optimal for all practical values of $P$. These results  illustrate the impact of high bandwidth point-to-point mm-wave links on the performance of the dual-band MARC, and can be useful in practical resource allocation in dual-band uplink scenarios.


%

\appendices
\section{Proof of Theorem \ref{thm1}}   \label{app1}
 {\textit{Outer Bounds}}: Assume that source $\mathsf{S}_k$ transmits $M_k, k \in \{1,2\}$. Since the destination knows $(\bm{H}_{\mathsf{D}}^n,\bm{\bar{H}}_{\mathsf{D}}^{n_1})$ where $\bm{H}_{\mathsf{D},i} \eqdef \{ H_{m\mathsf{D},i}\}_{m \in \{1,2,\mathsf{R}\} }, i=1,\dots,n$,  $\bm{\bar{H}}_{\mathsf{D},\ell} \eqdef \{ \bar{H}_{m\mathsf{D},\ell}\}_{m \in \{1,2,\mathsf{R}\} }, \ell=1,\dots,n_1$, from Fano's inequality  \begin{align} &nR_1  - n \epsilon_n \notag \\
&  \leq \text{I}(X_1^{n}, \hat{X}_1^{n_1}, \bar{X}_1^{n_1} ; Y_{\mathsf{D}}^{n}, \bar{Y}_{\mathsf{R}\mathsf{D}}^{n_1}, \bar{Y}_{1\mathsf{D}}^{n_1},  \bm{H}_{\mathsf{D}}^n, \bm{\bar{H}}_{\mathsf{D}}^{n_1}) \notag \\
& \stackrel{(a)}{\leq} \text{I}(X_1^{n}, \hat{X}_1^{n_1}, \bar{X}_1^{n_1} ; Y_{\mathsf{D}}^{n}, \bar{Y}_{\mathsf{R}\mathsf{D}}^{n_1}, \bar{Y}_{1\mathsf{D}}^{n_1}|  \bm{H}_{\mathsf{D}}^n, \bm{\bar{H}}_{\mathsf{D}}^{n_1}) \notag \\
&   \stackrel{(b)}{\leq} \text{I}(X_1^{n}, \hat{X}_1^{n_1}; Y_{\mathsf{D}}^{n} , \bar{Y}_{\mathsf{R}\mathsf{D}}^{n_1}| \bm{H}_{\mathsf{D}}^n, \bar{H}_{\mathsf{R}\mathsf{D}}^{n_1} )    + \textstyle \sum\limits_{\ell=1}^{n_1} \text{h}( \bar{G}^{1/2}_{1\mathsf{D},\ell} e^{j\bar{\Theta}_{1\mathsf{D},\ell}} \bar{X}_{1,\ell} + \bar{Z}_{1\mathsf{D},\ell}|\bar{G}_{1\mathsf{D},\ell},\bar{\Theta}_{1\mathsf{D},\ell}) -\text{h}(\bar{Z}_{1\mathsf{D},\ell}) \notag \\
&  \stackrel{(c)}{\leq} \text{I}(X_1^{n}, \hat{X}_1^{n_1}; Y_{\mathsf{D}}^{n} , \bar{Y}_{\mathsf{R}\mathsf{D}}^{n_1}| \bm{H}_{\mathsf{D}}^n, \bar{H}_{\mathsf{R}\mathsf{D}}^{n_1} )   +  \textstyle \sum\limits_{l=1}^{n_1}  \mathbb{E} [\log (1 + \bar{G}_{1\mathsf{D}} \bar{P}_{1,l})] \notag \\
&   \stackrel{(d)}{\leq} \text{I}(X_1^{n}, \hat{X}_1^{n_1}; Y_{\mathsf{D}}^{n} , \bar{Y}_{\mathsf{R}\mathsf{D}}^{n_1}|\bm{H}_{\mathsf{D}}^n, \bar{H}_{\mathsf{R}\mathsf{D}}^{n_1} ) + n_1 \mathbb{E} [\mathsf{C}(\bar{G}_{1\mathsf{D}} \bar{P}_{1})]   \label{appA1}
\end{align} where (a) follows since $(X_1^{n}, \hat{X}_1^{n_1}, \bar{X}_1^{n_1}) \indep  (\bm{H}_{\mathsf{D}}^n, \bm{\bar{H}}_{\mathsf{D}}^{n_1})$;  (b) follows by first expanding (a) into $4$ $\text{I}(.;.)$ terms using chain rule where two $\text{I}(.;.)$ terms turn out to be zero due to Markov chains (MC) $ \bar{X}_1^{n_1}   \rightarrow   (X_1^{n}, \hat{X}_1^{n_1}, \bm{H}_{\mathsf{D}}^n, \bar{H}_{\mathsf{R}\mathsf{D}}^{n_1})   \rightarrow   (Y_{\mathsf{D}}^{n}, \bar{Y}_{\mathsf{R}\mathsf{D}}^{n_1})$, and $ (X_1^{n}, \hat{X}_1^{n_1}, \bm{H}_{\mathsf{D}}^n, \bar{H}_{\mathsf{R}\mathsf{D}}^{n_1}, Y_{\mathsf{D}}^{n}, \bar{Y}_{\mathsf{R}\mathsf{D}}^{n_1})   \rightarrow   (\bar{X}_1^{n_1},  \bar{H}_{1\mathsf{D}}^{n_1})   \rightarrow   \bar{Y}_{1\mathsf{D}}^{n_1}$; the last two terms   follow from the Gaussian model  and applying chain rule and unconditioning to one of the remaining $\text{I}(.)$ terms; (c) follows from maximizing the first  $\text{h}(.)$ term in (b) by using $\bar{X}_{1,\ell} \sim \mathcal{CN}(0, \bar{P}_{1,\ell})$ where $\frac{1}{n_1} \mbox{\small$\sum\nolimits_{\ell=1}^{n_1}$}\bar{P}_{1,\ell} \leq \bar{P}_{1}$ and expectations are over $\bar{\Theta}_{1\mathsf{D},\ell} \sim \mathcal{U}[0,2\pi)$ i.i.d., $\bar{G}_{1\mathsf{D},\ell}$ i.i.d.; (d) follows by applying the Jensen's inequality. Bounding $R_2$ similarly, the following bounds
\begin{align} 	R_k \leq  \frac{1}{n}  \text{I}(X_k^{n}, \hat{X}_k^{n_1};Y_{\mathsf{D}}^{n}, \bar{Y}_{\mathsf{R}\mathsf{D}}^{n_1}|\bm{H}_{\mathsf{D}}^n, \bar{H}_{\mathsf{R}\mathsf{D}}^{n_1} ) + \notag  \frac{n_1}{n}   \mathbb{E} [\mathsf{C}(\bar{G}_{k\mathsf{D}} \bar{P}_{k})], \end{align} are found for $k \in \{1,2\}$, where expectations are over $\bar{G}_{k\mathsf{D}}$. Taking $n   \rightarrow   \infty$ such that $ {n_1}/{n}   \rightarrow   \alpha$ and $ \epsilon_n   \rightarrow   0$, then gives the bounds in Theorem \ref{thm1}, for some empirical probability mass function (pmf) distributed as
  \begin{align} 
	&\hspace*{-1mm} \prod\nolimits_{k=1}^2 p(x_k^n, \hat{x}_k^{n_1}, \bar{x}_k^{n_1})\prod\nolimits_{i=1}^{n} p(y_{\mathsf{R},i}, y_{\mathsf{D},i}| x_{1,i}, x_{2,i}, x_{\mathsf{R},i})   \prod\nolimits_{i=1}^{n} p(x_{\mathsf{R},i}|y_{\mathsf{R}}^{i-1}, \{\bar{y}_{k\mathsf{R}}^{n_1(i-1)}, {h}_{k\mathsf{R}}^{i-1}, \bar{h}_{k\mathsf{R}}^{n_1(i-1)}\}_{k=1}^2)  \notag \\ 
	&\hspace*{-1mm} \prod\nolimits_{\ell=1}^{n_1} p(\bar{x}_{\mathsf{R},\ell}|y_{\mathsf{R}}^{n(l-1)}, \{\bar{y}_{k\mathsf{R}}^{l-1}, h_{k\mathsf{R}}^{n(l-1)}, \bar{h}_{k\mathsf{R}}^{l-1}\}_{k=1}^2)  p(\bar{y}_{\mathsf{R}\mathsf{D},\ell}|\bar{x}_{\mathsf{R},\ell}) \notag \\
	&\hspace*{-1mm}\prod\nolimits_{\ell=1}^{n_1} p(\bar{y}_{1\mathsf{D},\ell}|\bar{x}_{1,\ell})  p(\bar{y}_{2\mathsf{D},\ell}|\bar{x}_{2,\ell}) p(\bar{y}_{1\mathsf{R},\ell}| \hat{x}_{1,\ell}) p(\bar{y}_{2\mathsf{R},\ell}| \hat{x}_{2,\ell}). \!\!  \label{decomp} \end{align}  

{  \textit{Achievability}}: We pick integers $(n, n_1)$ and a distribution that factors as \eqref{decomp}, and then code over $t$ blocks of symbols together. Define $ {U}_k \eqdef (X_k^{n}, \hat{X}_k^{n_1})$ and ${\bar{U}}_k \eqdef \bar{X}_k^{n_1}$ where $ U_1 \indep U_2$, and $ {\bar{U}}_k = \bar{X}_k^{n_1} \sim \mathcal{CN}(0, \bar{P}_k)$ i.i.d., $k=1,2$.  To encode $M_k \in \mathcal{M}_k$, we generate $2^{tnR_k}$ i.i.d. sequences $ {u}_k^t(M_k)$ and $ \bar{u}_k^t(M_k)$, distributed according to $p({u}_k^t)= \prod_{i=1}^t p( {u}_{k,i})=\prod_{i=1}^{t} p (x_{k,(i-1)n+1}^{in}, \hat{x}_{k,(i-1)n_1+1}^{in_1}) $ and $p(\bar{u}_k^t)= \prod_{i=1}^t p( \bar{u}_{k,i})$, $k=1,2$. To communicate $M_k$, we transmit $ {u}_k^t(M_k)$ and $\bar{u}_k^t(M_k)$ through the underlying RL-MARC and the $\mathsf{S}_k$-$\mathsf{D}$ direct links respectively. The relay assists each $(n, n_1)$ block of symbols, by producing codewords according to the relay-distribution in \eqref{decomp}, and forwarding them. The destination then decodes $M_k$ from the received signals, $(Y_{\mathsf{D}}^{nt}, \bar{Y}_{\mathsf{R}\mathsf{D}}^{n_1t}, \bar{Y}_{k\mathsf{D}}^{n_1t})$, using the CSI $(\bm{H}_{\mathsf{D}}^n, \bm{\bar{H}}_{\mathsf{D}}^{n_1})$. Applying standard random coding techniques as in \cite[Ch.~8.7]{cover_thomas},  the achievable rates are found to satisfy 
\begin{align}
R_k <  \frac{1}{n}  \text{I}(X_k^{n}, \hat{X}_k^{n_1};Y_{\mathsf{D}}^{n}, \bar{Y}_{\mathsf{R}\mathsf{D}}^{n_1}|\bm{H}_{\mathsf{D}}^n, \bar{H}_{\mathsf{R}\mathsf{D}}^{n_1} ) +  \frac{n_1}{n}  \mathbb{E} [\mathsf{C}(\bar{G}_{k\mathsf{D}} \bar{P}_{k})], \!\! \label{ach_rmarc_multiletter} 
\end{align}for $k \in \{1,2\}$. Finally, an achievable rate pair on the RL-MARC is given by the first term in \eqref{ach_rmarc_multiletter}, and its capacity $C_{RL}(\alpha)$ is the closure of the union of sets of all achievable rate pairs where the union is over all $(n, n_1)$ and pmfs factoring as \eqref{decomp} with $\bar{y}_{k\mathsf{D},\ell} = \bar{x}_{k,\ell} = \emptyset, k=1,2$. 	 

\section{Proof of Theorem \ref{thm2}}  \label{app2}
The achievable region is obtained by performing block Markov encoding over $B+1$ blocks with i.i.d. CSCG codewords and backward decoding at destination as follows (see \cite{patick_relay,kramer_relay} for details). 

\textit{Encoding}: Encoding for block $b \in \{ 1,\ldots,B+1\}$ proceeds as follows: ($\mathrm{i}$) the block lengths ($n$, $n_1$), and the input distributions $p_{m}(x), \bar{p}_{m}(\bar{x})$ and $\hat{p}_{k}(\hat{x}), m \in \{1,2,\mathsf{R}\} ,k \in \{1,2\}$ are chosen; ($\mathrm{ii}$) the message $M_{k,b} \in \mathcal{M}_k$ from $\mathsf{S}_k$ is encoded into codewords ${x}^{n}_k(M_{k,b})$ and ${\hat{x}}_k^{n_1}(M_{k,b})$, generated according to $\prod_{i=1}^{n}p_{k}(x_{k,i}(M_{k,b}))$ and $\prod_{\ell=1}^{n_1}\hat{p}_{k}(\hat{x}_{k,\ell}(M_{k,b}))$, $k \in \{1,2\}$, and  transmitted; ($\mathrm{iii}$) assuming that the relay estimated ($M_{1,b-1},M_{2,b-1}$) in block $b-1$ correctly, they are encoded into codewords ${x}^{n}_{\mathsf{R}}(M_{1,b-1}, M_{2,b-1})$ and $\bar{x}^{n_1}_{\mathsf{R}}(M_{1,b-1}, M_{2,b-1})$, generated according to $\prod_{i=1}^{n}p_{\mathsf{R}}(x_{\mathsf{R},i}(M_{1,b-1}, M_{2,b-1}))$ and $\prod_{\ell=1}^{n_1}\bar{p}_{\mathsf{R}}(\bar{x}_{\mathsf{R},\ell}(M_{1,b-1}, M_{2,b-1}))$, and  transmitted. The messages $M_{k,0}$ and $M_{k,B+1}$ are known at the destination, $k \in \{1,2\}$ as in \cite{patick_relay,kramer_relay}. 

\textit{Decoding at the Relay}: Assume that the message pair $(M_{1,b-1}, M_{2,b-1})$ was correctly decoded in block $b-1$. The relay then uses the side information ${x}^{n}_{\mathsf{R}}(M_{1,b-1}, M_{2,b-1})$ and $\bar{x}^{n_1}_{\mathsf{R}}(M_{1,b-1}, M_{2,b-1})$ and the CSI at block $b$, \ied $\{H_{k\mathsf{R}}^n(b), \bar{H}_{k\mathsf{R}}^{n_1}(b)\}_{k=1}^2$, and estimates $(M_{1,b}, M_{2,b})$ from the signals received in block $b$ as in \cite[Ch.~14.3.1]{cover_thomas}. Such decoding yields certain rate constraints on $R_1,R_2$ and $R_1+R_2$ which are then maximized by using i.i.d. CSCG codewords  $X_m \sim \mathcal{CN}(0,P_m),  \hat{X}_{k} \sim \mathcal{CN}(0,\hat{P}_k), m \in \{1,2,3\}, k \in \{1,2\}$. Finally, the achievable rates are obtained by averaging the resulting rate constraints over i.i.d. squared-magnitudes of fading coefficients $G_{k\mathsf{R}}$ and $\bar{G}_{k\mathsf{R}}$ (since rate constraints are independent of the phases), as given in \eqref{r1_relay}-\eqref{r1_r2_relay}.

\textit{Decoding at the Destination} (Backward decoding): Assuming that $(M_{1,b+1}, M_{2,b+1})$ were decoded correctly in block $b+1$, the decoder estimates $(M_{1,b}, M_{2,b})$ from the signals received in blocks $b$ and $b+1$ as in \cite[Ch.~14.3.1]{cover_thomas} by using the side information ${x}^{n}_k(M_{k,b+1})$ and $\hat{x}^{n_1}_k(M_{k,b+1}), k \in \{1,2\}$, and CSI in blocks $b$ and $b+1$, $( \{H_{m\mathsf{D}}^n(\ell), \bar{H}_{\mathsf{R}\mathsf{D}}^{n_1}(\ell)\}_{\ell = b}^{b+1}, m \in \{1,2,\mathsf{R}\})$. The resulting rate constraints are maximized by the same i.i.d. CSCG codewords as for the relay, and achievable rates are obtained by taking expectation over  $G_{k\mathsf{D}}$ and $\bar{G}_{\mathsf{RD}}$, as given by \eqref{r1_dest}-\eqref{r1_r2_dest}.  

\section{Proof of Theorem \ref{thm3}}  \label{app3}
For notational convenience, define $ \mathcal{U} \subseteq  \{1,2\}$ and $\mathcal{U}^c \eqdef \{1,2\} \setminus \mathcal{U} $ such that $X_{\mathcal{U}} \eqdef \{ X_k, k \in \mathcal{U}\}$. We derive the outer-bounds by applying the cut-set bounding technique (see \cite[Ch.~14.10]{cover_thomas} for details). Assume that source $\mathsf{S}_k$ transmits the message $M_k, k \in \{1,2\}$.  Since the destination knows $\bm{H}_{\mathsf{D}}^n$ and $\bar{H}_{\mathsf{R}\mathsf{D}}^{n_1}$ where $\bm{H}_{\mathsf{D},i} \eqdef \{H_{m\mathsf{D},i}\}_{m \in \{1,2,\mathsf{R}\}},i=1,\ldots,n$, by Fano's inequality,	   \begin{align}
&\textstyle \sum\nolimits_{k \in \mathcal{U}} n R_{\mathcal{U}} -  n \epsilon_n  \notag \\
& \stackrel{(a)}{\leq} \text{I}(M_{\mathcal{U}}; {Y}_{\mathsf{D}}^n, \bar{Y}_{\mathsf{R}\mathsf{D}}^{n_1}, \bm{H}_{\mathsf{D}}^n, \bar{H}_{\mathsf{R}\mathsf{D}}^{n_1},M_{\mathcal{U}^c})  \notag \\
&\stackrel{(b)}{=} \text{I}(M_{\mathcal{U}}; {Y}_{\mathsf{D}}^n| \bm{H}_{\mathsf{D}}^n, \bar{H}_{\mathsf{R}\mathsf{D}}^{n_1},M_{\mathcal{U}^c}\!) \! + \!\text{I}(M_{\mathcal{U}}; \bar{Y}_{\mathsf{R}\mathsf{D}}^{n_1}|{Y}_{\mathsf{D}}^n, \bm{H}_{\mathsf{D}}^n, \bar{H}_{\mathsf{R}\mathsf{D}}^{n_1},M_{\mathcal{U}^c}\!)  \notag \\
& \stackrel{(c)}{\leq} \textstyle \sum\limits_{i=1}^{n}   \text{h}({Y}_{\mathsf{D},i}| {Y}_{\mathsf{D}}^{i-1}, \bm{H}_{\mathsf{D}}^n, \bar{H}_{\mathsf{R}\mathsf{D}}^{n_1},M_{\mathcal{U}^c}, X_{\mathcal{U}^c,i})  - \text{h}({Y}_{\mathsf{D},i}|{Y}_{\mathsf{D}}^{i-1},\bm{H}_{\mathsf{D}}^n, \bar{H}_{\mathsf{R}\mathsf{D}}^{n_1},  M_{\mathcal{U}}, M_{\mathcal{U}^c},  X_{\mathcal{U},i}, X_{\mathcal{U}^c,i}, X_{\mathsf{R},i})  \notag \\
& \;\;+ \textstyle \sum\limits_{l=1}^{n_1}   \text{h}({\bar{Y}}_{\mathsf{R}\mathsf{D},l}| {\bar{Y}}_{\mathsf{R}\mathsf{D}}^{l-1}, Y_{\mathsf{D}}^n, \bm{H}_{\mathsf{D}}^n, \bar{H}_{\mathsf{R}\mathsf{D}}^{n_1}, M_{\mathcal{U}^c})  -\text{h}({\bar{Y}}_{\mathsf{R}\mathsf{D},l}|  {\bar{Y}}_{\mathsf{R}\mathsf{D}}^{l-1}, Y_{\mathsf{D}}^n,\bm{H}_{\mathsf{D}}^n, \bar{H}_{\mathsf{R}\mathsf{D}}^{n_1}, M_{\mathcal{U}^c}, M_{\mathcal{U}}, {\bar{X}}_{\mathsf{R},l})  \notag \\
& \stackrel{(d)}{=} \textstyle \sum\limits_{i=1}^{n}  \text{h}( \textstyle \sum\limits_{k \in \mathcal{U}}  G_{k\mathsf{D}}^{1/2} e^{j\Theta_{k\mathsf{D},i}} X_{k,i} + G_{\mathsf{R}\mathsf{D}}^{1/2} e^{j\Theta_{\mathsf{R}\mathsf{D},i}} X_{\mathsf{R},i} + Z_{\mathsf{D},i}|  \{G_{m\mathsf{D}, i},\Theta_{m\mathsf{D}, i}\}, m \in \{1,2,\mathsf{R}\}) -  \text{h}(Z_{\mathsf{D},i}) \notag \\
& \;\; + \textstyle \sum\limits_{l=1}^{n_1}  \text{h}(\bar{G}_{\mathsf{R}\mathsf{D}}^{1/2} e^{j\bar{\Theta}_{\mathsf{R}\mathsf{D},l}} \bar{X}_{\mathsf{R},l}+ \bar{Z}_{\mathsf{R}\mathsf{D},l}|\bar{G}_{\mathsf{R}\mathsf{D},l}, \bar{\Theta}_{\mathsf{R}\mathsf{D},l})- \text{h}(\bar{Z}_{\mathsf{R}\mathsf{D},l}) \notag \\
&\stackrel{(e)}{\leq}\textstyle \sum\limits_{i=1}^n  \mathbb{E} [\log \big(1 + G_{\mathsf{R}\mathsf{D}} P_{\mathsf{R},i}   + \textstyle \sum\limits_{k \in \mathcal{U}}  \normalsize ( G_{k\mathsf{D}} P_{k,i}  + 2 G_{k\mathsf{D}}^{1/2} G_{\mathsf{R}\mathsf{D}}^{1/2}  \mathfrak{Re}\{e^{j(\Theta_{k\mathsf{D}}-\Theta_{\mathsf{R}\mathsf{D}})}\mathbb{E}[X_{k,i}X_{\mathsf{R},i}^*]\})\big)]  \notag \\
&+ \textstyle \sum\limits_{l=1}^{n_1} \mathbb{E}  [\log (1 + \bar{G}_{\mathsf{R}\mathsf{D}} \bar{P}_{\mathsf{R},l})]   \notag \\
&\stackrel{(f)}{\leq} \textstyle \sum\limits_{i=1}^n  \mathbb{E} [\log(1 +  \textstyle \sum\limits_{k \in \mathcal{U}}   G_{k\mathsf{D}} P_{k,i} +  G_{\mathsf{R}\mathsf{D}} P_{\mathsf{R},i} )]  +  \textstyle \sum\limits_{l=1}^{n_1}  \mathbb{E} [\log (1 + \bar{G}_{\mathsf{R}\mathsf{D}} \bar{P}_{\mathsf{R},l} )] \notag \\
&\stackrel{(g)}{\leq} n \mathbb{E} [ \log (1 + \textstyle \sum\nolimits_{k \in \mathcal{U}}  G_{k\mathsf{D}} P_{\mathsf{D}} + G_{\mathsf{R}\mathsf{D}} P_{\mathsf{R}})]  + n_1 \mathbb{E} [\log (1 + \bar{G}_{\mathsf{R}\mathsf{D}} \bar{P}_{\mathsf{R}} )] \label{outer_bnd_phase} \end{align}  where (a) follows since including $M_{\mathcal{U}^c}$ does not reduce information; (b) follows by applying chain rule and noting that $I(M_{\mathcal{U}}; \bm{H}_{\mathsf{D}}^n, \bar{H}_{\mathsf{R}\mathsf{D}}^{n_1}, M_{\mathcal{U}^c})=0$ due to  $M_{\mathcal{U}} \indep (\bm{H}_{\mathsf{D}}^n, \bar{H}_{\mathsf{R}\mathsf{D}}^{n_1}, M_{\mathcal{U}^c})$; (c) follows from chain rule and the fact that conditioning with $X_{\mathcal{U},i}(M_{\mathcal{U}})$ and ${X}_{\mathcal{U}^c,i}(M_{\mathcal{U}^c})$ (deterministic functions of $M_{\mathcal{U}}$ and $M_{\mathcal{U}^c}$) do not alter entropy, while conditioning the negative $\text{h}(.)$ terms with $X_{\mathsf{R},i}$ and $\bar{X}_{\mathsf{R},l}$ does not decrease entropy; (d) follows from (c) by first unconditioning, next applying the MCs  due to the memoryless system model, \begingroup \small $({Y}_{\mathsf{D}}^{i-1}, \bm{H}_{\mathsf{D}}^{n\backslash i}, \bar{H}_{\mathsf{R}\mathsf{D}}^{n_1}, M_{\mathcal{U}}, M_{\mathcal{U}^c})   \rightarrow   (X_{\mathcal{U},i},  {X}_{\mathcal{U}^c,i}, X_{\mathsf{R},i}, \bm{H}_{\mathsf{D},i})   \rightarrow   {Y}_{\mathsf{D},i}$ \endgroup and \begingroup \small  $({\bar{Y}}_{\mathsf{R}\mathsf{D}}^{l-1}, Y_{\mathsf{D}}^n, \bm{H}_{\mathsf{D}}^{n}, \bar{H}_{\mathsf{R}\mathsf{D}}^{n_1\backslash l}, M_{\mathcal{U}}, M_{\mathcal{U}^c})   \rightarrow   ({\bar{X}}_{\mathsf{R},l}, \bar{H}_{\mathsf{R}\mathsf{D},l})   \rightarrow   {\bar{Y}}_{\mathsf{R}\mathsf{D},l}$ \endgroup, where a vector \begingroup \small $F^{m \backslash j} \eqdef \{F_i\}_{i = 1}^m \setminus F_i$\endgroup, and finally using the fading Gaussian model; (e) follows by maximizing the first $\text{h}(.)$ term of (d) by using $X_{k,i} \sim \mathcal{CN}(0,P_{k,i})$ \cite{kramer_relay}, with $P_{k,i} \eqdef \mathbb{E}[|X_{k,i}|^2], k \in \{ 1,2\}$, and $\mathbb{E}[X_{k,i}X_{\mathsf{R},i}^*]$ being the cross-correlation between $X_{k,i}$ and $X_{\mathsf{R},i}$ where the expectation are over column $i$ of the codebook, and $\mathfrak{Re}(.)$ denotes the real part; the third $\text{h}(.)$ term in (d) is similarly maximized by $\hat{X}_{k,l} \sim \mathcal{CN}(0,\hat{P}_{k,l})$; the outer expectation is over the i.i.d. fading magnitudes and phases; (f) follows since in the first term of (e), $\tilde{\Theta} \eqdef \Theta_{k\mathsf{D}}-\Theta_{\mathsf{R}\mathsf{D}} \sim  \mathcal{U}[0,2\pi)$, and thus each summand can be upper bounded by using ${\mathbb{E}}_{\tilde{\Theta}, G, B} \log(1+G+ 2 G^{1/2} B^{1/2} \mathfrak{R} \{e^{j\tilde{\Theta}} \rho \}) \leq {\mathbb{E}}_{A} \log(1+G)$ when $\tilde{\Theta} \sim \mathcal{U}[0,2\pi)$, $\rho \in \mathbb{C}$ \cite{kramer_relay}; and (g) follows from applying Jensen's inequality as in steps (c)-(d) of \eqref{appA1}. 

Thus as $n,n_1   \rightarrow   \infty$, we have
\begin{align}
  R_{ \mathcal{U}} \leq  \mathbb{E} [\mathsf{C} ( \textstyle \sum\limits_{k \in \mathcal{U}}  G_{k\mathsf{D}} P_{\mathsf{D}} + G_{\mathsf{R}\mathsf{D}} P_{\mathsf{R}})] + \alpha \mathbb{E} [\mathsf{C} ( \bar{G}_{\mathsf{R}\mathsf{D}} \bar{P}_{\mathsf{R}} )] \label{r1_ob}
\end{align} for $\mathcal{U} \subseteq \{1,2\}$, from which individual bounds on $R_1, R_2$ and $R_1+R_2$ are obtained by choosing $\mathcal{U} = \{1\}$, $\mathcal{U} = \{2\}$ and $\mathcal{U} = \{1,2\}$. Finally, under condition \eqref{r1_cond}-\eqref{r1_r2_cond}, the achievable region of Theorem \ref{thm2} reduces to bounds in \eqref{r1_dest}, \eqref{r2_dest} and \eqref{r1_r2_dest} which match the respective outer bounds, and thus achieves the capacity. 

\section{Solution of the Problem $[\mathcal{P}1]$}  \label{app4}
{  \textit{The KKT Conditions}}: We denote a feasible point by $\mathbf{x} \eqdef (p_1,q_1,p_2,q_2,R) \in  \mathbb{R}_+^5$, and use the equivalent objective, \emph{minimize} $-R$. Note that the objective is linear, and the equality constraints in \eqref{apc} are affine. Moreover, the constraint in \eqref{conv_1} is convex as its Hessian is a positive semidefinite matrix with  $\frac{\alpha \kappa}{2} ({d_1^2}/{(1+d_1p_1)^2},\; {r_1^2}/{(1+r_1q_1)^2},$ $ {d_2^2}/{(1+d_2 p_2)^2},\; {r_2^2}/{(1+r_2q_2)^2},\; 0)$  on its leading diagonal. Similarly, constraint \eqref{conv_2} is also convex. Furthermore, the feasible set is compact, and   $\mathbf{\tilde{x}} \eqdef \big(P - \epsilon,\; \epsilon, \; P - \epsilon, \; \epsilon, \; \sigma_{\mathsf{R}} \big)$  is strictly feasible for sufficiently small $\epsilon>0$. Hence $[\mathcal{P}1]$ is a convex optimization problem over a compact set that satisfies Slater's condition \cite{Boyd}, therefore it is solved using KKT conditions as in \cite[Chap.~5.5.3]{Boyd}.  The Lagrangian function for  $[\mathcal{P}1]$ is given by  \begin{align} 
&\mathfrak{L} =-R + \lambda_1 (R- \Sigma_{\mathsf{R}})  + \lambda_2  (R- \Sigma_{\mathsf{D}}) +   \mu_1 (p_1 + q_1 - P)   + \mu_2 (p_2 + q_2 - P) 	 - \rho_1 p_1 - \rho_2 q_1 - \rho_3 p_2 - \rho_4 q_2- \rho_5 R, \notag
\end{align}where $\{\lambda_k\}_{k=1}^2, \{\mu_k\}_{k=1}^2$ and $\{\rho_i\}_{i=1}^5$ are Lagrange multipliers corresponding to constraints  \eqref{conv_1}-\eqref{conv_2}, \eqref{apc}, and  $ (p_1,q_1,p_2,q_2,R) \succeq \mathbf{0}$ respectively, with $\Sigma_{\mathsf{R}}$ and $\Sigma_{\mathsf{D}}$  in \eqref{zr13}--\eqref{zr23}. 	With slight abuse of notation, we denote the optimal primal variables by $(p_1, q_1, p_2, q_2,R)$, and the optimal Lagrange multipliers (OLM) by $(\lambda_1, \lambda_2, \rho_1, \rho_2, \rho_3, \rho_4)$ and $(\mu_1, \mu_2)$, which satisfy the following KKT conditions
\begin{align}
&\lambda_1 + \lambda_2 = 1,  \label{sum_lam}\\ 
&\rho_1 = \mu_1 - \frac{\alpha}{2}\frac{d_1}{1+d_1 p_1}, \;\; \rho_2 = \mu_1 - \frac{\alpha}{2} \frac{\lambda_1 r_1}{1+r_1 q_1},  \rho_3 = \mu_2 - \frac{\alpha}{2} \frac{d_2}{1+d_2 p_2},  \;\; \rho_4 = \mu_2 - \frac{\alpha}{2} \frac{\lambda_1 r_2}{1+r_2 q_2} \label{rho4},\\ 
& p_1 +   q_1 = P, \;\;     p_2 +   q_2 = P, \label{apc25a}\\ 
& R - \Sigma_{\mathsf{R}} \leq 0, \;\; R - \Sigma_{\mathsf{R}} \leq 0, \notag \\
& \lambda_1(R-\Sigma_{\mathsf{R}}) = 0, \;\;  \lambda_2(R-\Sigma_{\mathsf{D}}) = 0 \label{lam2}, \\ 
&\rho_1 p_1 = 0,  \;\; \rho_2 q_1 = 0, \;\; \rho_3 p_2 = 0, \;\; \rho_4 q_2 = 0, \label{prod} \\ 
& (p_1, q_1, p_2, q_2, R) \succeq \mathbf{0}, \;\; (\lambda_1,\lambda_2, \rho_1, \rho_2, \rho_3, \rho_4) \succeq \mathbf{0}. \label{nonneg2}\end{align} with $\rho_5=0$ since $R \geq \min(\sigma_{\mathsf{D}},\sigma_{\mathsf{R}})>0$. 

\begin{table*}[t] 
	\renewcommand{\arraystretch}{1}
	\tabcolsep = 8pt
	\caption{ \small Set of $(\boldsymbol{\rho}, \boldsymbol{\lambda})$-tuples are partitioned into $18$ subsets and the LGR corresponding to each subset is provided.} 
	\label{table_LGR_subset_map}
	\centering
	\begin{tabular}{l| l l |l| l l |l| l l} \hlinewd{.5pt}
		&  $\boldsymbol{\lambda} \in \mathcal{L}_1$ & $\boldsymbol{\lambda} \in \mathcal{L}_2$ &   &  $\boldsymbol{\lambda} \in \mathcal{L}_1$ & $\boldsymbol{\lambda} \in \mathcal{L}_2$ &   &  $\boldsymbol{\lambda} \in \mathcal{L}_1$ & $\boldsymbol{\lambda} \in \mathcal{L}_2$ \\\hline  
		$\boldsymbol{\rho} \in \mathcal{I}_1 \cap \mathcal{J}_1$ &  $\mathcal{A}_{\sf r,r}$ &  $\mathcal{\tilde{A}}_{\sf r,r} \subseteq \mathcal{A}_{\sf r,r}$ & $\boldsymbol{\rho} \in \mathcal{I}_1 \cap \mathcal{J}_2$ &  $\mathcal{A}_{\sf r,d}$ &  $\mathcal{\tilde{A}}_{\sf r,d} \subseteq \mathcal{A}_{\sf r,d}$ 	& $\boldsymbol{\rho} \in \mathcal{I}_1 \cap \mathcal{J}_3$ &  $\mathcal{A}_{\sf r,rd}$ &  $\mathcal{S}_{\sf r,rd}$ 		\\   
		$\boldsymbol{\rho} \in \mathcal{I}_2 \cap \mathcal{J}_1$ &  $\mathcal{A}_{\sf d,r}$ &  $\mathcal{\tilde{A}}_{\sf d,r} \subseteq \mathcal{A}_{\sf d,r}$ & $\boldsymbol{\rho} \in \mathcal{I}_2 \cap \mathcal{J}_2$ &  $\mathcal{A}_{\sf d,d}$ &  $\mathcal{\tilde{A}}_{\sf d,d}$ is invalid & $\boldsymbol{\rho} \in \mathcal{I}_2 \cap \mathcal{J}_3$ &  $\mathcal{A}_{\sf d,rd}$ &  $\mathcal{S}_{\sf d,rd}$ 		\\   
		$\boldsymbol{\rho} \in \mathcal{I}_3 \cap \mathcal{J}_1$ &  $\mathcal{A}_{\sf rd,r}$ &  $\mathcal{S}_{\sf rd,r}$ & 		$\boldsymbol{\rho} \in \mathcal{I}_3 \cap \mathcal{J}_2$ &  $\mathcal{A}_{\sf rd,d}$ &  $\mathcal{S}_{\sf rd,d}$  &		$\boldsymbol{\rho} \in \mathcal{I}_3 \cap \mathcal{J}_3$ &  $\mathcal{A}_{\sf rd,rd}$ &  $\mathcal{S}_{\sf rd,rd}$ 		\\  \hlinewd{.5pt}
	\end{tabular}  \vspace*{-5mm}
\end{table*}

{ \textit{Partitioning the set of OLMs}}: We now partition the set of all $(\boldsymbol{\rho}, \boldsymbol{\lambda})$-tuples where $\boldsymbol{\rho} \eqdef (\rho_1, \rho_2, \rho_3, \rho_4)$ $ \succeq \mathbf{0}$ and $\boldsymbol{\lambda} \eqdef (\lambda_1, \lambda_2) \succeq \mathbf{0}$, into $18$ subsets. First, the set of $(\rho_1,\rho_2)$-tuples is partitioned into $3$ subsets, $\mathcal{I}_1 \eqdef \{(\rho_1, \rho_2): \rho_1 > 0, \rho_2 = 0 \}$, $ \mathcal{I}_2 \eqdef \{(\rho_1, \rho_2): \rho_1 = 0, \rho_2 > 0 \},$ and  $ \mathcal{I}_3 \eqdef \{(\rho_1, \rho_2): \rho_1 = 0, \rho_2 = 0 \}$, since subset  $\mathcal{I}_4 \eqdef \{(\rho_1, \rho_2): \rho_1 > 0, \rho_2 > 0 \}$ violates \eqref{apc25a} by requiring $p_1 = q_1 = 0$. The set of $(\rho_3,\rho_4)$-tuples is similarly partitioned into $3$ subsets $\mathcal{J}_k, k \in \{1,2,3\}$. Finally, the set of $\boldsymbol{\lambda}$-tuples  is partitioned into $2$ subsets $\mathcal{L}_1  \eqdef \{\boldsymbol{\lambda}: \lambda_1 = 1, \lambda_2=0\} $ and $\mathcal{L}_2 \eqdef \{\boldsymbol{\lambda}: \lambda_1 > 0, \lambda_2>0\} $, since subset $\mathcal{L}_3 \eqdef \{\boldsymbol{\lambda}: \lambda_1 = 0, \lambda_2=1\}$ violates the assumption $\gamma > 1$ in \oa by requiring $\Sigma_{\mathsf{D}} < \Sigma_{\mathsf{R}}$, and  $\mathcal{L}_4 \eqdef \{\boldsymbol{\lambda}: \lambda_1 = 0, \lambda_2 = 0\}$ violates \eqref{conv_1}--\eqref{conv_2} by requiring $R < \min(\Sigma_{\mathsf{D}}, \Sigma_{\mathsf{R}})$. Thus, the set of $(\boldsymbol{\rho}, \boldsymbol{\lambda})$-tuples are now partitioned into 18 subsets  $\mathcal{I}_k \cap \mathcal{J}_l \cap \mathcal{L}_m, k,l \in \{1,2,3\}, m \in \{1,2\}$. 	Note that a $(\boldsymbol{\rho},\boldsymbol{\lambda})$-tuple now satisfies the KKT conditions as well as the condition of the subset to which it belongs. When all conditions on $(\boldsymbol{\rho},\boldsymbol{\lambda})$ are expressed in terms of $(P, r_1, r_2, d_1, d_2, \gamma)$, each subset leads to an LGR as presented in Table \ref{table_LGR_subset_map}. However, only $14$ LGRs are valid, since $3$ are subsumed into an existing LGR ($\mathcal{\tilde{A}}_{(.,.)} \subseteq  \mathcal{A}_{(.,.)}$), and $\tilde{\mathcal{A}}_{\sf d,d}$ is invalid as it violates the assumption $\gamma > 1$.

{  \textit{Power Allocation in LGRs}}: Next, we express the conditions on $(\boldsymbol{\rho},\boldsymbol{\lambda})$ in each LGR in terms of $P$ and threshold powers in Table \ref{table_critical_power}. We also derive the expression of optimal powers in this process.

{LGR $\mathcal{A}_{\sf r,r}$:} Here, $\boldsymbol{\rho} \in \mathcal{I}_1 \cap \mathcal{J}_1$ and $\boldsymbol{\lambda} \in \mathcal{L}_1$. For $\boldsymbol{\rho} \in \mathcal{I}_1 \cap \mathcal{J}_1$, we have $\rho_1 > 0, \rho_2 = 0, \rho_3 > 0, \rho_4 = 0$, which require $ p_1=0, q_1 = P, p_2=0, q_2 = P$ from \eqref{apc25a}, \eqref{prod}-\eqref{nonneg2}. Now, $\boldsymbol{\lambda} \in \mathcal{L}_1$ requires $\Sigma_{\mathsf{R}} < \Sigma_{\mathsf{D}}$ that results in $P < \mathsf{P}_{\sf r,r}$ from \eqref{lam2}. The conditions for $\rho_1 > 0, \rho_3 > 0$ are derived by substituting $\rho_2 = \rho_4 = 0$ in \eqref{rho4} and eliminating $(\mu_1,\mu_2)$. Hence, the conditions for $\mathcal{A}_{\sf r,r}$ are given by $P  \leq  \mathsf{P}^{\prime}_{\sf d,d}  =  d_1^{-1} - r_1^{-1}, \;\; P  \leq  \hat{\mathsf{P}}^{\prime}_{\sf d,d} = d_2^{-1} - r_2^{-1}$, and $P  <  \mathsf{P}_{\sf r,r}$.

The conditions of the counterpart $\tilde{\mathcal{A}}_{\sf r,r}$ (with $\boldsymbol{\lambda} \in \mathcal{L}_2$ instead of $\boldsymbol{\lambda} \in \mathcal{L}_1$) is valid only for a set of measure zero at $P = \mathsf{P}_{\sf r,r}$ but the optimum powers are the same as in $\mathcal{A}_{\sf r,r}$, thus it is subsumed in $\mathcal{A}_{\sf r,r}$.

{LGR $\mathcal{A}_{\sf d,rd}$ and $\mathcal{S}_{\sf d,rd}$:} In $\mathcal{A}_{\sf d,rd}$,  $\boldsymbol{\rho} \in \mathcal{I}_2 \cap \mathcal{J}_3$ and $\boldsymbol{\lambda} \in \mathcal{L}_1$. For $\boldsymbol{\rho} \in \mathcal{I}_2 \cap \mathcal{J}_3$, we have $\rho_1 = 0, \rho_2 > 0, \rho_3 = 0, \rho_4 = 0$, which require $ p_1 = P, q_1 = 0, p_2 \geq 0, q_2 \geq 0$ from \eqref{apc25a}, \eqref{prod}-\eqref{nonneg2}. First, by substituting $\rho_3 = \rho_4 = 0,\lambda_1=1$ in \eqref{rho4}, we obtain $p_2 = 0.5(P + r_2^{-1} - d_2^{-1})$ and $q_2 = 0.5(P + d_2^{-1} - r_2^{-1})$, and conditions $(p_2,q_2)\succeq \mathbf{0}$ require $P \geq \hat{\mathsf{P}}^{\prime}_{\sf d,d}, P \geq \hat{\mathsf{P}}_{\sf d,d}$. The condition for $\rho_2 > 0$, found by substituting $\rho_1 = 0, \lambda_1=1$ in \eqref{rho4}, requires $P \leq \mathsf{P}_{\sf d,d} = r_1^{-1} - d_1^{-1}$. Finally, $\boldsymbol{\lambda} \in \mathcal{L}_1$ requires $\Sigma_{\mathsf{R}} < \Sigma_{\mathsf{D}}$, \ied $P < \mathsf{P}_{\sf d,rd} = (2\gamma -1)r_2^{-1} - d_2^{-1}$. Thus, the conditions for  $\mathcal{A}_{\sf d,rd}$ are $\min(\mathsf{P}_{\sf d,d}, \mathsf{P}_{\sf d,rd}) \geq P \geq \max( \hat{\mathsf{P}}_{\sf d,d}, \hat{\mathsf{P}}^{\prime}_{\sf d,d})$.

In $\mathcal{S}_{\sf d,rd}$,  $\boldsymbol{\rho} \in \mathcal{I}_2 \cap \mathcal{J}_3$, which still requires $ p_1 = P, q_1 = 0, p_2 \geq 0, q_2 \geq 0$. However, now $\boldsymbol{\lambda} \in \mathcal{L}_2$, \ied  $(\lambda_1,\lambda_2) \succeq \mathbf{0}$, which requires $\Sigma_{\mathsf{R}} = \Sigma_{\mathsf{D}}$, resulting in $q_2 = (\gamma -1)r_2^{-1}$, and $p_2 = P - (\gamma -1)r_2^{-1}$.  Due to $\gamma>1$, we have $q_2>0$, but $p_2>0$ additionally requires $P > \mathsf{P}_{\sf d,r} = (\gamma -1)r_2^{-1}$. Since $\lambda_1+\lambda_2=1$ in \eqref{sum_lam}, $(\lambda_1,\lambda_2) \succeq \mathbf{0}$ is equivalent to $1 > \lambda_1 > 0$. Solving for $\lambda_1$ by substituting $(p_2,q_2)$ above and $\rho_3 = \rho_4 = 0$ in \eqref{rho4}, the condition $1>\lambda_1 > 0$ requires $P > \mathsf{P}_{\sf d,rd}$. The condition for $\rho_2 > 0$, found by substituting $\rho_1 = 0$ in \eqref{rho4}, requires $P >  \overbar{\mathsf{P}}_{\sf d,rd}$ if $\boldsymbol{r} \in \mathcal{R}_{S2}$, and $P <  \overbar{\mathsf{P}}_{\sf d,rd}$ otherwise.  Therefore, the conditions of $\mathcal{S}_{\sf d,rd}$ are $P \geq \max( \mathsf{P}_{\sf d,r}, \mathsf{P}_{\sf d,rd}, \overbar{\mathsf{P}}_{\sf d,rd})$,  if $\boldsymbol{r} \in \mathcal{R}_{S2}$, and $ \max( \mathsf{P}_{\sf d,r}, \mathsf{P}_{\sf d,rd}) \leq P <  \overbar{\mathsf{P}}_{\sf d,rd}$, otherwise.

{LGR $\mathcal{A}_{\sf rd,r}$ and $\mathcal{S}_{\sf rd,r}$:} In $\mathcal{A}_{\sf rd,r}$, $\boldsymbol{\rho} \in \mathcal{I}_3 \cap \mathcal{J}_1$ and $\boldsymbol{\lambda} \in \mathcal{L}_1$. For $\boldsymbol{\rho} \in \mathcal{I}_3 \cap \mathcal{J}_1$, we have $\rho_1 = 0, \rho_2 = 0, \rho_3 > 0, \rho_4 = 0$, which require $ p_1 \geq 0, q_1 \geq 0, p_2 = 0, q_2 = P$ from from \eqref{apc25a}, \eqref{prod}-\eqref{nonneg2}. First, by substituting $\rho_1 = \rho_2 = 0,\lambda_1=1$ in \eqref{rho4} we find $p_1 = 0.5(P + r_1^{-1} - d_1^{-1})$ and $q_1 = 0.5(P + d_1^{-1} - r_1^{-1})$, and  $(p_1,q_1) \succeq \mathbf{0}$ require $P \geq \mathsf{P}^{\prime}_{\sf d,d}$ and $P \geq \mathsf{P}_{\sf d,d}$. The condition for $\rho_3 > 0$, found by substituting $\rho_4 = 0, \lambda_1=1$ in \eqref{rho4}, requires $P \leq \hat{\mathsf{P}}^{\prime}_{\sf d,d}$.  Also, $\boldsymbol{\lambda} \in \mathcal{L}_1$ (\ied $\Sigma_{\mathsf{R}} < \Sigma_{\mathsf{D}}$) requires $P < \mathsf{P}_{\sf rd,r}$. Thus, the conditions for  $\mathcal{A}_{\sf rd,r}$ are given by $P \geq \max( \mathsf{P}_{\sf d,d}, \mathsf{P}^{\prime}_{\sf d,d})$, and $P \leq \min(\hat{\mathsf{P}}^{\prime}_{\sf d,d}, \mathsf{P}_{\sf rd,r})$.

In $\mathcal{S}_{\sf rd,r}$, $\boldsymbol{\rho} \in \mathcal{I}_3 \cap \mathcal{J}_1$ still requires $ p_1 \geq 0, q_1 \geq 0, p_2 = 0, q_2 = P$, but $\boldsymbol{\lambda} \in \mathcal{L}_2$ now requires $\Sigma_{\mathsf{R}} = \Sigma_{\mathsf{D}}$, from which we solve for $\lambda_1$. Then, using $\lambda_1$ and $\rho_1 = \rho_2 = 0$ in \eqref{rho4}, we find $p_1 = P - r_1^{-1}( \gamma/(1+Pr_2) - 1)$ and $q_1 = r_1^{-1}( \gamma/(1+Pr_2) - 1)$, and $(p_1,q_1) \succeq \mathbf{0}$ require $\mathsf{P}_{\sf d,r} > P > \mathsf{P}_{\sf r,r}$. Conditions \eqref{sum_lam} and $\boldsymbol{\lambda} \in \mathcal{L}_2$ simplify to $1 > \lambda_1 > 0$ which requires $P > \mathsf{P}_{\sf rd,r}$ whereas the condition for $\rho_3 > 0$ requires $P < \overbar{\mathsf{P}}_{\sf rd,r}$. Thus, the conditions for  $\mathcal{S}_{\sf rd,r}$ are $\min(\mathsf{P}_{\sf d,r}, \overbar{\mathsf{P}}_{\sf rd,r}) \geq P \geq \max( \mathsf{P}_{\sf r,r}, \mathsf{P}_{\sf rd,r})$.

{LGR  $\mathcal{S}_{\sf rd,rd}$:}
Here, $\boldsymbol{\rho} \in \mathcal{I}_3 \cap \mathcal{J}_3$, \ied $\boldsymbol{\rho} = \mathbf{0}$, and $\boldsymbol{\lambda} \in \mathcal{L}_2$: this require $\Sigma_{\mathsf{R}} = \Sigma_{\mathsf{D}}$, from which we solve for $\lambda_1$. Conditions \eqref{sum_lam} and $\boldsymbol{\lambda} \in \mathcal{L}_2$ simplify to $1 > \lambda_1 > 0$ which requires $P > \mathsf{P}_{\sf rd,rd}$. Using the expression of $\lambda_1$ and $\boldsymbol{\rho} = \mathbf{0}$ in \eqref{rho4}, we find $q_2$ and $q_1$ as in the last and third to last rows of Table \ref{table_thm_1}. From \eqref{apc25a} we have $p_k= P -  q_k$, and $p_k > 0, k=1,2,$ requires $P > \max(\overbar{\mathsf{P}}_{\sf r,rd}, \overbar{\mathsf{P}}_{\sf rd,r})$. Finally, depending on the relay link gains, condition $q_k > 0, k=1,2,$ simplify to either $\max(\overbar{\mathsf{P}}_{\sf rd,d}, \overbar{\mathsf{P}}_{\sf d,rd}) < P$ for $\boldsymbol{r} \in \mathcal{R}_2 \cup \mathcal{R}_1$,  $\overbar{\mathsf{P}}_{\sf rd,d} < P < \overbar{\mathsf{P}}_{\sf d,rd}$ for $\boldsymbol{r} \in \mathcal{R}_{S2}$, or  $\overbar{\mathsf{P}}_{\sf d,rd} < P < \overbar{\mathsf{P}}_{\sf rd,d}$ for $\boldsymbol{r} \in \mathcal{R}_{S1}$, as in Table \ref{table_thm_1}. 

The optimal powers and conditions for $\mathcal{A}_{\sf rd,d}, \mathcal{S}_{\sf rd,d}, \mathcal{A}_{\sf r,rd}$ and $\mathcal{S}_{\sf r,rd}$ are derived from $\mathcal{A}_{\sf d,rd}, \mathcal{S}_{\sf d,rd}, \mathcal{A}_{\sf rd,r}$ and $\mathcal{S}_{\sf rd,r}$ by exchanging the roles of the direct and relay links, while those for $\mathcal{A}_{\sf d,d}$, $\mathcal{A}_{\sf d,r}$, $\mathcal{A}_{\sf r,d}$ and $\mathcal{A}_{\sf rd,rd}$  are derived through similar tedious algebraic manipulations. The details are omitted here.

\section*{Acknowledgment}

The authors are grateful to Meysam~Shahrbaf~Motlagh for improving the presentation of this paper.




\renewcommand{\baselinestretch}{1}
\bibliographystyle{IEEEtran}
\bibliography{IEEEabrv,majhi_bibfile}
\end{document}